%% file: QI-review.tex
\documentclass[10pt,journal,compsoc]{IEEEtran}

\ifCLASSOPTIONcompsoc
\usepackage[nocompress]{cite}
\else
\usepackage{cite}
\fi

\usepackage[table]{xcolor}
\usepackage{longtable}
\usepackage{makecell}
\usepackage{booktabs}
\usepackage{csquotes}
\usepackage{tabularx}
\usepackage{rotating}
\usepackage{multirow, hhline}
\usepackage{color} 
\usepackage{amssymb}
\usepackage[listings,skins,breakable]{tcolorbox}
\usepackage{threeparttable} 
\usepackage{caption}
\usepackage{standalone}
\usepackage{pgfplots}
\usepackage{url}
\usepackage{colortbl}
\usepackage{stfloats}
\usepackage{bm}

\newcommand{\rev}[1]{\textcolor{black}{#1}}
\definecolor{steel}{rgb}{0, 0.2, 0.9} 
\ifCLASSINFOpdf
\else
\fi
\newcolumntype{P}[1]{>{\centering\arraybackslash}m{#1}}
\newcolumntype{Y}{>{\centering\arraybackslash}X}

\newtheorem{relation}{Relation}

\begin{document}
	%
	
	
	\title{How to Evaluate Solutions in Pareto-based Search-Based Software Engineering? A Critical Review and Methodological Guidance}
	
	
	
	
	\author{
		Miqing~Li,~\IEEEmembership{}
		Tao~Chen,~\IEEEmembership{}
		Xin Yao,~\IEEEmembership{Fellow, IEEE} %
		
		\IEEEcompsocitemizethanks{
		    \IEEEcompsocthanksitem Miqing Li and Tao Chen contributed equally to this research. (Corresponding authors: Tao Chen and Xin Yao)
			\IEEEcompsocthanksitem Miqing Li is with the School of Computer Science, University of Birmingham, UK, B15 2TT. (email: m.li.8@bham.ac.uk )
			\IEEEcompsocthanksitem Tao Chen is with the Department of Computer Science, Loughborough University, UK, LE11 3TU. (email: t.t.chen@lboro.ac.uk)
			\IEEEcompsocthanksitem Xin Yao is with the Shenzhen Key Laboratory of Computational Intelligence (SKyLoCI), Department
			of Computer Science and Engineering, Southern University of Science and Technology, Shenzhen, P. R. China, and CERCIA, School of Computer Science, University of Birmingham, UK, B15 2TT. (email: xiny@sustc.edu.cn)
		}
		\thanks{}}

	\markboth{Manuscript Accepted by IEEE Transactions on Software Engineering, 2020}%
	{Shell \MakeLowercase{\textit{et al.}}: Bare Demo of IEEEtran.cls for Computer Society Journals}
	
	\IEEEtitleabstractindextext{%
		\begin{abstract}
			With modern requirements, there is an increasing tendency of considering multiple objectives/criteria simultaneously in many Software Engineering (SE) scenarios.
			Such a multi-objective optimization scenario comes with an important issue --- 
			how to evaluate the outcome of optimization algorithms, which typically is a set of incomparable solutions (i.e., being Pareto nondominated to each other).
			This issue can be challenging for the SE community, 
			particularly for practitioners of Search-Based SE (SBSE).
			On one hand, 
			multi-objective optimization could still be relatively new to SE/SBSE researchers,
			who may not be able to identify the right evaluation methods for their problems.
			On the other hand,
			simply following the evaluation methods for general multi-objective optimization problems may not be appropriate for specific SBSE problems,
			especially when the problem nature or decision maker's preferences are explicitly/implicitly known.
			This has been well echoed in the literature by various inappropriate/inadequate selection and inaccurate/misleading use of evaluation methods. 
			In this paper, we first carry out a systematic and critical review of quality evaluation for multi-objective optimization in SBSE.
			We survey 717 papers published between 2009 and 2019 from 36 venues in seven repositories, 
			and select 95 prominent studies, 
			through which we identify five important but overlooked issues in the area.
			We then conduct an in-depth analysis of quality evaluation indicators/methods and general situations in SBSE, 
			which, together with the identified issues, 
			enables us to codify a methodological guidance for selecting and using evaluation methods in different SBSE scenarios.
			 
		\end{abstract}

		}

	\maketitle

	\IEEEdisplaynontitleabstractindextext

	%
	\IEEEpeerreviewmaketitle

	\IEEEraisesectionheading{\section{Introduction}\label{sec:introduction}}
	
\IEEEPARstart{I}{n} software engineering (SE), 
	it is not uncommon to face a scenario where multiple objectives/criteria need to be considered simultaneously~\cite{Harman2012,DBLP:journals/corr/abs-2001-08236}.
	In such scenarios, 
	there is usually no single optimal solution but rather a set of Pareto optimal solutions 
	(termed a Pareto front in the objective space), 
	i.e., solutions that cannot be improved on one objective without degrading on some other objective.
	To tackle these multi-objective SE problems, 
	different problem-solving ideas have been brought up. 
	One of them is to generate a set of solutions to approximate the Pareto front. 
	This, in contrast with the idea of aggregating objectives (by weighting) into a single-objective problem,
	provides different trade-offs between the objectives, 
	from which the decision maker (DM) can choose their favorite solution. 
	
	In such Pareto-based optimization, 
	a fundamental issue is to evaluate the quality of solution sets (populations)
	obtained by computational search methods (e.g., greedy search, heuristics, and evolutionary algorithms) in order to know how well the methods perform.
	Since the obtained solution sets are typically not comparable to each other with respect to Pareto dominance\footnote{ 
		A solution set $\mathbf{A}$ is said to (Pareto) dominate a solution set $\mathbf{B}$ if for any solution in $\mathbf{B}$ there exists at least one solution in $\mathbf{A}$ dominating it, where the dominance relation between two solutions can be seen as a natural ``better'' relation of the objective vectors, i.e., better or equal on all the objectives, and better at least on one objective~\cite{Zitzler2003}.}, how to evaluate/compare them is non-trivial.
	A straightforward way is to plot the solution sets (by scatter plot) for an intuitive evaluation/comparison. 
	Yet this only works well for the bi-objective case, 
	and when the number of objectives reaches four, 
	it is impossible to show the solution sets by scatter plot.
	More importantly, 
	visual comparison cannot provide a quantitative comparative result between the solution sets. 
	
	Another way to evaluate the solution sets is to report their descriptive statistical results, 
	such as the best, mean, and median values on each objective from each solution set.  
	This has been profoundly used in Search-Based SE (SBSE)~\cite{Bavota2012Putting,DBLP:journals/infsof/ChenLY19,Bowman2010Solving,Fleck2017Model,Wada2012E,DBLP:conf/wosp/0001BWY18,DBLP:journals/jss/SobhyMBCK20}.
	However,
	some of these statistic indexes may easily give misleading evaluation results.
	That is, 
	a solution set which is evaluated better than its competitor could be never preferred by the DM under any circumstance. 
	This will be explained in detail in the text later (Section~\ref{sec:doe-issue}).

	Generic quality indicators, 
	which is arguably the most straightforward evaluation method that maps a solution set to a real number that indicates one or several aspects of the set's quality,
	have emerged in the fields of evolutionary computation and operational research \cite{Sayin2000,Zitzler2003,Knowles2006,Bozkurt2010}.
	Today analyzing and designing quality indicators has become an important research topic. 
	There are hundreds of them in literature~\cite{li2019quality},
	with some measuring closeness of the solution set to the Pareto front,
	some gauging diversity of the solution set,
	some considering a comprehensive evaluation of the solution set, 
	etc. 
	The SBSE community benefits from this prosperity.
	A common practice in SBSE is to use some well-established quality indicators,
	such as hypervolume ($HV$)~\cite{Zitzler1998} and inverted generational distance ($IGD$)~\cite{Coello2004},
	to evaluate the obtained solution sets.
	However, 
	some indicators may not be appropriate when it comes to practical SE optimization scenarios. 
	For example,
	since the Pareto front of a practical SBSE problem is typically unavailable,
	indicators that require a reference set that well represents the problem's Pareto front
	may not be well suited~\cite{li2019quality}, 
	such as $IGD$.  
	
	More importantly, 
	specific SBSE problems usually have their own nature and requirements.
	Simply following indicators that were designed for general Pareto-based optimization may fail to reflect these requirements.
	Take the software product line configuration problem as an example.
	In this problem, 
	the objective of a product's correctness is always prioritized above other objectives 
	(e.g., cost and richness of features).  
	Equally rating these objectives by using generic indicators like $HV$ 
	(which in fact has been commonly practiced in the literature~\cite{Sayyad2013Optimum,Sayyad2013On,Olaechea2014Comparison,DBLP:conf/icpads/KumarBCLB18})
	may return the DM meaningless solutions, 
	i.e., invalid products with good performance on the other objectives. 
	This situation also applies to the test case generation problem,
	where the DM may first favor the full code coverage and then others (e.g., low cost). 
	
	Moreover, 
	some SBSE problems may associate with the DM's explicit/implicit assumptions or preferences between the objectives.
	It is expected for researchers to select indicators bearing these assumptions/preferences in mind.
	For instance,
	in many SE scenarios, 
	the DM may prefer well-balanced trade-off solutions 
	(i.e., knee points on the Pareto front) between conflicting objectives. 
	An example is that when optimizing the conflicting non-functional quality of a software system 
	(e.g., latency and energy consumption), 
	knee points are typically the most preferred solutions, 
	as in such case, it is often too difficult, 
	if not impossible, 
	to explicitly quantify the relative importance between objectives.
	Under this circumstance, 
	quality indicators 
	that treat all points on the Pareto front equally (such as $IGD$)
	may not be able to reflect this preference, 
	despite the fact that they have been frequently used in such scenarios~\cite{Fleck2017Model,Mkaouer2016On}.

	Finally, 
	the study of quality indicator selection itself in multi-objective optimization 
	is in fact a non-trivial task.
	Each indicator has its own specific quality implication, 
	and the variety of indicators in literature can easily overwhelm the researchers and practitioners in the field.
	On the one hand,
	an accurate categorization of quality indicators is of high importance.
	Failing to do so can easily result in a misleading understanding of search algorithms' behavior, 
	see~\cite{li2018critical}. 
	On the other hand,
	even under the same category, 
	different indicators are of distinct quality implications,
	e.g., 
	$IGD$ prefers uniformly distributed solutions and 
	$HV$ is in favor of knee solutions. 
	A careful selection needs to be made to ensure the considered quality indicators 
	to be in line with the DM's preferences.
	In addition,
	many quality indicators involve critical parameters 
	(e.g., the reference point in the $HV$ indicator). 
	It remains unclear how to properly set these parameters 
	under different circumstances,
	particularly in the presence of the DM's preferences.

	Given the above,
	this paper aims to systematically survey and justify some of the overwhelming issues when evaluating solution sets in SBSE, and more importantly, to provide a systematic and methodological guidance of selecting/using evaluation methods and quality indicators 
	in various Pareto-based SBSE scenarios. 
	Such a guidance is of high practicality to the SE community, 
	as research from the well-established community of multi-objective optimization may still be relatively new to SE researchers and practitioners. This is, to the best of our knowledge, the first work of its kind to specifically target the quality evaluation of solution sets in SBSE based on a theoretically justifiable methodology. 
	
	\rev{It is worth mentioning that recently there are some attempts from the perspective of empirical studies 
	to provide guidelines for quality indicator selection~\cite{wang2016practical,Ali2020quality}.
	Wang et al.~\cite{wang2016practical} proposed a practical guide for SBSE researchers 
	based on the observations from experimental results in three SBSE real-world problems.
	Ali et al.~\cite{Ali2020quality} significantly extended that work and provided a set of guidelines 
	based on the observations from experimental results in nine SBSE problems
	from industrial, real-world and open-source projects.
	However,
	observations drawn from an empirical investigation 
	on specific SBSE scenarios may not be generalizable.
	Indeed, different DMs may prefer different trade-offs between objectives, 
	even for the same optimization problem, 
	as nondominated solutions are in essence incomparable. 
	Observations obtained on one (or some) scenario(s) is therefore difficult to be transferred into other scenarios.
	As a result, 
	a general and theoretically sound guidance based upon the DM's preferences is needful 
	since the fundamental goal of multi-objective optimization is to supply 
	the DM a set of solutions which are the most consistent with their preferences.}

	\rev{For the rest of the paper, 
	we start by providing some background knowledge of multi-objective optimization and quality evaluation (Section 2).
	Then, we conduct a systematic survey of the SE problems 
	that involve Pareto-based search (hence termed Pareto-based SBSE problems) across all phases 
	in the classic Software Development Life Cycle (SDLC)~\cite{ruparelia2010software}, 
	along with their problem nature, the DM's preferences, the quality indicators and evaluation methods used (Sections 3 and 4). 
	The survey has covered 717 searched papers published between 2009 and 2019, 
	on 36 venues from seven repositories, 
	leading to 95 prominent primary studies in the SBSE community.
	This is followed by a critical review on the evaluation method selection and use in those primary studies, 
	based on which we identify five important issues that have been significantly overlooked (Section 5).
	Then, 
	we carry out an in-depth analysis of frequently-used quality indicators in the area (Section 6), 
	in order to make it clear which indicators fit in which situation.
	Next, to mitigate the identified issues in the future work of SBSE,
	we provide a methodological guidance and procedure of selecting, 
	adjusting, and using evaluation methods in various SBSE scenarios (Section 7).
	The last three sections are devoted to threats to validity, related work, and conclusion, respectively.}

	\section{Preliminaries on Multi-Objective Optimization}
	\label{sec:bg}
	
	\rev{Multi-objective optimization is an optimization scenario 
	that considers multiple objectives/criteria simultaneously.
	Apparently, 
	when comparing solutions\footnote{For simplicity,
		we refer to an objective vector as a \textit{solution} 
		and the outcome of a multi-objective optimizer as a \textit{solution set}.} in multi-objective optimization, 
	we need to consider all the objectives of a given optimization problem. 
	There are two commonly used terms to define the relations between solutions, Pareto dominance and weak Pareto dominance.}
	
	\rev{Without loss of generality, 
	let us consider a minimization scenario. 
	For two solutions $\mathbf{a}, \mathbf{b} \in Z$ 
	($Z \subset \mathbb{R}^m$, where $m$ denotes the number of objectives),  
	solution $\mathbf{a}$ is said to \textit{weakly dominate} $\mathbf{b}$ 
	(denoted as $\mathbf{a}\preceq \mathbf{b}$) 
	if $\mathbf{a}_i \leq \mathbf{b}_i$ for $1 \leq i \leq m$.
	If there exists at least one objective $j$ on which $\mathbf{a}_j < \mathbf{b}_j$, 
	we say that $\mathbf{a}$ \textit{dominates} $\mathbf{b}$ (denoted as $\mathbf{a}\prec \mathbf{b}$). 
	A solution $\mathbf{a} \in Z$ is called \textit{Pareto optimal} 
	if there is no $\mathbf{b} \in Z$ that dominates $\mathbf{a}$. 
	The set of all Pareto optimal solutions of a multi-objective optimization problem is called its \textit{Pareto front}.}
	
	\rev{The above relations between solutions can immediately be extended to between sets. 
	Let $\mathbf{A}$ and $\mathbf{B}$ be two solution sets.}
	
	\rev{\begin{relation}\textit{[Dominance between two sets~\cite{Zitzler2003}] We say that $\mathbf{A}$ \textit{dominates} $\mathbf{B}$ 
			(denoted as $\mathbf{A}\prec \mathbf{B}$) 
			if for every solution $\mathbf{b}\in \mathbf{B}$ there exists at least one solution $\mathbf{a}\in \mathbf{A}$ 
			that dominates $\mathbf{b}$.}
	\end{relation}}
	
	\rev{\begin{relation}\textit{[Weak Dominance between two sets~\cite{Zitzler2003}] We say that $\mathbf{A}$ \textit{weakly dominates} $\mathbf{B}$ 
			(denoted as $\mathbf{A}\preceq \mathbf{B}$) 
			if for every solution $\mathbf{b}\in \mathbf{B}$ there exists at least one solution $\mathbf{a}\in \mathbf{A}$ 
			that \textit{weakly dominates} $\mathbf{b}$.} 
	\end{relation}}

	\rev{We can see that the \textit{weak dominance} relation between two sets does not rule out their equality, 
	while the \textit{dominance} relation does but it also rules out the case that there exist same solutions with respect to the two sets. 
	Thus, 
	we may need another relation to define that $\mathbf{A}$ is \textit{generally} better than $\mathbf{B}$.}
	
	\rev{\begin{relation}\textit{[Better relation between two sets~\cite{Zitzler2003}] We say that $\mathbf{A}$ is \textit{better} than $\mathbf{B}$ (denoted as $\mathbf{A} \vartriangleleft \mathbf{B}$) 
			if for every solution $\mathbf{b}\in \mathbf{B}$ there exists at least one solution $\mathbf{a}\in \mathbf{A}$ 
			that \textit{weakly dominates} $\mathbf{b}$, 
			but there exists at least one solution in $\mathbf{A}$ 
			that is not weakly dominated by any solution in $\mathbf{B}$.}
	\end{relation}}

	\rev{The \textit{better} relation $\vartriangleleft$ reflects the most general assumption 
	of the DM's preferences to compare solution sets.
	However, 
	the \textit{better} relation may leave many solution sets incomparable 
	since it is very likely that there exist some solutions being nondominated with each other in the sets. 
	As typically the size of the Pareto front of a multi-objective optimization problem 
	can be prohibitively large or even infinite, 
	a solution set that can well represent the Pareto front is preferred, 
	especially when the DM's preferences are unavailable.
	This leads to four quality aspects that we often care about~\cite{li2019quality} --- 
	Convergence, how close the solution set is to the Pareto front;
	Spread, how large the region that the set covers is;
	Uniformity, how evenly the solutions are distributed in the set;  
	Cardinality, how many (unique) nondominated solutions are in the set.
	Over the last three decades, 
	numerous quality evaluation methods have been developed for these four aspects.
	Among them, quality indicators are the most popular ones~\cite{li2019quality}.
	They typically map a solution set to a real number 
	that indicates one or more of the four quality aspects, 
	defining a total order among solution sets for comparison.}

	\section{Review Methodology}
	
	Despite that we have randomly witnessed several inappropriate evaluations of solution sets through our own experiences working in SBSE, 
	the key challenge in this work remains \textit{to systematically understand what the trends of issues on the way to evaluate solution sets in the SBSE community are, if any,} so that a clarified guidance can be drawn thereafter. To this end, \rev{we at first conduct a systematic literature review covering the studies published between 2009 and 2019. 
		A reason that we consider this period is that one of the most well-known SBSE surveys (Harman et al.~\cite{Harman2012}) has covered the SBSE work from 1976 to 2008, 
		and we try to cover the period that has not been reviewed by that work. 
		In addition, 
		since 2010 or so, 
		there is a rapidly increasing interest in using Pareto-based optimization techniques to deal with SBSE. 
		Given these two reasons, we have chosen 2009 as the starting year of our review.
		Having said that, 
		we do not aim to provide a complete review
		on all parts of the SBSE work, but specifically on the aspects related to the major
		trends of evaluating solution sets.
}
	
	Our review methodology follows the best practice of a systematic literature review for software engineering~\cite{Kitchenham2009Systematic}, consisting of search protocol, 
	inclusion/exclusion criteria, 
	formal data collection process, 
	and pragmatic classification. 
	Specifically, the review aims to answer three research questions (RQs): 
	
	\begin{itemize}
	    \item \textbf{RQ1:} What evaluation methods have been used to evaluate solution sets in SBSE? (\textbf{What})
	    \item \textbf{RQ2:} What are the reasons and practice of using the generic quality indicators?  (\textbf{Why} and \textbf{How})
	    \item \textbf{RQ3:} In what domain and context the evaluation methods have been used? (\textbf{Where})
	\end{itemize}
	
	
	\begin{figure}[t!]
		\centering
		\includegraphics[width=\columnwidth]{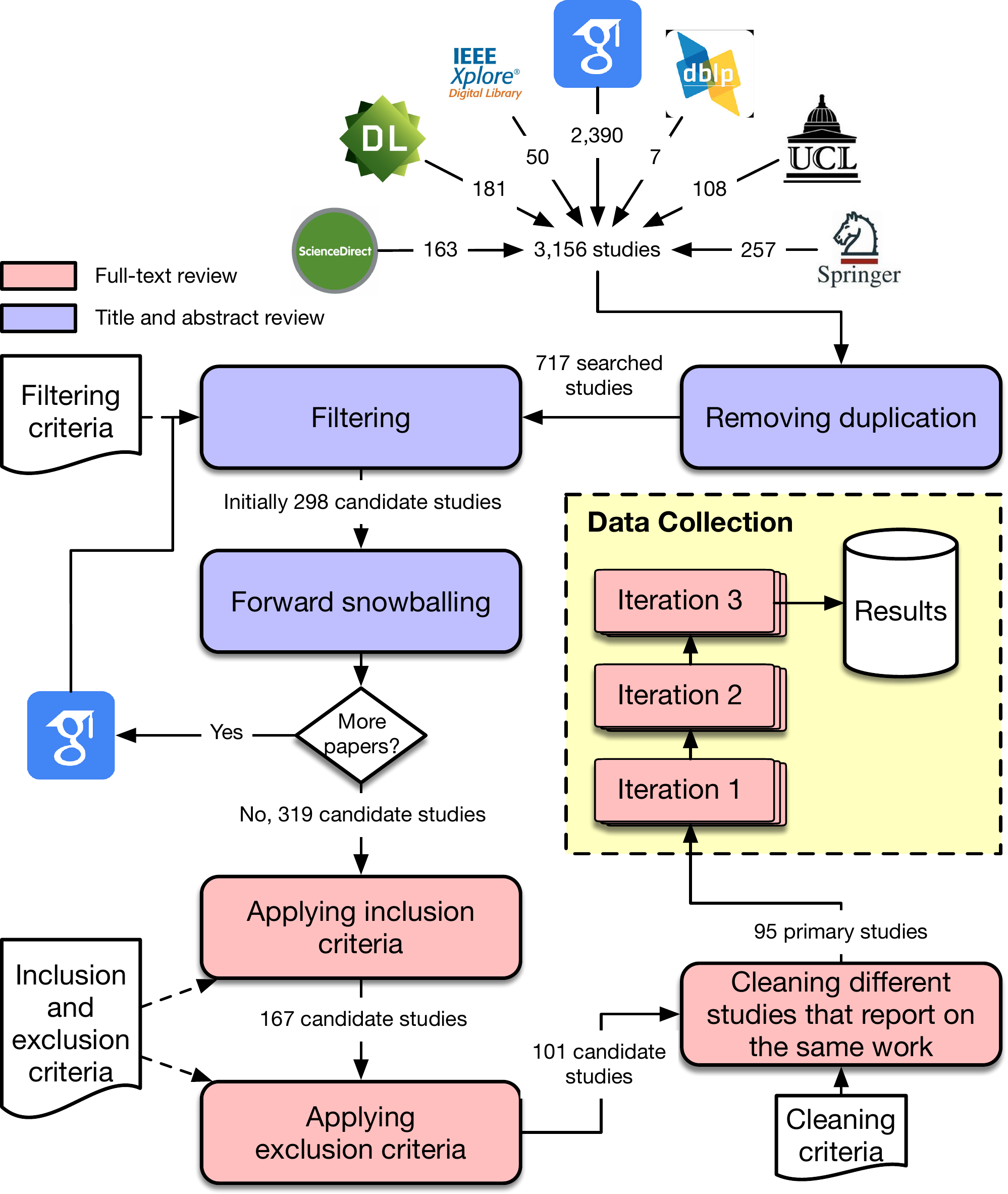}
		\caption{Systematic literature review protocol.}
		\label{Fig:slr}
	\end{figure}

	\subsection{Overview of the Literature Review Protocol}
	
	\rev{As shown in Figure~\ref{Fig:slr}, our literature review protocol obtains inputs from various sources via automatic search, which we will elaborate into details in Section~\ref{sec:scope}. This gives us 3,156 studies including duplication. We then removed any duplicated studies by automatically matching their titles\footnote{Patents, citation entries, inaccessible papers, and any other non-English documents were also eliminated.}, leading to 717 \textbf{\emph{searched studies}}. Next, we filtered the searched studies by reading through their titles and abstracts using two simple filtering criteria:}
	\begin{itemize}
	  \item \rev{The paper is not relevant to SBSE.}
	  \item \rev{The paper does not investigate or compare multi-objective search/optimization.}
	\end{itemize}
	
	\rev{A study was ruled out if it meets any of the above two filtering criteria. The aim of filtering is to reduce the found studies to a much smaller and more concise set, namely the \textbf{\emph{candidate studies}}. As can be seen, the process resulted in 298 candidate studies prior to manual search. Starting from the 298 studies, we adopted an iterative forward snowballing as suggested by Felizardo et al.~\cite{DBLP:conf/esem/FelizardoMKSV16}, where the newly included studies (after filtering) were placed into the next snowballing round. The reason why we did not do backward snowballing is 
		because we have set strict time scale on the studies searched within the last decade, and thus backward snowballing would too easily violate such a requirement of timeliness. To avoid a complicated process, we relied on Google Scholar as the single source for forward snowballing, as it returns most of the searched results as shown in Figure~\ref{Fig:slr} and has been followed by software engineering surveys~\cite{DBLP:journals/tse/GalsterWTMA14}. This snowballing stopped when no new studies can be found and it eventually led to 319 candidate studies, upon which the procedure for the full-text review begins.}
	
	\rev{At the next stage, we reviewed all the 319 studies and temporarily included studies using the inclusion criteria from Section~\ref{sec:in-ex}, which resulted in 167 candidate studies. We then applied the exclusion criteria (see Section~\ref{sec:in-ex}) to extract the temporarily included studies, leading to 101 candidate studies. By using the cleaning criteria specified in Section~\ref{sec:in-ex}, a further cleaning process was conducted to prune different studies that essentially report on the same work, e.g., conference extended journal papers. All the processes finally produced 95 \textbf{\emph{primary studies}} for data analysis and collection.}
	
	\rev{On these 95 primary studies, we conducted systematic and pragmatic data collection via three iterations, whose details are given in Sections~\ref{sec:item-class} and~\ref{sec:data-collection}. The summarized results were reported thereafter.}

	
	\subsection{Scope and Search String}
	\label{sec:scope}
	
	\rev{From 12th to 19th Aug 2019, we conducted an automatic search over a wide range of scientific literature sources, including ACM Library, IEEE Xplore, ScienceDirect, SpringerLink, Google Scholar, DBLP and the SBSE repository maintained by the CREST research group at UCL\footnote{http://crestweb.cs.ucl.ac.uk/resources/sbse\_repository}.}

	We used a search string that aims to cover a variety of problem nature and application domains with respect to multi-objective optimization. Synonyms and keywords were properly linked via logical operators (AND, OR) to build the search term. The final search string is shown as below:
	
	\begin{tcolorbox}[left=5pt,right=5pt,top=5pt,bottom=5pt]
	\begin{displayquote} 
		\emph{(``multi objective" OR ``multi criteria" OR ``Pareto based" OR ``non dominated" OR ``Pareto front") AND ``search based software engineering" AND optimization}
	\end{displayquote}
	\end{tcolorbox}
	
	\rev{We conducted a full-text search on ACM Library, IEEE Xplore, ScienceDirect, SpringerLink, and Google Scholar, but rely on searching the title only for DBLP and UCL's SBSE repository, due to their restricted feature. Since DBLP's search feature cannot handle the whole search string, we paired each term in the first bracket with ``\texttt{search based software engineering}" to run the search independently and collect all results returned. We omitted ``\texttt{optimization}" as it rarely appears together with ``\texttt{search based software engineering}" in the title, and having it along would produce many irrelevant results. Due to the similar reason, for the UCL's SBSE repository, we searched each term from the first bracket independently, as it is known that all the studies in this source are SBSE related.}
	
	\rev{On all the sources except DBLP and UCL's SBSE repository, we tried two versions of the search string: one with a hyphen between the commonly used terms (e.g., ``\texttt{multi(-)objective}") and another without. The returned results with the highest number of items were used\footnote{This is because their results are not mutually exclusive, e.g., on Google Scholar, ``\texttt{multi objective}" would also return all the studies that contain ``\texttt{multi-objective}" but not the other way around.}. In particular, when searching on UCL's SBSE repository, the results of these two versions were combined, for example, ``\texttt{multi objective}" and ``\texttt{multi-objective}" would lead to different results. We recorded all the results returned under semantically equivalent terms.}

	\subsection{Inclusion, Exclusion, and Cleaning Criteria}
	\label{sec:in-ex}
	For all the candidate studies identified, we first extract the primary studies by using the inclusion criteria as below; studies meeting all of the criteria were temporarily chosen as the primary studies:
	
	\begin{enumerate}
		
		\item The study primarily focuses on (or has a section that discusses) 
		a Pareto-based multi-objective solution to the SBSE problem. 
		This means we do not consider papers that utilize the multi-objective treatment that relies on objective aggregation (e.g., weighted sum), unless they have explicitly compared the solution against a Pareto-based multi-objective solution. This is reasonable as if a clear aggregation of objectives can be defined, then there would be almost no need to select quality indicators but rely on the said aggregation to obtain a utility value for comparison.
		
		\item The study explicitly or implicitly discusses (or at least makes assumptions about) the DM's preferences/contextual information between the objectives for the SBSE problem in hand. \rev{By implicit discussion, 
			we mean that the study does not clearly state the assumptions, 
			but such assumptions can be easily interpreted from the paper. 
			For example, 
			in the software product line configuration problem, 
			some studies do not explicitly study the assumptions, 
			but the number of valid products (one objective to be optimized) is solely used as an indicator to compare the peer approaches, 
			which gives a clear indication that it is more important than the other objectives.} 
		Note that this also includes the assumption of no preferences/contextual information.


		\item The SBSE problem in hand can be framed into at least one phase of the classic SDLC~\cite{ruparelia2010software}.
		
		\item The study uses at least one search algorithm to solve the problem.
		
		\item The study includes quantitative experimental results with clear instructions on how the results were obtained.
		
		\item The study uses at least one method to evaluate the experimental results. 
		
	\end{enumerate}
	
	Subsequently, studies meeting any of the exclusion criteria below are filtered out from the temporary primary studies:
	
	\begin{enumerate}
		
		\item The study neither explicitly nor implicitly mentions SBSE, where the computational search is the key.
		
		\item The study is not ``highly visible'' or widely followed. We used the citation information from Google Scholar as a single metric to (partially) assess the impact of a study\footnote{Admittedly, 
			there is no metric that is able to well quantify the impact of a paper.
			Nevertheless, 
			the citation count can indicate something about a paper, e.g., its popularity.
		}. In particular, we follow a pragmatic strategy that: a study has 5 citations per year from its year of publication is counted in, e.g., a 2010 study would expect to have at least 45 citations\footnote{All the citations were counted by 23rd Nov 2019.}. The only exception is for the work published in the year of writing this article (i.e., 2019), we consider those that were published for shorter than 6 months and have been cited by more than once, together with the pre-press ones that have not yet been given an issue number regardless of their citation counts. The reasons behind this setting are three-folds:  	
	   
	  	\rev{\hspace{1em} (a) We do not attempt to provide a comprehensive survey on the whole SBSE field, but rather to gather the major trends on how solution sets are evaluated, which can at least provide some sources for detailed analysis and discussion. Therefore, some metrics are required to ensure a trade-off between the trend coverage and a reasonably required effort for detailed data collections. This is similar to a sampling of the literature with the aim to gather the ``representative'' samples.
	  		This approach was adopted by many works, such as~\cite{DBLP:journals/infsof/FuMS16}
	  		where they used the citation count from Google Scholar as a threshold to select studies for review, 
	  		as we did in this work.}

	  	\rev{\hspace{1em} (b) It is not uncommon to see that software engineering surveys are conducted using some metrics to measure the ``impact" of a work. For example, some restrict their work only at what the authors believe to be premium venues~\cite{DBLP:journals/tse/GalsterWTMA14}, others use a threshold on the impact factors of the published journals, e.g., Cai and Card~\cite{cai2008analysis} use $0.55$ and Zou et al.~\cite{8466000} use $2.0$. 
	  		In our case, it may not be a best practice to apply a metric at the venue level as the SBSE work is often multi-disciplinary (as we will show in Table~\ref{table:papers-count}) --- it is difficult to quantify the ``impact" across communities. We, therefore, have taken a measurement at the paper level based on the citation counts from Google Scholar, 
	  		which has been used as the sole metric to differentiate between the studies in some prior work~\cite{ten-year-sbse,DBLP:journals/tse/GalsterWTMA14,DBLP:journals/infsof/FuMS16}.}
		   
				\rev{\hspace{1em} (c) Indeed, 
				there is no rule to set the citation threshold. 
				The settings in this work were taken from the (rounded) average figure within the population of the candidate studies. These may seem very high at the first glance probably due to two reasons: (i) by publication date, we meant the official date that the work appears on the publisher's webpage (for journal work, this means it has been given an official issue number). Yet, it is not uncommon that many studies are made citable as pre-prints before the actual publication, 
				e.g., ICSE often has around 6 months gap between notification and official publication, 
				and there is an even larger gap for some journals. 
				This has helped to accumulate citations. (ii) Google Scholar counts the citations made by any publicly available documents and self-citation, which can still be part of the impact but implies their citation count may be higher 
				than those purely made by peer-reviewed publications. 
				Nevertheless, this could indeed pose a threat of construct validity, which we will discuss in Section~\ref{sec:tov}.}
		
		\item The study is a short paper, i.e., shorter than 8 pages (double column) or 15 pages (single column).
		
		\item The study is a review, survey, or tutorial.
		
		\item The study is published in a non-peer-reviewed public venue, e.g., arXiv.
		
	\end{enumerate}
	
	Finally, if multiple studies of the same research work are found, we applied the following cleaning criteria to determine if they should all be considered. The same procedure is applied if the same authors have published different studies for the same SBSE approach, and thereby only significant contributions are analyzed for the review.
	
	\begin{itemize}
		\item All studies are considered if they report on the same SBSE problem but different solutions.
		
		\item All studies are considered if they report on the same SBSE problem and solutions but have different assumptions about the DM's preference, nature of the problem, or new findings of the problem.
		
		\item When the above two points do not hold, only the latest version or the extended journal version is considered.
		
	\end{itemize}
	
	\input{Tables/data-item}
	
	\subsection{Data Items Analysis and Classification Strategy}
	\label{sec:item-class}          
	
	\rev{The items to be collected when reviewing the details of the primary studies have been shown in Table~\ref{tb:items}. We now describe their design rationales and the procedure to extract and classify the data from each item.}
	
	\rev{The data for $I_1$ to $I_5$ is merely used as the meta-information of the primary studies. $I_6$, which answers \textbf{RQ1}, is the key item of our review. The evaluation method(s) used can be easily identified in a study, most commonly at the \textit{Experiment} section. In general, apart from identifying the evaluation methods used in each study, we also seek to classify them into the following four categories:}

	\begin{itemize}
	  \item \rev{\textit{Generic Quality Indicator:} This refers to indicators that are designed to evaluate the quality of solution sets for generic multi-objective optimization problems (e.g., $HV$, $IGD$ and $Spread$), as documented by Li and Yao~\cite{li2019quality}. Formally, a quality indicator is a metric that maps a set of solutions (i.e., solution vectors) to a real number that indicates one or several aspects of the solution set quality~\cite{li2018critical,li2019quality},
	e.g., to indicate how close the set is to the Pareto front and how evenly solutions are distributed in the set.}

	   \item \rev{\textit{Solution Set Plotting ($SSP$):} This is a straightforward way to evaluate solution sets --- visualizing the results by plotting them.}
	   
	  \item \rev{\textit{Descriptive Objective Evaluation ($DOE$):} This resorts to the direct statistical results of objective values, e.g., the best/mean/median of the solution set on each objective.}
	   
	  \item \rev{\textit{Problem Specific Indicator ($PSI$):} This refers to indicators that are not used for generic multi-objective optimization problems, but specifically designed for a given SBSE problem.  
}
	\end{itemize}
	
	\rev{$I_7$ is heavily relevant to $I_6$, but requiring more detailed inspection to the studies. By this means, we aim to collect information about when a generic quality indicator is used, what quality aspect the study seeks to evaluate by it (for \textbf{RQ2}), which is the key reason of why such an indicator is chosen. $I_7$ is classified based on the four quality aspects of a solution set as concluded by Li and Yao~\cite{li2019quality}, i.e., Convergence, Spread, Uniformity, and Cardinality. For each study, we first looked for the section where the generic quality indicators are explained, if no information found, we then searched for every place where the generic quality indicators are mentioned. We classify each indicator into the quality aspects based on whether their keywords have been clearly mentioned, otherwise, the indicator is marked as \textit{Unknown} under $I_7$ of a study.}
	
	\rev{For $I_8$, we wish to understand how the generic quality indicators are used, as some of them requiring a reference point (e.g., $HV$) or a reference Pareto front (e.g., $GD$ and $IGD$) in order to be used correctly (for \textbf{RQ2}). This is again following a similar procedure to that of $I_7$; when no such information can be found for an indicator that requires a reference, we marked \textit{Unknown} under the indicator for $I_8$ of the study.}
	
	\rev{To answer \textbf{RQ3}, $I_9$ is rather straightforward, 
		and understanding it can help us to know whether some evaluation methods are used appropriately, as some of them have limitations in terms of the number of objectives to be optimized. $I_{10}$ is also relatively easy to identify, most commonly from the \textit{Introduction} section and we classify the Pareto-based SBSE problems into the SDLC phases in a classic waterfall model according to~\cite{ruparelia2010software}. Note that we choose this model by no mean to rely on its usefulness, but only because it is one of the oldest models which consists of very generic phases that allow us to showcase the categories of SBSE problems.}
	
	\rev{Finally, to complete \textbf{RQ3}, $I_{11}$ is crucial as it enables us to assess whether the evaluation methods are used correctly, given the DM's preferences over the objectives and/or the contextual information, which is one of the core initiatives of this paper. To classify the preferences and contextual information, we followed a pragmatic classification coding:}
	
	\begin{itemize}
	  \item \rev{\textit{Contextual information: } Every problem has its own nature and characteristics; there is no exception for Pareto-based SBSE problems. In general, such nature and characteristics of the problem form the contextual information, which is precise, clear, and explicitly stated as a fact in a study. For example, in the software product line configuration problem, many studies state that there is no doubt that the correctness objective has higher priority than any others, as an invalid product has no value in practice.}
	  \item \rev{\textit{DM's preferences:} The DMs often have preferences over certain objectives or are able to provide information about their relative importance and expectation. This may be for example, ``\texttt{objective A is preferred as long as objective B has at least reached b}"; or well-balanced solutions (a.k.a. knee solutions) are preferred. When the DM's preferences are aligned with contextual information, they are indeed similar. However, the key difference is that the contextual information is clear, and it is a hard requirement that is well acknowledged on the given Pareto-based SBSE problem regardless of whether the DM explicitly states it or not. In contrast, the DM's preferences are often vague and imprecise, 
	  	and it cannot be generalized to all the scenarios of the given Pareto-based SBSE problem.}
	  \item \rev{\textit{Not specified:} When neither of the above categories can be applied, the preference and contextual information is marked as \emph{Not specified}.}
	   
	\end{itemize}
	
	\rev{Extracting the data for $I_{11}$ focuses on understanding exactly what DM's preferences and contextual information are assumed in each study. This was achieved by inspecting the sections relevant to \textit{Problem Statement} and \textit{Approach Design}. If no information can be found, we then looked for insights from the \textit{Experiment} section. For example, when a single objective, which belongs to part of the search, is explicitly discussed and used to compare the peer approaches in the evaluation, 
		it often reflects the assumptions of contextual information and/or DM's preferences in the study.}
	
	\subsection{Data Collection}
	\label{sec:data-collection}
	
	\rev{For each primary study identified, the data items from Table~\ref{tb:items} were collected and classified based on the coding from Section~\ref{sec:item-class}. The first two authors of this paper reviewed the primary studies independently. The data and classification extracted by one author were checked by the other. Disagreements and discrepancies were resolved by discussing between the two authors or by consulting an additional researcher.}

	\rev{Following the strategy recommended in a recent survey~\cite{8466000}, we adopted three iterations for the data collection process, which are detailed as below:}
		
	\rev{\textit{\underline{Iteration 1:}} This iteration aims to conduct an initial data collection to summarize the data and perform preliminary classification. During the process, a notable difficulty between the authors was that the evaluations using descriptive statistics and problem-specific indicators are hard to be distinguished. This is due to the fact that most of them are not clearly stated in the studies and there is a wide variety of problem-specific indicators across all Pareto-based SBSE problems (we found 34 of them in our review). 
	Therefore, any study, which the authors suspected that these two types of methods might have been used but could not be certain, was placed into a \emph{bin} for further investigation in the next iteration. There were 26 studies in the \emph{bin} when this iteration finished.}

	\rev{\textit{\underline{Iteration 2:}} In this iteration, the two authors checked the data and classification from each other to ensure consistency. A study was discussed during the process if either author has any concern about the data extracted. Any unresolved studies from the \emph{bin} were also checked by the other author again. Particularly, for each study in the \emph{bin}, a common agreement on the descriptive statistics and problem-specific indicators used was reached via either discussion between the authors or counseling external researchers. Further reading to understand the nature of an evaluation method (for problem-specific indicators) was conducted when necessary. Apart from this, other major discussions raised were concerned with certain generic quality indicators, due to two reasons: (i) some studies have indeed used generic quality indicators, but the actual name of the indicator is missing, although some detailed calculation has been provided, e.g.,~\cite{Yoo2010Using,Harman2009Search}; (ii) some other studies have used completely different names to refer to the same quality indicator (or even invented their own name), e.g.~\cite{Zhang2013Empirical,Li2014Robust}. These cases require both the authors to thoroughly inspect the detailed calculation of those indicators before reaching an agreement. Overall, a total of 60 studies were discussed in this iteration.}

	\rev{\textit{\underline{Iteration 3:}} The process of the final iteration is similar to that of \textit{Iteration 1}, but its goal is to eliminate any typo, missing labels, and errors. The extracted data for 11 studies, which contain errors, were corrected during the process.}

\input{Tables/venue}

	\section{Results of the Review}
	
	
		 \input{Tables/acronyms}
	
	A breakdown of the studies identified with respect to the venues where they were published has been shown in Table~\ref{table:papers-count}. As can be seen, the studies come from a wide range of conferences and journals, which are all respectful\footnote{The raw data and minutes recorded during the discussion in the data collection process are publicly available at: \url{https://github.com/taochen/sbse-qi}.}. It is worth noting that the results do not only include studies published in software engineering venues, but also those published in service, system and cloud engineering conferences/journals as well as those in the computational intelligence venues, as long as they are related to problems in the software engineering domain and comply with the inclusion/exclusion criteria.
	
	Next, we report on the results collected from our systematic literature review, which would further motivate the remaining of our work.

	\begin{figure}[!t]
 \centering
	\includegraphics[width=\columnwidth]{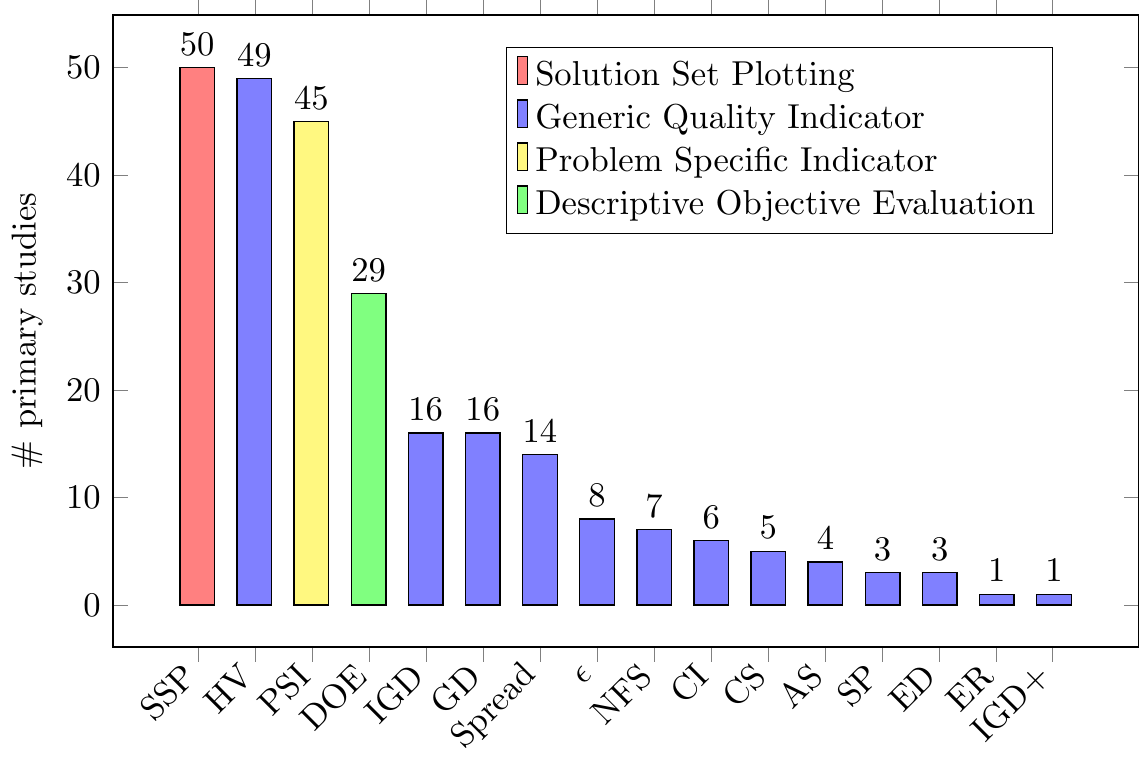}
  \caption{Usage of evaluation methods in primary studies.}
 \label{fig:qi-count}
\end{figure}

  \subsection{RQ1: What Evaluation Methods?}
	
	
	The usage of evaluation methods has been presented in Figure~\ref{fig:qi-count} along with the details for every single primary study presented in Table A1 (at appendix). As can be seen, a total of 13 generic quality indicators have been used in the primary studies (i.e., $HV$, $IGD$, $GD$, $Spread$, $\epsilon$, $NFS$, $CI$, $CS$, $AS$, $SP$, $ED$, $ER$ and $IGD^+$). Explanations of all the acronyms can be found in Table~\ref{tb:acronym}. In particular, $HV$, $IGD$ and $GD$ are the top three most widely used generic quality indicators across almost all the SBSE problems, due presumably to their popularity as well as ``inertia'' (i.e., researchers tend to use indicators which were used before even though they are not the best fit)~\cite{li2019quality}.

	
	There are also 45 primary studies using $PSI$; for example, MoJoFM~\cite{Kumari2016Hyper} is a commonly used symmetric indicator for the software modularization problem, which aims to compare two resulted partitions of classes (i.e., two solutions), thus inapplicable to other optimization scenarios. Note that since the $PSI$ is highly domain-dependent and they are not explicitly designed for evaluating solution sets, we do not specify the usage details for every single one of them. We do however present which particular $PSI$ is used under which context, as shown in Table~\ref{table:papers-preferences} which we will elaborate in Section~\ref{sec:pref-con}. In summary, we found a total of 34 different $PSI$ over all the Pareto-based SBSE problems.
	
	Apart from the quality indicators, $SSP$ and $DOE$ have also been overwhelmingly used by 50 and 29 primary studies respectively to evaluate solution sets. For $SSP$, we found only two sub-types: the Parallel Coordinate plot which shows the solutions' objective values upon $n$ parallel lines, where $n$ is the number of objectives; and the Scatter Plot that plots the solutions in the objective space. 
	$DOE$ involves more diverse forms, as shown in Table~\ref{table:DOE}, including 27 cases to compare the mean, best, worst, median, or statistical results of each objective in the evaluation along with the remaining five cases that use other three forms.
	 \input{Tables/doe}

	\input{Tables/qi-summary}

	\subsection{RQ2: Why and How Generic Quality Indicators are Used?}
	
	As shown in Table~\ref{tb:qi-summary}, for those primary studies that made use of generic quality indicators, most commonly there is a clear statement about what quality aspect(s) they selected an indicator for, as appeared in 73 cases. This is also the reason and evidence that these studies used to justify their choices. However, there is still a considerable amount of cases (59) that were marked as \textit{Unknown}, i.e., no clear and explicit rationale of the choice has been provided. For the reference front used for $GD$ and $IGD$, whilst most of the cases the best Pareto front found by all algorithms (i.e., nondominated solutions of the set consisting of the solutions produced by all algorithms) is used, many still do not explicitly declare such information. On the reference point used by $HV$, we see a diverse way of obtaining such a point, including using the worst objective value of all the solutions found, 
	the boundary of the optimization problem in SBSE, 
	and the nadir point from the Pareto front of all the solutions found.

	\subsection{RQ3: What Context?}

	\subsubsection{Number of Objectives}
	
	Table~\ref{tb:obj-summary} shows the number of objectives considered by the evaluation methods. On the generic quality indicators, we see that $IGD$ has been used for the highest number, i.e., 15 objectives, and all of them (except $SP$ and $ER$) have ever been used on the bi-objective cases, which is the minimum number required to build a Pareto front. Whilst most of the generic quality indicators have been used under the bi- and tri-objectives cases, a considerable amount of them have been used on the objective number over three.
	
	As for $PSI$, $DOE$, and $SSP$, we can observe that they are used on a relatively wider range of objective numbers as compared with most of the generic quality indicators.

	\subsubsection{Pareto-based SBSE Problems}
	
	From Table~\ref{table:papers-sdlc}, it is clear that our systematic review has revealed 21 distinct Pareto-based SBSE problems, which are spread across all the six common phases in the SDLC\footnote{Note that \cite{1600-7617} studies three different problems.}. Notably, we can see that certain problems have attracted more attention than the others, as evidenced by the much higher number of primary studies included, such as the software product line and the white/black-box test case generation problems. Among others, the software testing phase, as well as the deployment and maintenance phase contain much more diverse problems than the other SDLC phases. This is probably because the nature of those problems, which are usually in later phases of the SDLC, fits the requirements of search-based optimization well.
	
	Indeed, some of the Pareto-based SBSE problems can arguably fit into more than one phase of the SDLC; but in this work, we classify those problems according to which phases can be better matched with the detailed formalization of the problem and the hypotheses that the authors made. Further, certain problems in the deployment and maintenance phase are not classic software engineering problems (e.g., resource management and service composition); however, they have recently attracted more and more attention from software engineering researchers and have been increasingly considered as important issues in the software engineering domain~\cite{Harman2013Cloud}.
	
	\input{Tables/obj-summary}
	\input{Tables/problem}
	
  \input{Tables/preference}

	\subsubsection{Assumptions on Preferences and Contextual Information of Problems and Their Evaluation Methods}
	\label{sec:pref-con}
	
	In Table~\ref{table:papers-preferences}, we summarize the assumptions on the DM's preferences and contextual information about the objectives for each Pareto-based SBSE problem reviewed, and the evaluation methods used to compare different solution sets under these contexts. As we can see, there are 17 cases, covering 25 primary studies, have made assumptions on DM's preferences. The contextual information, in contrast, has been used in five cases over 20 primary studies. The others do not clearly state the assumptions in this regard and hence noted as \emph{Not specified}. One of the most notable observations is that a particular SBSE problem may have multiple, distinct assumptions about the DM's preferences and contextual information. In fact, most of the problems have more than one assumption on the preferences/contextual information, and particularly, the project scheduling problem and software configuration/adaptation problem involve up to four different types of assumptions. This reflects the fact that many problems are complex and the actual preferences can be situation-dependent. Yet, it does bring the requirements that all those situations need to be catered for.
	In contrast, problems such as effort estimation and requirement assignment 
	have assumed only one type of preferences/contextual information, 
	which implies a relatively more straightforward selection and use in the quality evaluation process on the solution sets.


	\section{Issues on Quality Evaluation in Pareto-based SBSE}

	
	Based on our systematic literature review, 
	this section provides a systematic analysis of five issues of quality evaluation, 
	classified into two categories, 
	from state-of-the-art Pareto-based SBSE work.
	
	\subsection{Problematic Use of Illustrative and Descriptive Statistic Evaluation Methods}
 
	As shown in Figure~\ref{fig:qi-count}, Tables~\ref{table:papers-preferences} and A1 (at appendix), 
	there exist many SBSE studies, 
	particularly in early days, 
	that relied on plotting the solution set returned ($SSP$) and/or on reporting some $DOE$ results to reflect the quality of solution sets. 
	Despite these two methods being simple to apply,
	they may easily lead to inaccurate evaluations and conclusions.


	\subsubsection{\textsc{ISSUE I:} Inadequacy of Solution Set Plotting ($SSP$)}
	
	A straightforward way to evaluate/compare the quality of solution sets returned by search algorithms is to plot solution sets and judge intuitively how good they are.
	Such visual comparison is among the most frequently used methods in SBSE, 
	but it may not be very practical in many cases. 
	
	First, 
	it cannot scale up well ---
	when the number of objectives is larger than three, 
	the direct observation of solution sets (by scatter plot) is unavailable.
	Second, 
	the visual comparison fails to quantify the difference between solution sets.
	Finally,
	when an algorithm involves stochastic elements, 
	different runs usually result in different solution sets. 
	So, it may not be easy to decide which run should be considered. 
	Printing the solution sets obtained in all the runs can easily clutter the picture.  
	As such, 
	plotting solution sets does not suffice to the quality evaluation in Pareto-based SBSE, 
	despite the fact that it has been used solely to compare solution sets in a considerable amount of the primary studies, e.g., 
	\cite{Gueorguiev2009Software, Zhang2007The, Heaven2011Simulating, busari2017radar, Li2010SLA, Martens2010Automatically}, as shown in Table A1 (at appendix).
	Nevertheless, 
	it is worth mentioning that $SSP$ is useful as an extra evaluation method in addition to quality indicators, particularly in bi- and tri-objective cases. 
	This will be discussed in the guidance section (Section~\ref{sec:guidance}) later on.
	

	\subsubsection{\textsc{ISSUE II:} Inappropriate Use of Descriptive Objective Evaluation ($DOE$)}
	\label{sec:doe-issue}
	
\begin{figure}[tbp]
		\centering
		\includegraphics[width=0.6\columnwidth]{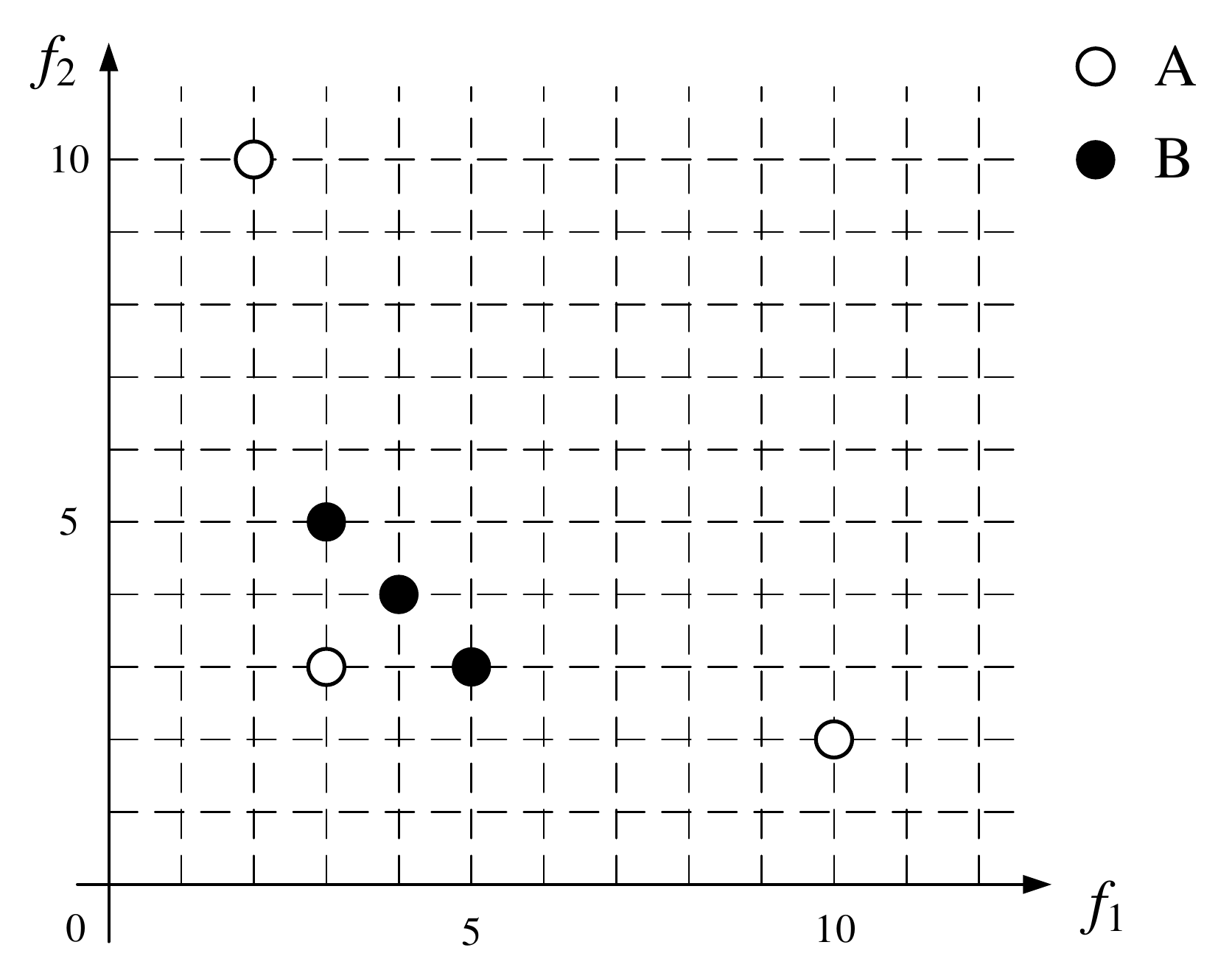}
		\vspace{-0.3cm}
		\caption{An example that comparing the mean on each objective fails to reflect the quality of solution sets. 
			In this minimization problem, 
			solution set $\mathbf{A}$ dominates solution set $\mathbf{B}$ (i.e., any solution in $\mathbf{B}$ is dominated by at least one solution in $\mathbf{A}$), 
			thus always being favored by the DM. 
			Yet, 
			the mean of $\mathbf{A}$ on either objective $f_1$ or $f_2$ is $5$, 
			larger than that of $\mathbf{B}$ ($4$); 
			thus $\mathbf{A}$ is regarded as inferior to $\mathbf{B}$.}
		\label{Fig:Mean}
		\vspace{-0.3cm}
	\end{figure}

	Many Pareto-based SBSE studies evaluate solution sets by $DOE$ --- statistical objective values in the obtained solution set(s). 
	For example, 
	as it can be seen in Table~\ref{table:DOE}, 
	the mean objective value was considered in~\cite{Sarro2017Adaptive,Bowman2010Solving,Simons2010Interactive,Simons2012Elegant,Sayyad2013Optimum,Bavota2012Putting,Pascual2015Applying,pradhan2019,main-sbse,Boukharata2019};
	the median value in~\cite{Bavota2012Putting,Fleck2017Model,Kalboussi2013Preference}; 
	the best value in~\cite{Minku2013Software,Bowman2010Solving,Kumari2016Hyper,Praditwong2011Software,Mao2016Sapienz,Wu2015Deep,Wada2012E,Durillo2014Multi,PubSub10849_Chen,White2011Evolutionary,Panichella2015Reformulating,Wagner2012Multi};
	the worst value in~\cite{Bowman2010Solving,Wada2012E};
	the statistical significance of the differences between distinct solution sets' objective values in~\cite{Li2017Zen,Wang2015Cost,Segura2016Multi,Abdeen2014Multi}. 
	Such $DOE$ measures need to be used in line with the DM's preference. 
	For example, 
	comparing the best value of each objective can well evaluate solution sets if the DM prefers the extreme points (solutions),
	but may not be well-suited when balanced points are wanted,
	which, unfortunately, was practiced in some studies such as~\cite{Wang2015Cost} shown in Table~\ref{table:papers-preferences}. 
	Worse still, 
	many $DOE$ measures may give a misleading evaluation,
	including those comparing the mean, median, and worst values of each objective and comparing statistically significant differences on each objective.
	That is to say, 
	by a $DOE$ measure a solution set is evaluated better than another set, 
	but in fact, the latter is always preferred by the DM under any circumstances. 
	Figure~\ref{Fig:Mean} gives such an example (minimization) with respect to calculating the mean of each objective. 
	As shown, 
	the mean of the solution set $\mathbf{A}$ on either objective $f_1$ or $f_2$ is $5$, 
	larger than that of the solution set $\mathbf{B}$ ($4$), 
	thus $\mathbf{A}$ being regarded as inferior to $\mathbf{B}$.
	Yet, 
	$\mathbf{A}$ will always be favored by the DM since there is one solution in $\mathbf{A}$ better than  
	any solution in $\mathbf{B}$.
	
	On the other hand, recalled from Table~\ref{table:DOE}, 
	some work in the primary studies considers selecting one
	particular solution (by using a decision-making method)
	from the whole solution set produced by the Pareto-based search
	for comparison.
	For example, 
	the studies in~\cite{Li2017Zen,Wang2015Cost} considered Mean Fitness Value ($MFV$) and the studies in~\cite{Shen2016Dynamic}\cite{shen2018q} considered Analytic Hierarchy Process (AHP)~\cite{fulop2005introduction}. 
	However, 
	one question is that if we know clear weighting between objectives of the DM (thus being able to take only one solution from the whole solution set into account), 
	why not directly integrate this information into the problem model, 
	thus converting a multi-objective problem into an easier single-objective problem in the first place.

	\subsection{Problematic Use of Generic Quality Indicators}
	
	As disclosed in Tables~\ref{tb:obj-summary}, \ref{table:papers-preferences} and A1 (at appendix), 
	it has been commonly seen in Pareto-based SBSE studies that select or use quality indicators
	that cannot accurately reflect the quality of solution sets.
	This is virtually because the SBSE researchers/practitioners may not be very clear about indicators' behavior, role, and characteristics. 
	This leads them either to fail to select appropriate indicators to evaluate the generic quality of solution sets, 
	or to fail to align the considered indicators with the DM's preferences or the problem's contextual information. 
	
	\subsubsection{\textsc{ISSUE III:} Confusion of the Quality Aspects Covered by Generic Quality Indicators}
	
	
	As mentioned, 
	the generic quality of a solution set in Pareto-based optimization can be interpreted as how well it represents the Pareto front.
	It can be broken down into four aspects: 
	convergence, spread, uniformity, and cardinality~\cite{li2019quality}.
	It is expected that when the DM's preferences are unknown \textit{a priori},
	an indicator (or a combination of indicators) can cover all the four quality aspects 
	since a solution set with these qualities can well represent the Pareto front and have a great probability of being preferred
	by the DM.
	
	Unfortunately, as shown in Tables~\ref{tb:obj-summary} and A1 (at appendix), in SBSE many studies only consider part of these quality aspects. 
	For example, 
	the studies in \cite{Harman2009Search,Li2014Robust} used the convergence indicator $GD$~\cite{Veldhuizen1998} as the sole indicator to compare the solution sets.	  
	The study in \cite{Yoo2010Using} considered both $PFS$~\cite{Henard2015Combining} and $CI$ which however are merely for convergence and cardinality.
	In addition,
	some indicators were used to evaluate certain quality aspect(s) of solution sets which, unfortunately, 
	were not designed for, 
	as shown in~\cite{li2018critical}. 
	For example,
	the indicator $\mathcal{C}$~\cite{Zitzler1998}, 
	designed for convergence evaluation,
	was considered for evaluating spread in~\cite{wang2016practical}.
	$SP$~\cite{Schott1995}, 
	which can only reflect the uniformity of solution sets, 
	was used to evaluate the diversity (i.e., both spread and uniformity) in~\cite{Ramirez2016Comparative}.
	$PFS$, which counts nondominated solutions in the set, 
	was placed into the category of diversity indicators in~\cite{Henard2015Combining,wang2016practical},
	and $\epsilon$-indicator,
	which is able to reflect all the quality aspects of solution sets, 
	was placed into the category of convergence indicators in~\cite{wang2016practical}.
	
	As can be seen from Tables \ref{tb:qi-summary} and \ref{tb:obj-summary}, 
	some indicators were used incorrectly,  
	For example,
	the indicator \textit{Spread} (i.e., $\Delta$ in~\cite{Deb2002}) as well as its variants (e.g., $GS$~\cite{Zhou2006}), 
	which is only effective in the bi-objective case, was frequently used in optimization problems of SBSE with three or more objectives, 
	such as in~\cite{shen2018q,Shen2016Dynamic,Sayyad2013Optimum,Sayyad2013On,Ouni2017Search,Henard2015Combining,abdessalem2018}. 
	Another example is the setting of the critical parameter reference point in the $HV$ indicator, 
	which has experienced various versions. 
	For example, some studies set it to the worst value obtained for each objective during all runs~\cite{shen2018q,Shen2016Dynamic,Li2017Zen,Olaechea2014Comparison,Ferrer2012Evolutionary,Pascual2015Applying,Tan2018Evolutionary}; 
	some did it to precisely the boundaries of the optimization problem in SBSE~\cite{Hierons2016SIP,Xiang2018Configuring,Durillo2014Multi};
	some did it to the nadir point of the Pareto front~\cite{Panichella2015Improving}. 
	The first two settings may overemphasize the boundary solutions (as the reference point may be far away from the set to be evaluated), 
	while the last one may lead to the boundary solutions to contribute nothing to the $HV$ value. 
	
	It is worth mentioning that as usually the problem's Pareto front in SBSE is unavailable,
	for indicators which need the Pareto front for reference, 
	a common practice is to collect the nondominated set of all the solutions produced as an estimated Pareto front, as we have shown in Table~\ref{tb:qi-summary}. 
	However, 
	different indicators have different sensitivity to this practice~\cite{li2019quality}. 
	For example, 
	$IGD$ and \textit{Spread} require the Pareto front consisting of uniformly-distributed points, 
	while $HV$ and $\epsilon$-indicator do not~\cite{li2019quality}.
	Therefore, 
	$IGD$ and \textit{Spread} may not be very suitable in SBSE, 
	despite the fact they were frequently used, 
	e.g., in~\cite{shen2018q,Shen2016Dynamic,Sarro2017Adaptive,Ferrucci2013Not,Durillo2009A,Zhang2013Empirical,Henard2015Combining}.

	\begin{figure}[tbp]
	\centering
	\includegraphics[width=0.6\columnwidth]{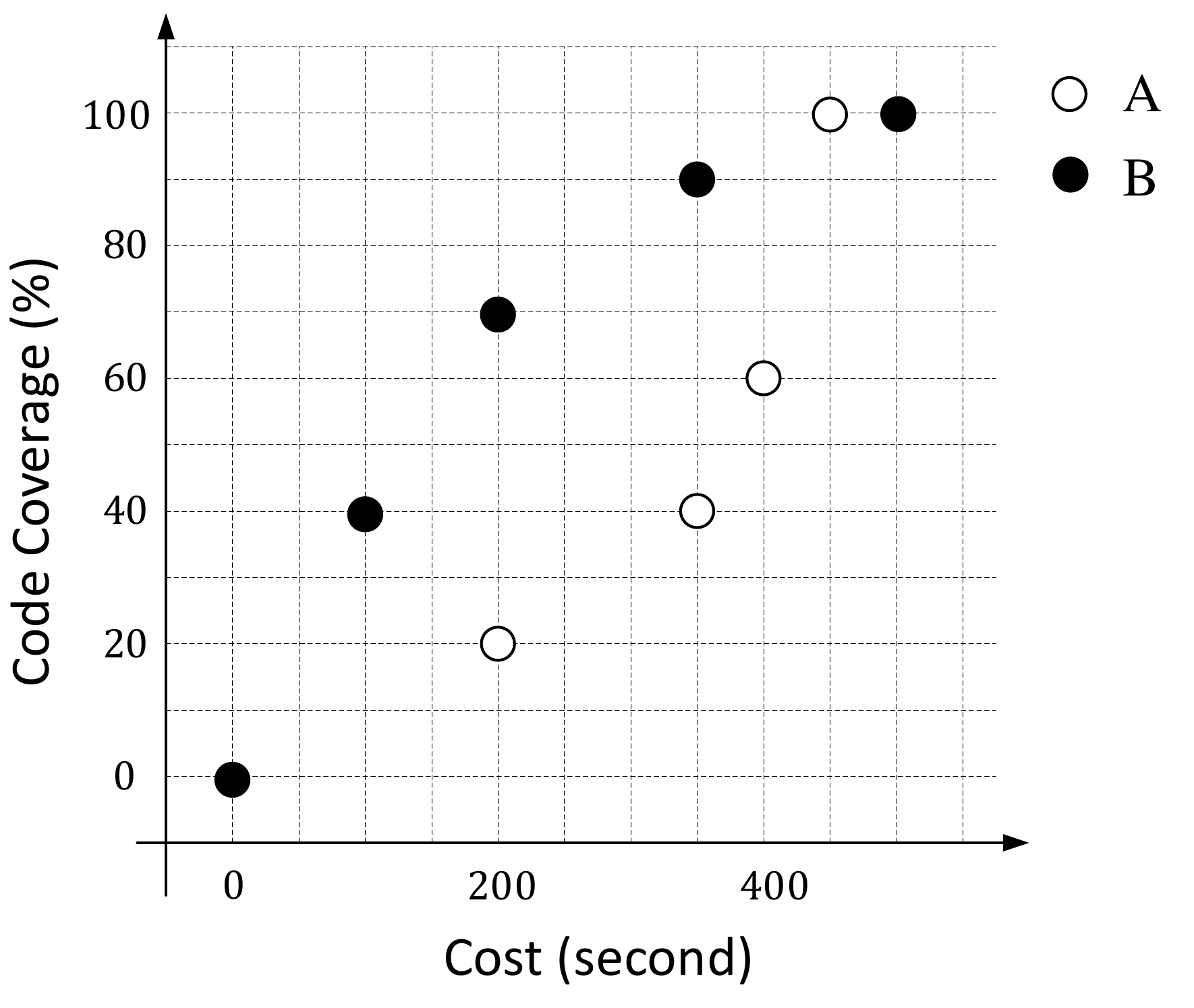}
	\vspace{-0.3cm}
	\caption{An example where lack of considering contextual information 
		may give unwanted evaluation results~\cite{li2018critical} . 
		Considering two solutions sets ($\mathbf{A}$ and $\mathbf{B}$) 
		for optimizing the code coverage and the cost of testing time on the software test case generation problem, 
		where $\mathbf{A} = \{(200,0.2),(350,0.4),(400,0.6),(450,1.0)\}$ and $\mathbf{B} = \{(0,0),(100,0.4),(200,0.7),(350,0.9),(500,1.0)\}$.
		$\mathbf{B}$ is evaluated better than $\mathbf{A}$ on eight frequently-used indicators:
		$GD(\mathbf{B})=0.02 < GD(\mathbf{A})=0.26, ED(\mathbf{B})=0.5 < ED(\mathbf{A})=0.89, \epsilon(\mathbf{B})=0.1 < \epsilon(\mathbf{A})=0.3, 
		GS(\mathbf{B})=0.15 < GS(\mathbf{A})=0.46, PFS(\mathbf{B})=5 > PFS(\mathbf{A})=4, IGD(\mathbf{B})=0.02 < IGD(\mathbf{A})=0.27,
		HV(\mathbf{B})=0.77 > HV(\mathbf{A})=0.43, \mathcal{C}(\mathbf{B})=0.8 > \mathcal{C}(\mathbf{A})=0.25.$ 
		However, the DM may be more interested in $\mathbf{A}$ (specifically solution $(450,1.0)$) 
		if they favor the full code coverage and then
		possible low cost.}
	\label{Fig:FullCoverage}
	\vspace{-0.3cm}
\end{figure}	
	
	\subsubsection{\textsc{ISSUE IV:} Oblivion of Context Information}
	
	As shown in Table~\ref{table:papers-preferences}, in Pareto-based SBSE, 
	many studies compare solution sets without bearing in mind the contextual information with respect	to the considered optimization problem. 
	They typically adopt commonly-used quality indicators to directly evaluate the set of all the solutions obtained, 
	although some of these solutions may never or rarely be of interest to the DM.
	\mbox{Figure~\ref{Fig:FullCoverage}} shows such an example, 
	under a scenario of optimizing the code coverage and the cost of testing time on the software test case generation problem, borrowed from \cite{li2018critical}.
	As can be seen,
	the set $\mathbf{B}$ is evaluated better than the set $\mathbf{A}$ by 
	all eight commonly used quality indicators ($GD$~\cite{Veldhuizen1998}, $ED$~\cite{Cochrane1973}, $\epsilon$-indicator~\cite{Zitzler2003}, $GS$~\cite{Zhou2006}, $PFS$~\cite{Henard2015Combining}, $IGD$~\cite{Coello2004}, $HV$~\cite{Zitzler1998} and $\mathcal{C}$~\cite{Zitzler1998}) in SBSE~\cite{wang2016practical}. 
	However, 
	depending on the context (as shown in Table~\ref{table:papers-preferences}), 
	the DM might first favor the full code coverage and then possible low cost~\cite{Zheng2016Multi}. 
	This will lead to set $\mathbf{A}$ to be of more interest,  
	as it has the solution ($450,1.0$) 
	that achieves full coverage and lower cost than the one in $\mathbf{B}$ ($500,1.0$).
	
	Similar observations have been seen in the optimal product selection in software product line~\cite{Sayyad2013Scalable,Sayyad2013On,Henard2015Combining,tan2015optimizing,1600-7617,Guo2019,lee2019,saber2017IsSeeding,lian2017}
	where the correctness of configurations is regarded as one objective and equally rated as other objectives 
	(e.g., richness of features and cost). 
	This may lead to an invalid product to be evaluated better than a valid product if the former performs better in other objectives, 
	which is apparently of no value to the DM.
	In addition, 
	in many SBSE problems, 
	cost could be an objective to minimize, 
	but solutions with zero cost are trivial, e.g., the solution with zero cost and zero coverage in Figure~\ref{Fig:FullCoverage}. 
	However, 
	these solutions may largely affect the evaluation results. 
	Therefore, 
	it is necessary to remove solutions that would never be interested by the DM before the evaluation,
	which, unfortunately, has been rarely practiced in Pareto-based SBSE.  
	
	\subsubsection{\textsc{ISSUE V:} Noncompliance of the DM's Preferences}
	
\begin{figure}[tbp]
		\centering
		\includegraphics[width=0.6\columnwidth]{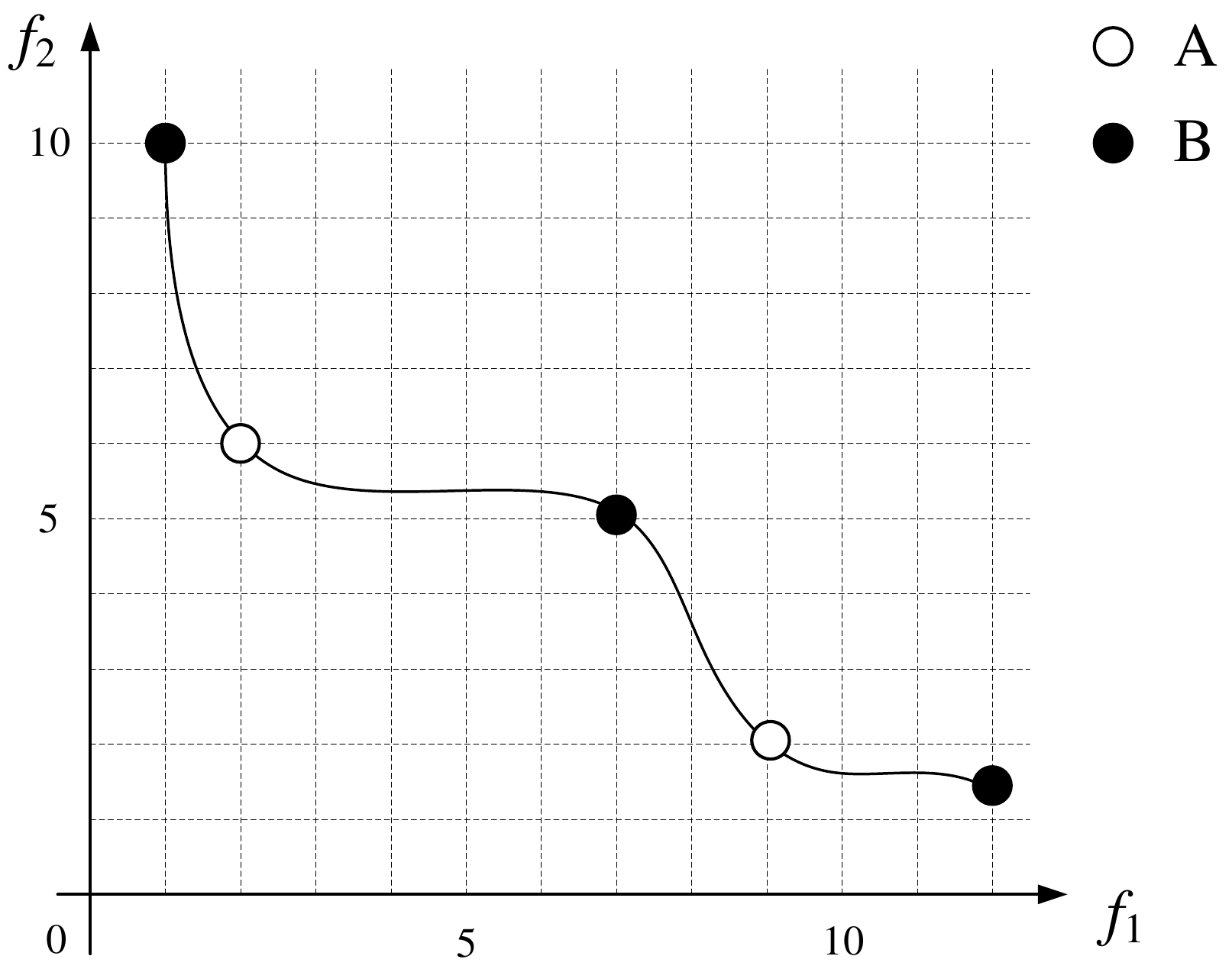}
		\vspace{-0.3cm}
		\caption{An example that preferring knee points while using the indicator $IGD$ (as well as $GD$ and $CI$) can lead to misleading results.
			Consider two solution sets ($\mathbf{A}$ and $\mathbf{B}$) for a bi-objective minimization scenario, 
			where $\mathbf{A} = \{(2,6),(9,2)\}$ are the two knee points of the Pareto front, 
			and $\mathbf{B} = \{(1,10),(7,5),(12,1.5)\}$ are three well-distributed non-knee points on the Pareto front.
			Apparently, if the DM prefers knee points then solutions in $\mathbf{A}$ will certainly be selected. 
			Yet, $\mathbf{A}$ is evaluated worse than (or as equal as) $\mathbf{B}$ by $IGD$, $GD$ and $CI$: 
			$IGD(\mathbf{A})=2.154 > IGD(\mathbf{B})=1.433$, 
			$GD(\mathbf{A}) = GD(\mathbf{B})=0$, and 
			$CI(\mathbf{A})=0.4 < CI(\mathbf{B})=0.6$.
			In contrast, 
			the indicator $HV$ can reflect this preference, 
			$\mathbf{A}$ being evaluated better than $\mathbf{B}$: $HV(\mathbf{A})=71.0 > HV(\mathbf{B})=45.5$ 
			(the reference point is $(13,11)$).}
		\label{Fig:Knee}
		\vspace{-0.3cm}
	\end{figure}

	Although every quality indicator is designed to reflect certain quality aspect(s) of solution sets (i.e., convergence, 
	spread, uniformity, cardinality, or their combination), 
	they do have their own implicit preferences. 
	For example, 
	the indicators $HV$ and $IGD$, 
	both designed to cover all of the four quality aspects,
	have rather distinct preferences.
	$HV$ prefers knee points of a solution set, 
	while $IGD$ is in favor of a set of uniformly distributed solutions.
	Therefore, 
	it is important to select indicators whose preferences are in line with the DM's. 
	Neglecting this can lead to misleading evaluation results. 
	Figure~\ref{Fig:Knee} gives such an example --- when preferring knee points, 
	considering the indicator $IGD$ could return a misleading result. 
	That is, the set having knee points is evaluated worse than that having no knee point.  
	Similar observations also apply to the indicators $GD$ and $CI$, 
	as shown in the same figure.

	Unfortunately, as what has been revealed in Table~\ref{table:papers-preferences}, 
	such misuse of indicators is not uncommon in the SBSE community. 
	For example, 
	preferring knee points yet using $IGD$ in~\cite{Fleck2017Model,Mkaouer2016On};
	preferring knee points yet using $GD$ and $CI$ in~\cite{Ferrucci2013Not,Sarro2017Adaptive};
	and preferring extreme solutions yet using $HV$ and $IGD$ in~\cite{Tan2018Evolutionary,Wada2012E}.
	$HV$ can be somehow in favor of extreme solutions if the reference point is set far away from the considered set, 
	but $IGD$ certainly does not prefer extreme solutions.  
	Therefore, 
	it is of high importance to understand the behavior, role, and characteristics of the considered indicators, 
	which may not be very clear to the community. 
	In the next section, 
	we will detail widely used quality evaluation methods in the area (as well as other useful indicators) and 
	explain the scope of their applicability.

\section{Revisiting Quality Evaluation for Pareto-based Optimization}\label{sec:QI_revisit}
	
	In Pareto-based optimization,
	the general goal for the algorithm designer is to supply the DM a set of solutions 
	from which they can select their preferred one. In general, the actual preferences can be either articulated by the DM or derived from the contextual information of the problem, which may differ depending on the situation.
	Having said that, 
	the Pareto dominance relation is apparently the foremost criterion in any case,
	provided that the concept of optimum is solely based on the direct comparison of solutions' objective values (other than on other criteria, e.g., robustness and implementability with respect to decision variables).
	That is to say, 
	the DM would never prefer a solution to the one that dominates it.

	As discussed in Section~\ref{sec:bg}, 
	the \textit{better} relation ($\vartriangleleft$) represents 
	the most general and weakest form of superiority between two sets. 
	That is, 
	for two solution sets $\mathbf{A}$ and $\mathbf{B}$, 
	$\mathbf{A} \vartriangleleft \mathbf{B}$ indicates that $\mathbf{A}$ is at least as good as $\mathbf{B}$, 
	while $\mathbf{B}$ is not as good as $\mathbf{A}$.
	It meets any preference potentially articulated by the DM. 
	If $\mathbf{A} \vartriangleleft \mathbf{B}$, 
	then it is always safe for the DM only to consider solutions in $\mathbf{A}$.
	Apparently, 
	it is desirable that a quality evaluation method is able to capture this relation; 
	that is to say, for any two solution sets $\mathbf{A}$ and $\mathbf{B}$, 
	if $\mathbf{A} \vartriangleleft \mathbf{B}$, 
	then $\mathbf{A}$ is evaluated better than $\mathbf{B}$. 
	Unfortunately, 
	there are very few quality evaluation methods holding this property. 
	$HV$ is one of them~\cite{Zitzler2007a}.
	There is a weaker property called being \textit{Pareto compliant}~\cite{Zitzler2003,Knowles2006}, 
	which is more commonly used in the literature. 
	That is, a quality evaluation method is said to be \textit{Pareto compliant} if and only if 
	``at least as good'' in terms of the dominance relation implies 
	``at least as good'' in terms of the evaluation values 
	(i.e., $\forall \mathbf{A}, \mathbf{B}: \mathbf{A} \preceq \mathbf{B} \Rightarrow I(\mathbf{A}) \leq I(\mathbf{B})$, 
	where $I$ is the evaluation method, assuming that the smaller the better). 
	Despite the relaxation, 
	many quality indicators are not Pareto compliant, 
	including widely used ones, 
	such as $GD$, $IGD$, \textit{Spread}, $GS$, and $SP$.
	Pareto compliant indicators are mainly those falling into the category of evaluating convergence of solution sets (e.g., $\mathcal{C}$ and $CI$) 
	and the category of evaluating comprehensive quality of solution sets (e.g., $HV$, $\epsilon$-indicator, $IPF$~\cite{Bozkurt2010}, $R2$~\cite{Hansen1998}, and $PCI$~\cite{Li2015b}).
	$DCI$~\cite{Li2014d} is the only known diversity indicator compliant with Pareto dominance when comparing two sets. 
	In addition,
	some non-compliant indicators can become Pareto compliant after some modifications. 
	For example, $GD$ and $IGD$ can be transformed into two Pareto compliant indicators (called $GD$$^+$ and $IGD$$^+$) if considering ``superiority'' distance instead of Euclidean distance between points~\cite{Ishibuchi2015}. 
	Overall, 
	it is highly recommended to consider (at least) Pareto compliant quality indicators to evaluate solution sets; 
	otherwise, it may violate the basic assumption of the DM's preferences.
	That is, recommending the DM a solution set $\mathbf{B}$ over $\mathbf{A}$, 
	where each solution in $\mathbf{B}$ is inferior to or can be replaced by (in the case of equality) some solution in $\mathbf{A}$. 
	This is what the $DOE$ evaluation method that compares the mean on each objective 
	does in the example of Figure~\ref{Fig:Mean}.

	Now, one may ask why not directly use the \textit{better} relation to evaluate solution sets. 
	The reason is that the \textit{better} relation may leave many solution sets incomparable 
	since in most cases there exist some solutions from different sets being nondominated to each other.
	Therefore, 
	we need stronger assumptions about the DM's preferences,  
	which are reflected by quality evaluation methods.
	However, 
	stronger assumptions (than the \textit{better} relation) cannot guarantee that the favored set (under the assumptions)
	is certainly preferred by the DM, 
	as in different situations the DM indeed may prefer different trade-offs between objectives.  
	Consequently, 
	it is vital to ensure the considered evaluation methods in line with 
	the DM's explicit or implicit preferences.
	
	Back to the example in \mbox{Figure~\ref{Fig:FullCoverage}} where optimizing the objectives code coverage and cost of testing time,
	essentially these two solution sets are not comparable with respect to the \textit{better} relation 
	despite the fact that most solutions in $\mathbf{A}$ are dominated by some solution in $\mathbf{B}$.
	As stated, 
	the DM may be more interested in full code coverage and then possible lower cost, 
	thus preferring $\mathbf{A}$ to $\mathbf{B}$. 
	However, 
	the considered eight indicators fail to capture this information and give opposite results.  
	This clearly indicates the importance of understanding quality evaluation methods 
	(including what kind of assumptions they imply).  
	\rev{Next, 
	we will review several quality evaluation methods
	which are commonly used in the SBSE community (as we have shown in Table~\ref{fig:qi-count}) and at the same time 
	are very representative to reflect certain aspect(s) of solution set quality.}

	\subsection{Descriptive Objective Evaluation ($DOE$)}
	
	As stated before, 
	the $DOE$ methods evaluate a solution set 
	(or several sets obtained by a search algorithm in multiple runs) by directly reporting statistical results of objective values of its solutions, such as the mean, median, best, worst, and statistical significance (in comparison with other sets). 
	Unfortunately, 
	such methods are rarely being Pareto compliant and unlikely to be associated with the DM's preferences. 
	However, 
	an exception is the method that considering the best value of some objective(s) in a solution set, 
	since it is Pareto compliant and able to directly reflect the DM's preferences in the case that they prefer extreme solutions. 
	Overall, 
	the $DOE$ methods are not recommended,
	unless the DM explicitly expresses their preferences in line with them.

	\subsection{Contribution Indicator ($CI$)}
	
	
	The $CI$ indicator~\cite{Meunier2000}, 
	which was designed to compare the convergence of two solution sets,
	has been frequently used in SBSE, e.g., in~\cite{Sarro2016Multi,Ferrucci2013Not,Sarro2017Adaptive,Yoo2010Using,Wu2015Deep,zhang2018}. 
	$CI$ calculates the ratio of the solutions of a set that are not dominated by any solution in the other set. 
	Formally, 
	given two sets $\mathbf{A}$ and $\mathbf{B}$,
	\begin{small}
	\begin{equation}
	CI(\mathbf{A},\mathbf{B}) = \frac{|\mathbf{A}\cap \mathbf{B}|/2 + |\mathbf{A}_{\prec B}| + |\mathbf{A}_{\npreceq \cap \nsucc B}|}{|\mathbf{A}\cap \mathbf{B}| + |\mathbf{A}_{\prec B}| + |\mathbf{A}_{\npreceq \cap \nsucc B}| + |\mathbf{B}_{\prec A}| + |\mathbf{B}_{\npreceq \cap \nsucc A}|}
	\label{eq:CI}
	\end{equation} 
	\end{small}
	where $\mathbf{A}_{\prec B}$ stands for the set of solutions in $\mathbf{A}$ that dominate some solution of $\mathbf{B}$ (i.e., $\mathbf{A}_{\prec B} = \{\mathbf{a} \in \mathbf{A}| \exists \mathbf{b} \in \mathbf{B}: \mathbf{a} \prec \mathbf{b}\}$), 
	and $\mathbf{A}_{\npreceq \cap \nsucc B}$ stands for the set of solutions in $\mathbf{A}$ that do not weakly dominate any solution in $\mathbf{B}$ and also are not dominated by any solution in $\mathbf{B}$ (i.e., $\mathbf{A}_{\npreceq \cap \nsucc B} = \{\mathbf{a} \in \mathbf{A}| \nexists \mathbf{b} \in \mathbf{B}: \mathbf{a} \preceq \mathbf{b} \vee \mathbf{b} \prec \mathbf{a}\}$). 
	
	The $CI$ value is in the range of $[0,1]$. 
	A higher value is preferable.
	It is apparent that $CI(\mathbf{A},\mathbf{B}) + CI(\mathbf{B},\mathbf{A}) = 1$. 
	A clear strength of the indicator $CI$ is that it holds the \textit{better} relation\footnote{Note that it is not unusual that binary indicators (i.e., those directly comparing two sets) holds the \textit{better} relation~\cite{li2019quality}.}, 
	i.e., if $\mathbf{A} \vartriangleleft \mathbf{B}$ then $CI(\mathbf{A},\mathbf{B}) > CI(\mathbf{B},\mathbf{A})$. 
	Moreover, 
	if $\mathbf{A} \prec \mathbf{B}$, 
	then $CI(\mathbf{B},\mathbf{A}) = 0$.
	In addition, 
	apart from comparing the convergence of solution sets, 
	$CI$ can reflect their cardinality to some extent. 
	A set having a larger number of solutions is likely to be favored by the indicator.
	
	A clear weakness of $CI$ is that it relies completely on the dominance relation between solutions, 
	thus providing little information about to what extent one set outperforms another. 
	Moreover, they may leave many solution sets incomparable if all solutions from the sets are nondominated to each other. 
	This may happen frequently in many-objective optimization, 
	where more objectives are considered.
	
	There is another well-known dominance-based quality indicator (called $\mathcal{C}$ or $CS$)~\cite{Zitzler1998}, 
	used in e.g.~\cite{Shen2016Dynamic,Assun2014A,Chen2017Self,zhang2017constraint}. 
	It measures the proportion of solutions in a set that is weakly dominated by some solution in the other set; 
	in other words, 
	the percentage of a set that is \textit{covered} by its opponent.
	The details of the indicator $\mathcal{C}$ can be found in~\cite{li2019quality}.
	$\mathcal{C}$ tends to be more popular in the multi-objective optimization community, 
	despite sharing the above strengths and weaknesses with $CI$.  
	Finally, 
	it is worth mentioning that despite only partially reflecting the convergence of solution sets, 
	such dominance-based indicators are useful since most problems in SBSE are combinatorial ones, 
	where the size of the Pareto front may be relatively small and it is likely to have comparable solutions (i.e., dominated/duplicate solutions) from different sets~\cite{li2019quality}.

	\subsection{Generational Distance ($GD$)}
	
	As one of the most widely used convergence indicators in SBSE (used in e.g., \cite{Sarro2016Multi,Shen2016Dynamic,Sarro2017Adaptive,Ferrucci2013Not,Zhang2013Empirical,Ouni2017Search,Assun2014A,Abdessalem2016Testing,Chen2017Self,Pascual2015Applying}), 
	$GD$~\cite{Veldhuizen1998} is to measure how close the obtained solution set is from the Pareto front. 
	Since the Pareto front is usually unknown \textit{a priori}, 
	a reference set, $\mathbf{R}$, 
	which consists of nondominated solutions of the collection of solutions obtained by all search algorithms considered,
	is typically used to represent the Pareto front in practice.
	Formally, 
	given a solution set $\mathbf{A} = \{\mathbf{a}_1, \mathbf{a}_2,...,\mathbf{a}_n\}$,
	$GD$ is defined as
	\begin{equation}
	GD(\mathbf{A}) = \frac{1}{n} \left(\sum_{i=1}^{n} (\min_{\mathbf{r}\in \mathbf{R}}d(\mathbf{a}_i, \mathbf{r}))^p\right)^{1/p}
	\label{eq:GD}
	\end{equation}
	where $d(\mathbf{a}_i, \mathbf{r})$ means the Euclidean distance between $\mathbf{a}_i$ and $\mathbf{r}$, 
	and $p$ is a parameter determining what kind of mean of the distances is used, e.g., the quadratic mean and arithmetic mean.
	
	The $GD$ value is to be minimized and the ideal value is zero, 
	which indicates that the set is precisely on the Pareto front.
	In the original version, 
	the parameter $p$ was set to 2. 
	Unfortunately, 
	this would make the evaluation value rather sensitive to outliers and also affected by the size of the solution set (when $N \rightarrow \infty$, $GD$ $\rightarrow 0$ even if the set is far away from the Pareto front~\cite{Schutze2012}).
	Setting $p=1$ has now been commonly accepted. 
	
	Compared to those dominance-based convergence indicators (e.g., $CI$ and $\mathcal{C}$), 
	$GD$ is more accurate in terms of measuring the closeness of solution sets to the Pareto front 
	due to it considering the distance between points.
	However, 
	a clear weakness of $GD$ is not being Pareto compliant~\cite{Knowles2002,Zitzler2003}. 
	This is very undesirable since $GD$, as a convergence indicator, 
	fails to provide reliable evaluation results with respect to the weakest assumption of the DM's preferences. 
	A simple example was given in~\cite{Zitzler2003}: 
	consider two solution sets $\mathbf{A} = \{(2, 5)\}$ and $\mathbf{B} = \{(3, 9)\}$ on a bi-objective minimization scenario, 
	where the reference set is $\mathbf{R} = \{(1,0),(0,10)\}$. 
	Clearly, 
	$\mathbf{A}$ dominates $\mathbf{B}$, 
	but $GD$ returns an opposite result: $GD(\mathbf{A}) = \sqrt{26} > GD(\mathbf{B}) = \sqrt{10}$.
	Recently, 
	a modified $GD$ was proposed to overcome this issue, called $GD^+$~\cite{Ishibuchi2015},
	where the Euclidean distance between $\mathbf{a}_i$ and $\mathbf{r}$ in Equation~(\ref{eq:GD})
	is modified by only considering the objectives where $\mathbf{r}$ is superior to $\mathbf{a}_i$. 
	Specifically, 
	\begin{equation}
	d^{+}(\mathbf{a}_i, \mathbf{r}) = (\sum_{j=1}^{m} (\max\{\mathbf{a}_{ij} - \mathbf{r}_{j}, 0\})^2)^{1/2}
	\end{equation}
	where $m$ denotes the number of objectives, 
	and $\mathbf{a}_{ij}$ denotes the value of solution $\mathbf{a}_{i}$ on the $j$th objective.
	This modification makes the indicator compliant with Pareto dominance. 
	Going back to the above example,
	now we have the evaluation results of $\mathbf{A}$ better than $\mathbf{B}$
	($GD^+(\mathbf{A}) = 2 < GD^+(\mathbf{B}) = 3$.
	Finally,
	note that for both $GD$ and $GD^+$, 
	normalization of solution sets is needed as their calculation involves objective blending~\cite{li2019quality}.

	\subsection{Spread ($\Delta$)}
	
	
	The indicator $Spread$ (aka $\Delta$)~\cite{Deb2002} and its variants~\cite{Zhou2006,Shen2016Dynamic} 
	have been commonly adopted to evaluate the diversity (i.e., spread and uniformity) of solution sets in the field, e.g., in~\cite{shen2018q,Shen2016Dynamic,Sayyad2013Optimum,Sayyad2013On,Ouni2017Search,Henard2015Combining,Durillo2009A,Zhang2013Empirical}. 
	Specifically, 
	the indicator $\Delta$ of a solution set $\mathbf{A}$ (assuming the set only consisting of nondominated solutions) 
	in a bi-objective scenario is defined as follows.
	\begin{equation}
	\Delta(\mathbf{A}) = \frac{d_{upper} + d_{bottom} + \sum_{i=1}^{n-1} |d_i - \overline{d}|}{d_{upper} + d_{bottom} + (n-1)\overline{d}}
	\label{eq:Spread}
	\end{equation}
	where $n$ denotes the size of $\mathbf{A}$, $d_i$ ($i=1,2,...n-1$) is the Euclidean distance between consecutive solutions in the $\mathbf{A}$, and $\overline{d}$ is the average of all the distances $d_i$.
	$d_{upper}$ and $d_{bottom}$ are the Euclidean distance between the two extreme solutions of $\mathbf{A}$ and the two extreme points of the Pareto front, respectively.
	
	A small $\Delta$ value is preferred, 
	which indicates a good distribution of the set in terms of both spread and uniformity. 
	When $\Delta = 0$ means that solutions in the set are equidistantly spaced and their boundaries reach the Pareto front extremes.  
	
	A major weakness of $\Delta$ (including its variants) is that it only works reliably on bi-objective problems as where nondominated solutions are located consecutively on either objective.
	With more objectives, 
	the neighbor of a solution on one objective may be far away on another objective~\cite{Li2014}. 
	This issue applies to any distance-based diversity indicator~\cite{li2019quality}. 
	For problems with more than two objectives,
	region division-based diversity indicators are more accurate~\cite{li2019quality}. 
	They typically divide the space into many equal-sized cells and then consider cells instead of solutions (e.g., counting the number of these cells). 
	This is based on the fact that a set of more diversified solutions usually populate more cells. 
	However, 
	such indicators may suffer from the curse of dimension as they typically need to record information of every cell. 
	In this regard,
	the diversity indicator $DCI$ \cite{Li2014d} may be a pragmatic option since its calculation only involves non-empty cells, 
	thus independent of the number of cells (linearly increasing computational cost in objective dimensionality).

	In addition, 
	another indicator $Spacing$ (aka $SP$) \cite{Schott1995} has also been used to evaluate the diversity of solution sets in e.g., \cite{Ramirez2016Comparative,Shen2016Dynamic}.
	However, 
	this indicator can only reflect the uniformity (not spread) of solution sets~\cite{li2019quality}.

	\subsection{Nondominated Front Size ($NFS$)}
	
	Used in e.g.~\cite{Henard2015Combining,Panichella2015Improving,Yoo2010Using}, 
	the $NFS$ (also called Pareto Front Size, $PFS$) is to simply count 
	how many nondominated solutions are in the obtained solution set. 
	However, 
	this indicator may not be very practical as in many cases all solutions in the obtained set are nondominated to each other, 
	particularly in many-objective optimization. 
	In addition, 
	as by definition duplicate solutions are nondominated to each other,
	a set full of duplicate solutions would be evaluated well by $NFS$ if there is no other solution in the set dominating them.
	
	As such, 
	a measure that only considers unique nondominated solutions which are not dominated by any other set seems more reasonable. 
	Specifically, 
	one can consider the ratio of the number of such solutions in each set to the size of the reference set 
	(which consists of unique nondominated solutions of the collections of solutions obtained by the algorithms). 
	In other words, 
	we quantify the contribution of each set to the combined nondominated front of all the sets.
	Formally,
	let $\mathbf{A_{unf}}$ be the unique nondominated front of a given solution set $\mathbf{A}$ (i.e., $\mathbf{A_{unf}} \subseteq \mathbf{A} \wedge \mathbf{A_{unf}}\preceq \mathbf{A} \wedge \forall \mathbf{a_i}\in \mathbf{A_{unf}}, \nexists \mathbf{a_j}\in \mathbf{A_{unf}}, \mathbf{j}\neq \mathbf{i}: \mathbf{a_j}\preceq \mathbf{a_i}$). 
	Then,
	the indicator, 
	denoted as Unique Nondominated Front Ratio ($UNFR$), 
	is defined as
	\begin{equation}
	UNFR(\mathbf{A}) = \frac{|\mathbf{a}\in \mathbf{A_{unf}}| \nexists \mathbf{r}\in \mathbf{R_{unf}}: \mathbf{r} \prec \mathbf{a}|}{|\mathbf{R_{unf}}|}
	\label{eq:UNFR}
	\end{equation}
	where $\mathbf{R_{unf}}$ denotes the reference set which consists of the unique nondominated solutions of the collections of all solutions produced.
	
	The $UNFR$ value is in the range of $[0,1]$. 
	A high value is preferred.
	Being zero means that for any solution in $\mathbf{A}$ there always exists some solution better in the other sets. 
	Being one means that for any solution in the other sets there always exists some solution in $\mathbf{A}$ better than (or at least equal to) it
	(i.e., the reference set is precisely comprised by solutions of $\mathbf{A}$).  
	In addition, 
	$UNFR$ is Pareto compliant.
	
	\begin{figure*}[tbp]
		\begin{center}
			\footnotesize
			\begin{tabular}{cc}
				\includegraphics[scale=0.30]{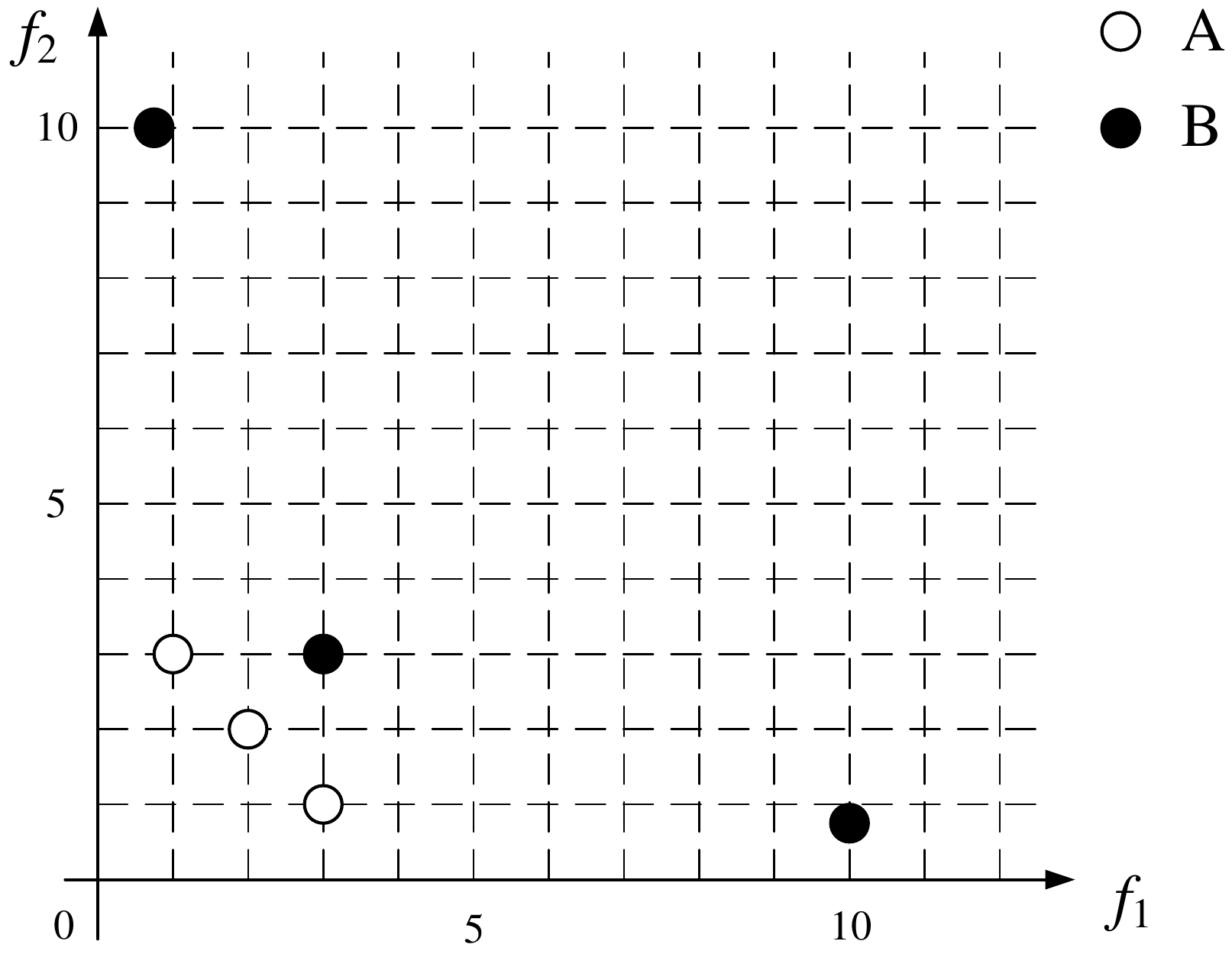}~~~~~~~~&~~~~~~~~
				\includegraphics[scale=0.30]{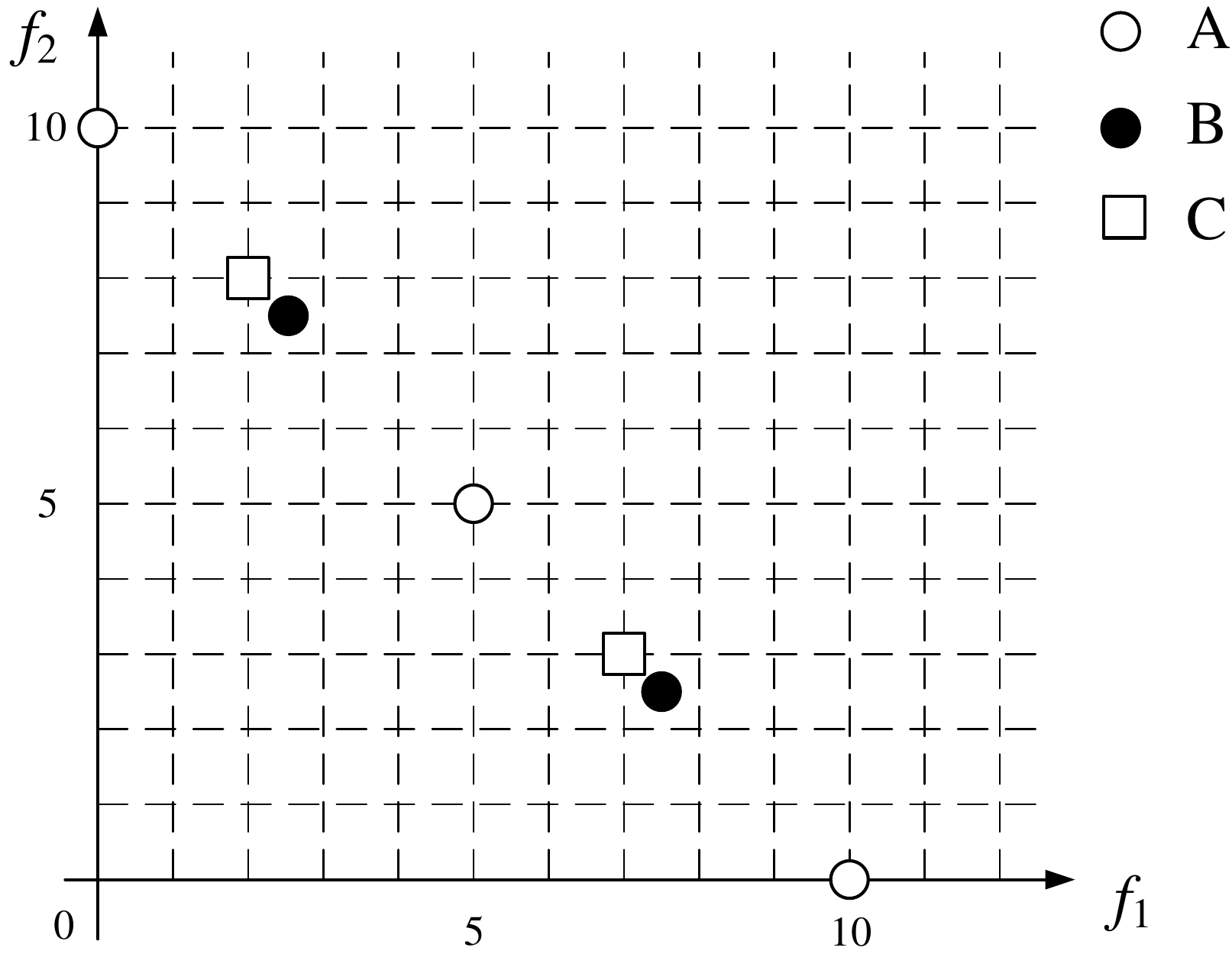}\\
				(a) &(b) \\
			\end{tabular}
		\end{center}
		\caption{Two examples that the collection of solution sets as the reference set may lead to misleading evaluations for $IGD$. 
			(a) For two bi-objective sets $\mathbf{A}=\{(1,3),(2,2),(3,1)\}$ and $\mathbf{B}=\{(0.75,10),(3,3),(10,0.75)\}$,
			$\mathbf{A}$ should be highly likely to be preferred to $\mathbf{B}$ as solutions of $\mathbf{B}$ are either dominated by some solution in $\mathbf{A}$ or slightly better on one objective but significantly worse on the other objective.
			but $IGD$ gives opposite results: $IGD(\mathbf{A}) \approx 2.80 > IGD(\mathbf{B}) \approx 1.08$.
			(b) For three bi-objective sets $\mathbf{A}=\{(0,10),(5,5),(0,10)\}$, 
			$\mathbf{B}=\{(2.5,7.5),(7.5,2.5)\}$ and $\mathbf{C}=\{(2,8),(7,3)\}$,
			in general $\mathbf{A}$ may be likely to be preferred by the DM than $\mathbf{B}$ and $\mathbf{C}$ 
			as it provides better spread and cardinality,
			but $IGD$ gives opposite results: $IGD(\mathbf{A}) \approx 1.82 > IGD(\mathbf{B}) \approx 1.72 > IGD(\mathbf{C}) \approx 1.61$.    
		}
		\label{Fig:RefSet}
	\end{figure*}
	
	\subsection{Inverted Generational Distance ($IGD$)}
	
	$IGD$ \cite{Coello2004} is a well-known indicator in the field (e.g., in \cite{shen2018q,Henard2015Combining,Fleck2017Model,Mansoor2016Multi,Assun2014A,Epitropakis2015Empirical,Frey2013Search,Gerasimou2016Search,Calinescu2017Designing,Tan2018Evolutionary,Mkaouer2016On,Mkaouer2014High}). 
	As the name suggests,
	$IGD$, an inversion of $GD$, 
	is to measure how close the Pareto front is to the obtained solution set.
	Formally, 
	given a solution set $\mathbf{A}$ and a reference set $\mathbf{R}$,
	$IGD$ is calculated as
	\begin{equation}
	IGD(\mathbf{A}) = \frac{1}{|\mathbf{R}|} \sum_{\mathbf{r}\in \mathbf{R}} \min_{\mathbf{a}\in \mathbf{A}} d(\mathbf{r}, \mathbf{a})
	\label{eq:IGD}
	\end{equation} 
	where $d_(\mathbf{r}, \mathbf{a})$ is the Euclidean distance between $\mathbf{r}$ and $\mathbf{a}$.
	A low $IGD$ value is preferable.
	
	$IGD$ is capable of reflecting the quality of a solution set in terms of all the four aspects: convergence, spread, uniformity, and cardinality. 
	However, 
	a major weakness of $IGD$ is that the evaluation results heavily depend on the behavior of its reference set. 
	A reference set of densely and uniformly distributed solutions along the Pareto front is required; 
	otherwise, it could easily return misleading results~\cite{li2019quality}. 
	This is particularly problematic in SBSE 
	since the reference set is created normally from the collection of all the obtained solutions; 
	its distribution cannot be controlled.

	Consider an example in \mbox{Figure~\ref{Fig:RefSet}(a)}, 
	where comparing two solution sets $\mathbf{A}$ and $\mathbf{B}$.
	The reference set is comprised of all the nondominated solutions, 
	i.e., the three solutions of $\mathbf{A}$ and the two boundary solutions of $\mathbf{B}$.
	As can be seen, 
	$\mathbf{B}$ performs significantly worse than $\mathbf{A}$ in terms of convergence, 
	with its solutions being either dominated by some solution in $\mathbf{A}$ or slightly better on one objective but much worse on the other objective; 
	thus $\mathbf{B}$ unlikely to be preferred by the DM.
	However, 
	$IGD$ gives an opposite evaluation: 
	$IGD(\mathbf{A}) \approx 2.80 > IGD(\mathbf{B}) \approx 1.08$.

	In addition,
	the way of how the reference set is created makes IGD prefer a specific distribution pattern 
	consistent with the majority of the considered solution sets~\cite{li2019quality}.
	In other words, 
	if a solution set is distributed very differently from others, 
	then the set is likely to assign a poor $IGD$ value 
	whatever its actual distribution is.
	\mbox{Figure~\ref{Fig:RefSet}(b)} is such an example.
	When comparing $\mathbf{A}$ with $\mathbf{B}$ (the reference set comprised of these two sets),
	we will have $\mathbf{A}$ evaluated better than $\mathbf{B}$
	($IGD(\mathbf{A}) \approx 1.41 < IGD(\mathbf{B}) \approx 2.12$).
	But if adding another set $\mathbf{C}$ 
	which has the similar distribution pattern to $\mathbf{B}$ into the evaluation, 
	and now the reference set is comprised of the three sets,
	we will have $\mathbf{A}$ worse than $\mathbf{B}$ 
	($IGD(\mathbf{A}) \approx 1.82 > IGD(\mathbf{B}) \approx 1.72$).
	A potential way to deal with this issue is to cluster crowded solutions in the reference set first 
	and then to consider these well-distributed clusters instead of arbitrarily-distributed points,
	as did in the indicator $PCI$~\cite{Li2015b}.
	Yet, this could induce another issue --- 
	how to properly cluster the solutions in the reference set subject to potentially highly irregular solution distribution.

	\subsection{Hypervolume ($HV$)}
	Like $IGD$, $HV$ \cite{Zitzler1998} evaluates the quality of a solution set in terms of all the four aspects.  
	Due to its desirable practical usability and theoretical properties, 
	$HV$ is arguably the most commonly used indicator in SBSE, 
	e.g., used in \cite{shen2018q,Shen2016Dynamic,Chicano2011Using,Olaechea2014Comparison,Ferrer2012Evolutionary,Pascual2015Applying,Tan2018Evolutionary,Hierons2016SIP,Xiang2018Configuring,Durillo2014Multi,Panichella2015Improving,Ouni2017Search,Fleck2017Model,wang2010multi}.
	For a solution set, 
	its $HV$ value is the volume of the union of the hypercubes determined by each of its solutions and a reference point. 
	It can be formulated as
	\begin{equation}
	HV(\mathbf{A}) = \lambda(\bigcup_{\mathbf{a}\in \mathbf{A}} \{\mathbf{x}|\mathbf{a}\prec \mathbf{x} \prec \mathbf{r}\})
	\label{eq:HV}
	\end{equation}
	where $\mathbf{r}$ denotes the reference point and $\lambda$ denotes the Lebesgue measure.
	A high $HV$ value is preferred.
	
	A limitation of the $HV$ indicator is its exponentially increasing computational time with respect to the number of objectives. 
	Many efforts have been made to reduce its running time, theoretically and practically (see \cite{li2019quality} for a summary), 
	which makes the indicator workable on a solution set with more than 10 objectives (under a reasonable set size). 
	
	\begin{figure}[tbp]
		\begin{center}
			\footnotesize
			\begin{tabular}{@{}c@{}c@{}}
				\includegraphics[scale=0.25]{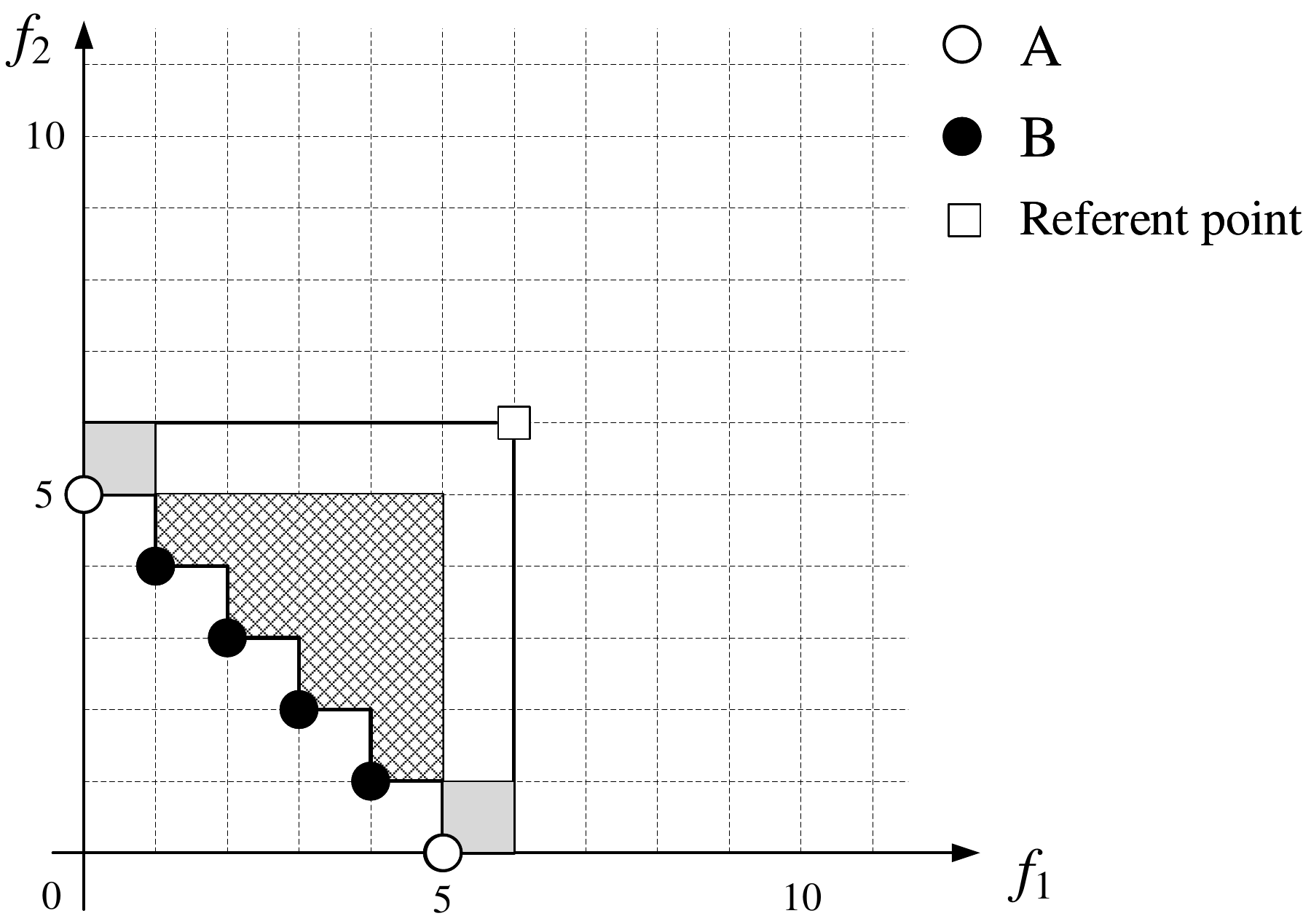}&
				\includegraphics[scale=0.25]{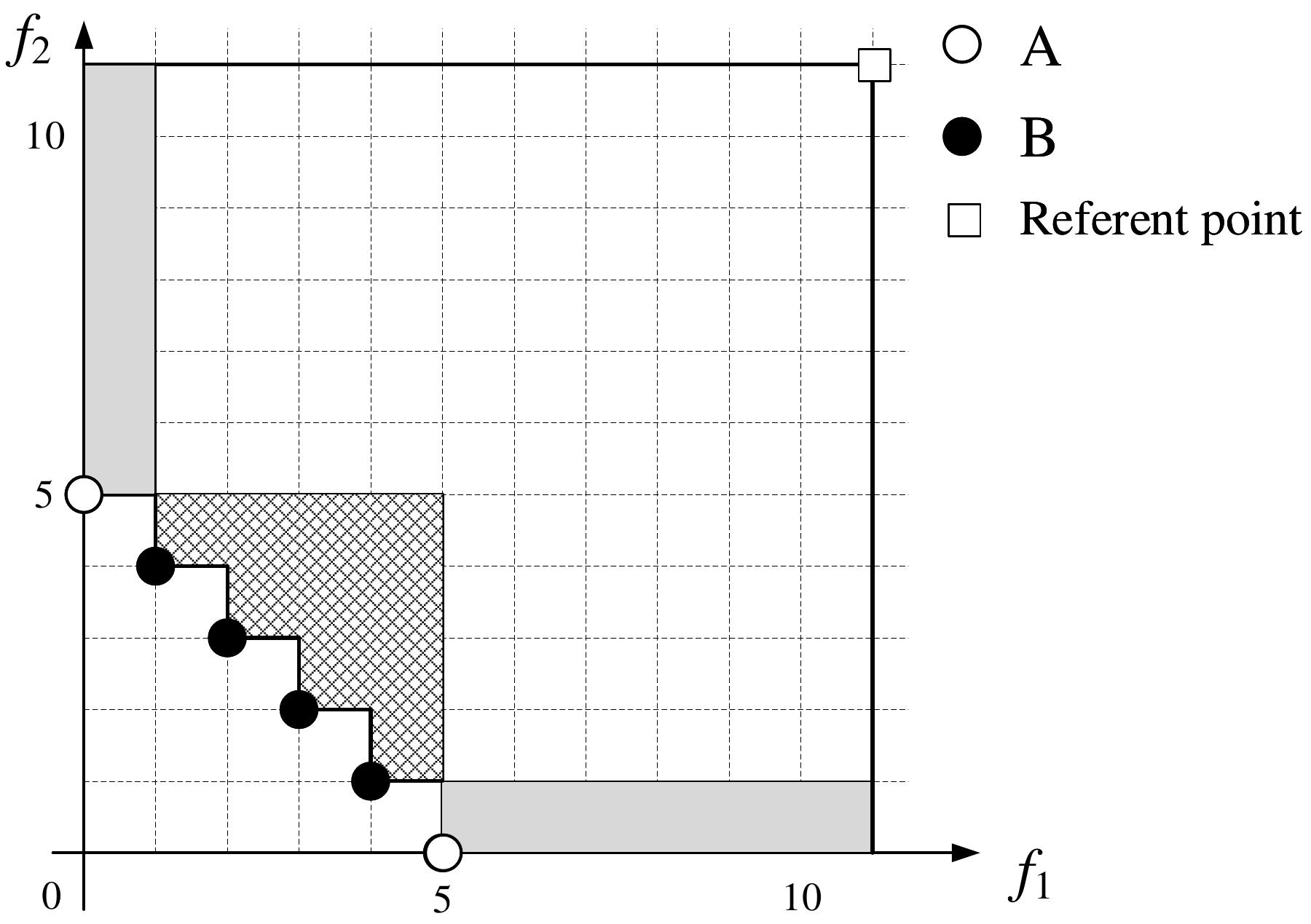}\\
				(a) &(b) \\
			\end{tabular}
		\end{center}
		\caption{An example that distinct reference points lead to that $HV$ prefers different solution sets,
			where the set $\mathbf{A}$ consists of two boundary solutions ($\mathbf{A}=\{(0,5),(5,0)\}$), 
			and the set $\mathbf{B}$ consists of four uniformly distributed inner solutions ($\mathbf{B}=\{(1,4),(2,3),(3,2),(4,1)\}$).
			The grey area $\subset HV(\mathbf{A})$ but $\nsubseteq HV(\mathbf{B})$ and the hatched area $\subset HV(\mathbf{B})$ but $\nsubseteq HV(\mathbf{A})$.
			In (a) where the reference point is $(6,6)$,
			$\mathbf{A}$ is evaluated worse than $\mathbf{B}$: $HV(\mathbf{A}) = 11 < HV(\mathbf{B}) = 19$.
			In (b) where the reference point is $(11,11)$,
			$\mathbf{A}$ is evaluated better than $\mathbf{B}$: $HV(\mathbf{A}) = 96 > HV(\mathbf{B}) = 94$.    
		}
		\label{Fig:HV}
	\end{figure}
	
	As stated previously, 
	$HV$ is in favor of knee points of a solution set, 
	thus a good choice when the DM prefers knee points of the problem's Pareto front. 
	In addition, 
	the settings of the reference point can affect its evaluation results. 
	Consider the two solution sets $\mathbf{A}$ and $\mathbf{B}$ in Figure~\ref{Fig:HV},
	where $\mathbf{A}$ consists of two boundary solutions, 
	and $\mathbf{B}$ consists of four uniformly distributed inner solutions.
	When the reference point is set to $(6,6)$ (Figure~\ref{Fig:HV}(a)), 
	$\mathbf{A}$ is evaluated worse than $\mathbf{B}$.
	When the reference point is set to $(11,11)$ (Figure~\ref{Fig:HV}(b)),
	$\mathbf{A}$ is evaluated better than $\mathbf{B}$.
	Fortunately, we can make good use of such a behavior of $HV$ to enable the indicator to reflect the DM's preferences. 
	If the DM prefers the extreme points, 
	then a reference point can be set to be fairly distant from the solution sets' boundaries, 
	e.g., doubling the Pareto front's range, namely,
	$\mathbf{r}_i = nadir_i + l_i$ where $nadir_i$ is the nadir point of the Pareto front (or the reference set, i.e., the combined nondominated front) on its $i$th objective,
	and $l_i$ is the range of the Pareto front (or the reference set) on the $i$th objective.
	If there is no clear preference from the DM, 
	unfortunately, 
	no consensus regarding how to set the reference point has been reached in the multi-objective optimization field.
	A common practice is to set it 1.1 times of the range of the combined nondominated front (i.e., $\mathbf{r}_i = nadir_i + l_i/10$).
	Some recent studies~\cite{Ishibuchi2018} suggested to set it as $\mathbf{r}_i = nadir_i + l_i/h$, 
	where $h$ is an integer subject to $C_{m-1}^{h+m-1} \leq n < C_{m-1}^{h+m}$ ($m$ and $n$ being the number of objectives and the size of the considered set, respectively).
	In any case,
	the reference point setting is non-trivial ---
	an appropriate setting needs to consider not only the number of objectives and the size of the solution set, 
	but also the actual dimensionality of the set, its shape, etc.

	\subsection{$\epsilon$-indicator}
	$\epsilon$-indicator is another well-established comprehensive indicator frequently appearing in SBSE, e.g., \cite{Henard2015Combining,Epitropakis2015Empirical,Gerasimou2016Search,Calinescu2017Designing,Wagner2012Multi}.
	It measures the maximum difference between two solution sets and can be defined as
	\begin{equation}
	\epsilon(\mathbf{A},\mathbf{B}) = \max_{\mathbf{b}\in \mathbf{B}} ~\min_{\mathbf{a}\in \mathbf{A}} \max_{i\in \{1...m\}} \mathbf{a}_i - \mathbf{b}_i
	\label{eq:addepsilon}
	\end{equation}
	where $\mathbf{a}_{i}$ denotes the objective of $\mathbf{a}$ for the $i$th objective
	and $m$ is the number of objectives. 
	A low value is preferred. 
	$\epsilon(\mathbf{A},\mathbf{B}) \leq 0$ implies that $\mathbf{A}$ weakly dominates $\mathbf{B}$.
	When replacing $\mathbf{B}$ with a reference set that represents the Pareto front, 
	the $\epsilon$-indicator becomes a unary indicator, 
	measuring the gap of the considered set to the Pareto front.
	
	$\epsilon$-indicator is Pareto compliant and user friendly (parameter-free and quadratic computational effort).
	Yet, 
	the calculation of $\epsilon$-indicator only involves one particular objective of one particular solution in either set (where the maximum difference is),
	rendering its evaluation omitting the difference on other objectives and other solutions. 
	This may lead to different solution sets having same/similar evaluation results, 
	as reported in \cite{Liefooghe2016}.
	In addition, 
	in some studies~\cite{Ravber2017}, 
	$\epsilon$-indicator has been empirically found to behave very similarly as $HV$ in ranking solution sets.

	\subsection{Summary}

	\input{Tables/qi-revisit}

	Table~\ref{Table:Summary} summarizes the above 12 indicators on several aspects, namely, 
	(i) what kind of quality aspect(s) they are able to reflect, 
	(ii) if they are Pareto compliant, 
	(iii) what we need to take care of when using them, 
	and (iv) what situation they are suitable for. 
	The following guidelines can be derived from the table. 
	
	\begin{itemize}
		\item If the DM wants to know the convergence quality of a solution set to the Pareto front, 
		$GD^+$ (instead of $GD$) could be an ideal choice --- 
		it is Pareto compliant and the reference set required can be set as the combined nondominated front of all the considered sets, not necessarily a set of uniformly-distributed points. 
		If the DM wants to know the relative quality between two solution sets in terms of the Pareto dominance relation, 
		$CI$ (or $\mathcal{C}$) could be a choice.

		\item If the DM wants to know the diversity quality (both spread and uniformity) of a solution set, 
		for bi-objective cases, 
		$\Delta$ is a good choice; 
		for problems with more objectives, 
		$DCI$ can be used.
		$SP$ can only reflect the uniformity of a solution set which may not be very useful --- uniformly distributed solutions concentrating in a tiny area typically not in the DM's favor.

		\item The indicator $UNFR$ should replace $NFS$ to measure the cardinality of solution sets.

		\item Regarding comprehensive evaluation indicators, 
		$HV$ can generally be the first choice, 
		especially when the DM prefers knee points.
		In addition, 
		if the DM prefers extreme solutions, 
		the reference point needs to be set fairly distant from the solution sets' boundaries.
		$\epsilon$-indicator is user-friendly, 
		but is less sensitive to solution sets' quality difference than $HV$ since its value only lies upon one particular solution on one particular objective. 
		$IGD$ may not be very practical as it requires a Pareto front representation 
		consisting of densely and uniformly distributed points.

	\end{itemize}

	\section{Methodological Guidance to Quality Evaluation in Pareto-based SBSE}
	\label{sec:guidance}
	
	In this section,
	we provide guidance on how to select and use quality evaluation methods in Pareto-based SBSE. 
	As discussed previously,
	selecting and using quality evaluation methods needs to be aligned with the DM's preferences\footnote{For convenience, 
	from this point forwards we use DM's preference information as a general term, 
	which refers to not only the preferences that the DM articulates but also the requirements derived from problem nature and contextual information.}. 
	A solution set being evaluated better means nothing but its solutions having a bigger chance to be picked out by the DM.
	However, 
	to different problems or even to the same problem but under different circumstances, 
	the articulation of the DM's preferences may differ. 
	In some cases, 
	the DM is confident to articulate their preferences (or can be easily derived from contextual information); 
	e.g., they see one objective more important than others. 
	In some cases, 
	the DM may experience difficulty in precisely articulating their preferences;
	e.g., they are only able to provide some vague preference information such as a fuzzy region around one point.
	In some other cases,
	the DM's preferences may not be available at all; 
	e.g., when the DM wants to see what the whole Pareto front looks like before articulating their preferences.
	Therefore, 
	quality evaluation needs to be conducted in accordance with different cases.
	Next, 
	we consider four general cases of quality evaluation with respect to the DM's preferences.

	\subsection{When the DM's Preferences Are Clear}
	\label{sec:p1}
	
	The case of the DM's preferences being clear can often fall into two categories over Pareto-based SBSE problems. 
	The first is when relative importance/weighting among the objectives considered can be explicitly expressed and quantified, 
	e.g., in~\cite{Shen2016Dynamic}. 
	It is worth noting that the weighting between objectives may not need to be fixed \textit{a priori}. 
	For example, 
	in the case of interactive Pareto-based SBSE for software modeling and architecting problems~\cite{Simons2012Elegant}, 
	the DM is asked to explicitly rank the relative importance of the objectives as the search proceeds. 
	Under this circumstance, 
	the sum of the weighted objectives can be used to find the fittest solution from a solution set, 
	and then determine the quality of the set.
	
	The other category concerns when the DM prefers some objective to some other (i.e., a clear priority can be assumed, which is a unique situation that often implies some clear contextual information of a hard requirement under the SBSE problem),
	or when the DM is only interested in solutions which is up to scratch on some objective (which could be seen as a constraint). 
	This happens frequently in the software product line configuration problem~\cite{Sayyad2013Optimum,Sayyad2013Scalable,Sayyad2013On,Hierons2016SIP,Henard2015Combining,Olaechea2014Comparison,Xiang2018Configuring}, 
	where the correctness of the products (i.e., the feature model's dependency compliance) is always of higher priority than other objectives such as the richness and the cost of the model --- 
	only the solutions (products) that achieve full dependency compliance are of interest. 
	This is obvious, 
	as a violation of dependency implies faulty and incorrect configuration, 
	thus valueless in practice.
	A similar situation applies to the test case generation problem~\cite{Zheng2016Multi,Panichella2015Improving,Shahbazi2016Black,Hierons2019Many,parejo2016multi}
	where the DM is typically interested in test suites with full coverage.
	In addition, 
	the DM may only be interested in solutions that reach a certain level on some objective.
	For example,
	in software deployment and maintenance, 
	it is not uncommon to have a statement like ``\texttt{The software service shall be available for at least 95\% of the time}''. 
	In such a case, 
	it is rather clear that any value of availability less than 95\% is unacceptable, 
	while anything beyond 95\% can be considered.
	

	An appropriate way to perform evaluation under the above circumstance is to transfer the DM's preferences into the solution set to be evaluated.
	This can be done by first removing solutions that are irrelevant from the set.
	After that,
	the set of the remaining solutions is evaluated, 
	subject to two situations: 
	if the remaining solutions are of the same value on the objective(s) where the DM articulates their preferences,
	then the quality evaluation is performed only on the other objectives; 
	otherwise, the evaluation is done on all the objectives. 
	The former has been commonly seen when the DM is only interested in solutions which achieve the best of the objective, 
	such as the solutions with full coverage for the test case generation problem,
	whereas the latter often applies when the DM is interested in a particular threshold of solution quality on the objective, 
	such as in the software deployment and maintenance case mentioned above, only the solutions with availability values not less than 95\% would be evaluated.


	\subsection{When the DM's Preferences Are Vague/Rough}
		\label{sec:p2}

	
	
	
	
	It is not uncommon that there exist important, 
	yet imprecise preferences in the SDLC. 
	In general, 
	they are mainly derived from the non-functional requirements recorded in documentations, notes, and specifications, which are often vague in nature, 
	as in~\cite{Heaven2011Simulating}~\cite{busari2017radar}~\cite{Martens2010Automatically}~\cite{Letier2014Uncertainty}~\cite{Abdeen2014Multi}.
	For example, in the software configuration and adaptation problem,
	some statements may be rather ambiguous like ``\texttt{the first objective should be reasonable and the others are as good as possible}''.
	In such a situation, 
	one may not be able to integrate the preferences into the quality evaluation 
	since it is not possible to quantify qualitative descriptions like ``reasonable''.  
	As such,
	a safe choice is to treat them as a general multi-objective optimization case (i.e., without specific preferences).  
	
	In other situations,
	the SBSE researchers/practitioners may give some preference information around some values/thresholds on one (or several) objective. 
	This, in contrast to the case of the DM's preferences being clear, 
	allows some tolerances on the specified value/threshold. 
	For example,
	the software may have a requirement stating that ``\texttt{the cost shall be low while the product shall support ideally up to 3000 simultaneous users}''. 
	This typically happens for SME where the budget of a software project (e.g., money for buying required Cloud resources or the consumption of data centers) is low, 
	and thus it is more realistic to set a threshold point such that certain level of performance (e.g., $3000$ simultaneous users) would be sufficient (anything beyond is deemed as equivalent). 
	However, while the requirement gives a clear cap of the best performance expected, 
	it does not constrain on the worst case, 
	implying that it allows tolerances when the $3000$ users goal cannot be met.
	
	Despite not impossible, 
	it can be a challenging task to find a quality indicator that is able to reflect such preference information. 
	First, 
	the quality indicator should be capable of accommodating such preference information in the sense that the evaluation results can embody it. 
	Second, 
	the introduction of the preferences should neither compromise the general quality aspect that the indicator reflects,
	nor violate properties that the indicator complies with (e.g., being Pareto compliant).
	In this regard, 
	the indicator $HV$~\cite{Zitzler1998} could be a good choice since it (i) can relatively easily integrate the DM's preference information~\cite{Zitzler2007a,Wagner2010} 
	and (ii) can still be Pareto compliant after a careful introduction of the DM's preference information~\cite{Zitzler2007a}. 
	
	To integrate preferences into $HV$, 
	one approach, called the weighted $HV$ presented in~\cite{Zitzler2007a}, 
	is to interpret the $HV$ value as the volume of the objective space enclosed by the attainment function~\cite{Fonseca2001} and the axes. 
	Here, 
	the attainment function is to give for each vector in the objective space the probability that it is weakly dominated by the outcome of the solution set.
	Then, to give different weights to different regions by a weight distribution function, 
	the weighted $HV$ is calculated as the integral over the product of the weight distribution function and the attainment function~\cite{Zitzler2007a}. 
	This essentially transforms the preference information into a weight distribution function to unequalize the $HV$ contribution from different regions. 
	\rev{However, 
	it is not trivial to construct a weight distribution function that is able to reflect preferences expressed by the SBSE researchers/practitioners. 
	Even for the situation that the preference information is clear (e.g., clear weighting between objectives), 
	the preferences cannot be used as the weight distribution function 
	because of the interaction of the weight distribution function 
	and the attainment function in the calculation.}
	
	\begin{figure*}[tbp]
		\begin{center}
			\footnotesize
			\begin{tabular}{cccc}
				\includegraphics[scale=0.28]{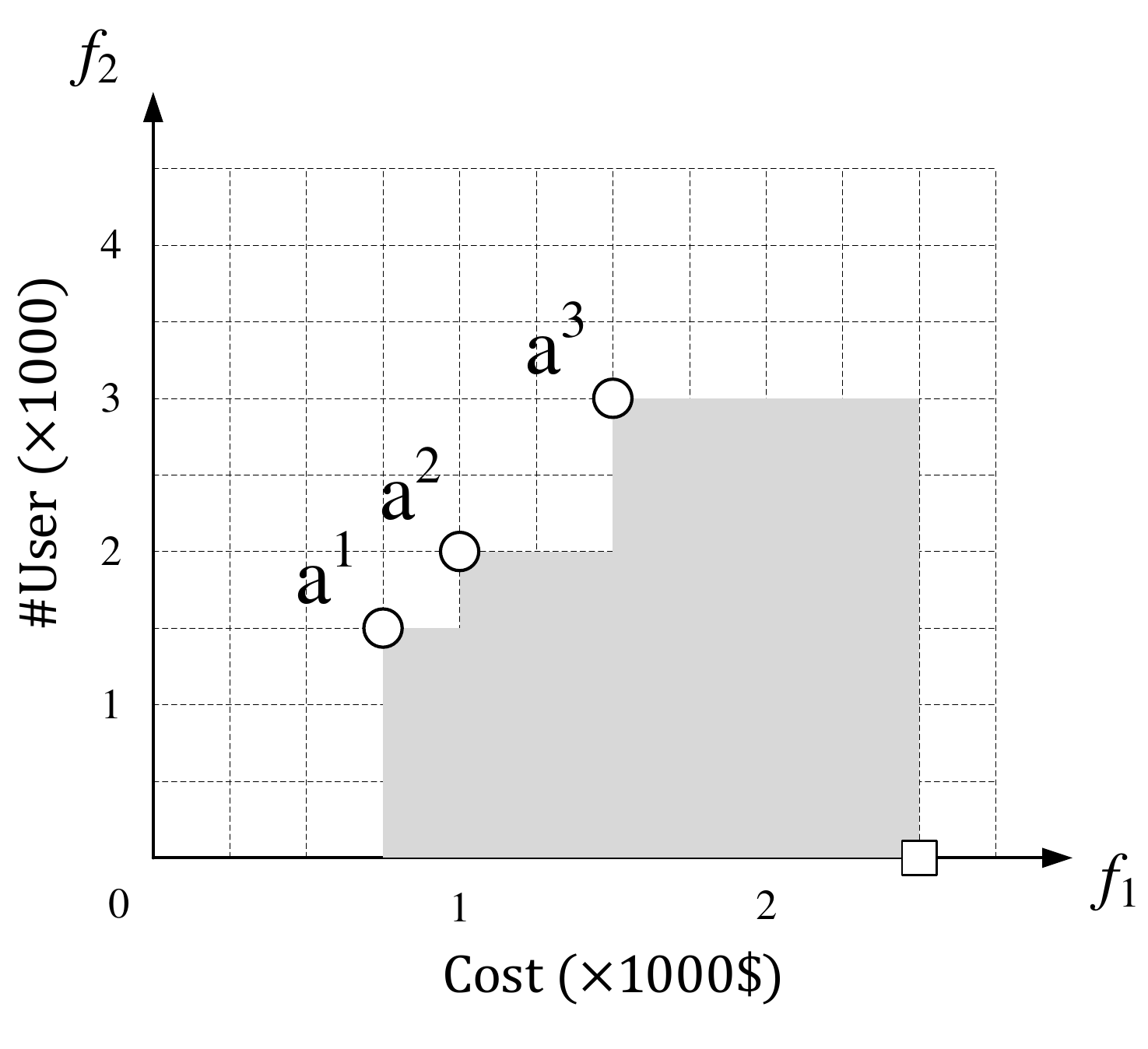}&
				\includegraphics[scale=0.28]{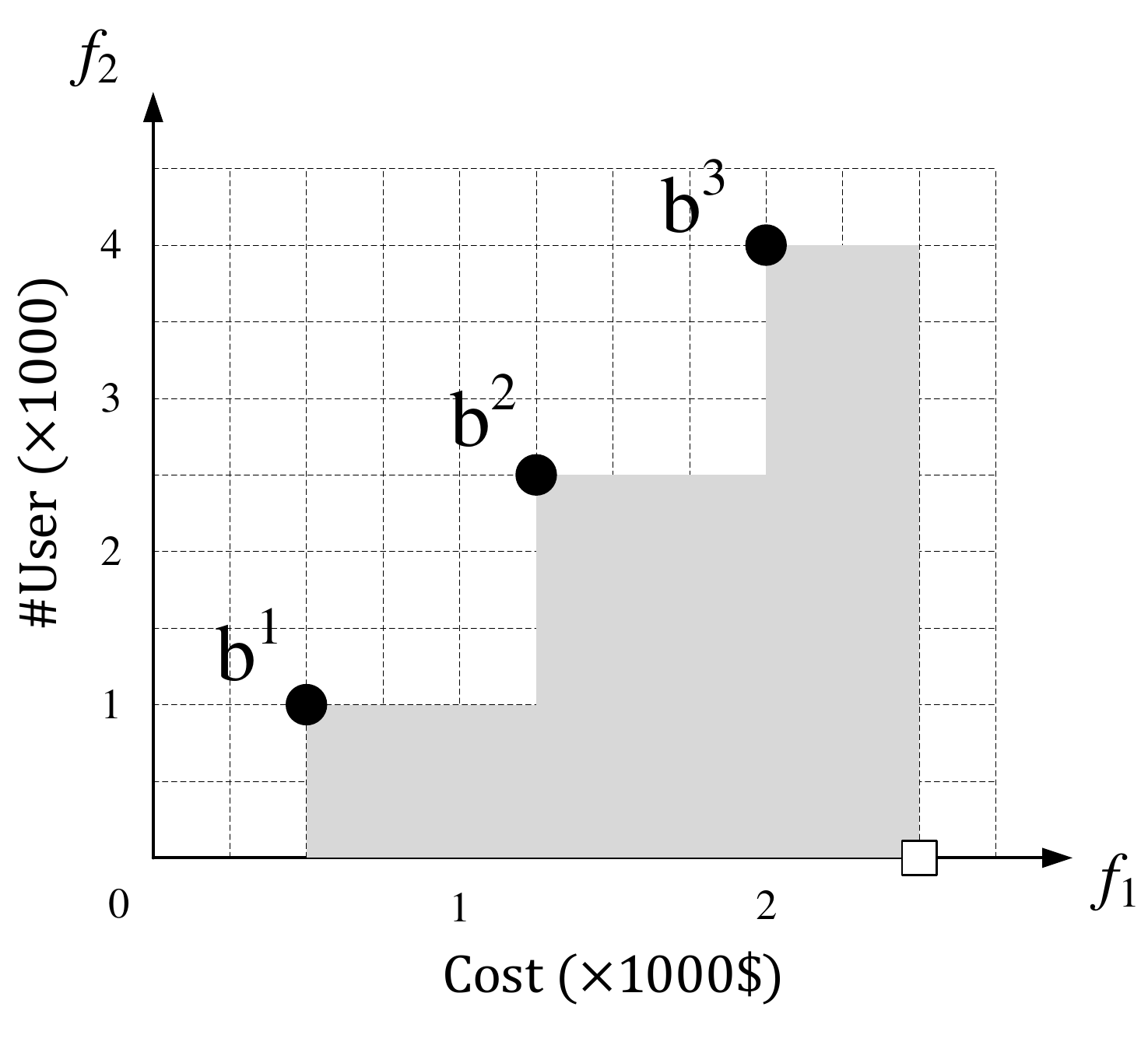}&
				\includegraphics[scale=0.28]{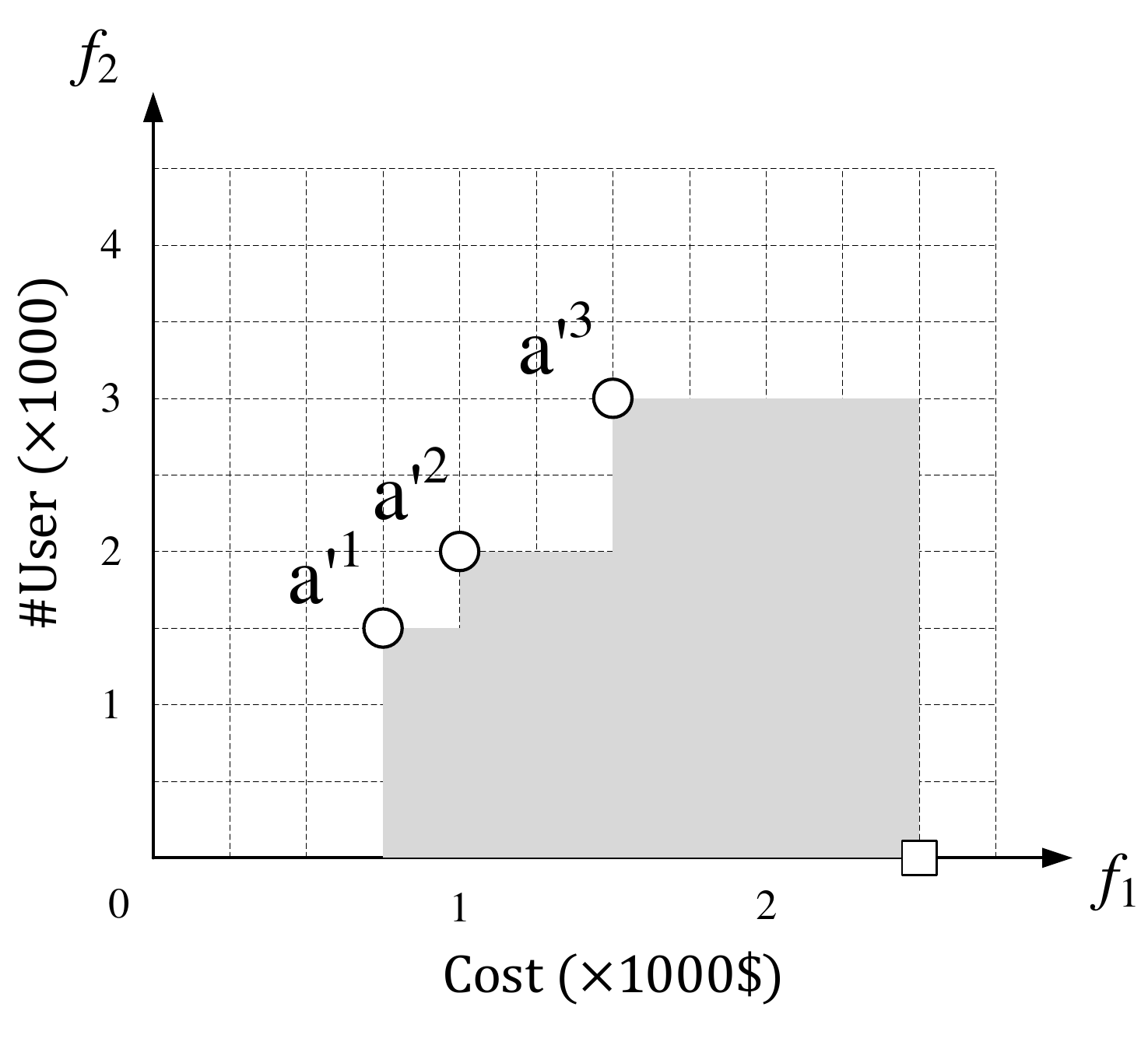}&
				\includegraphics[scale=0.28]{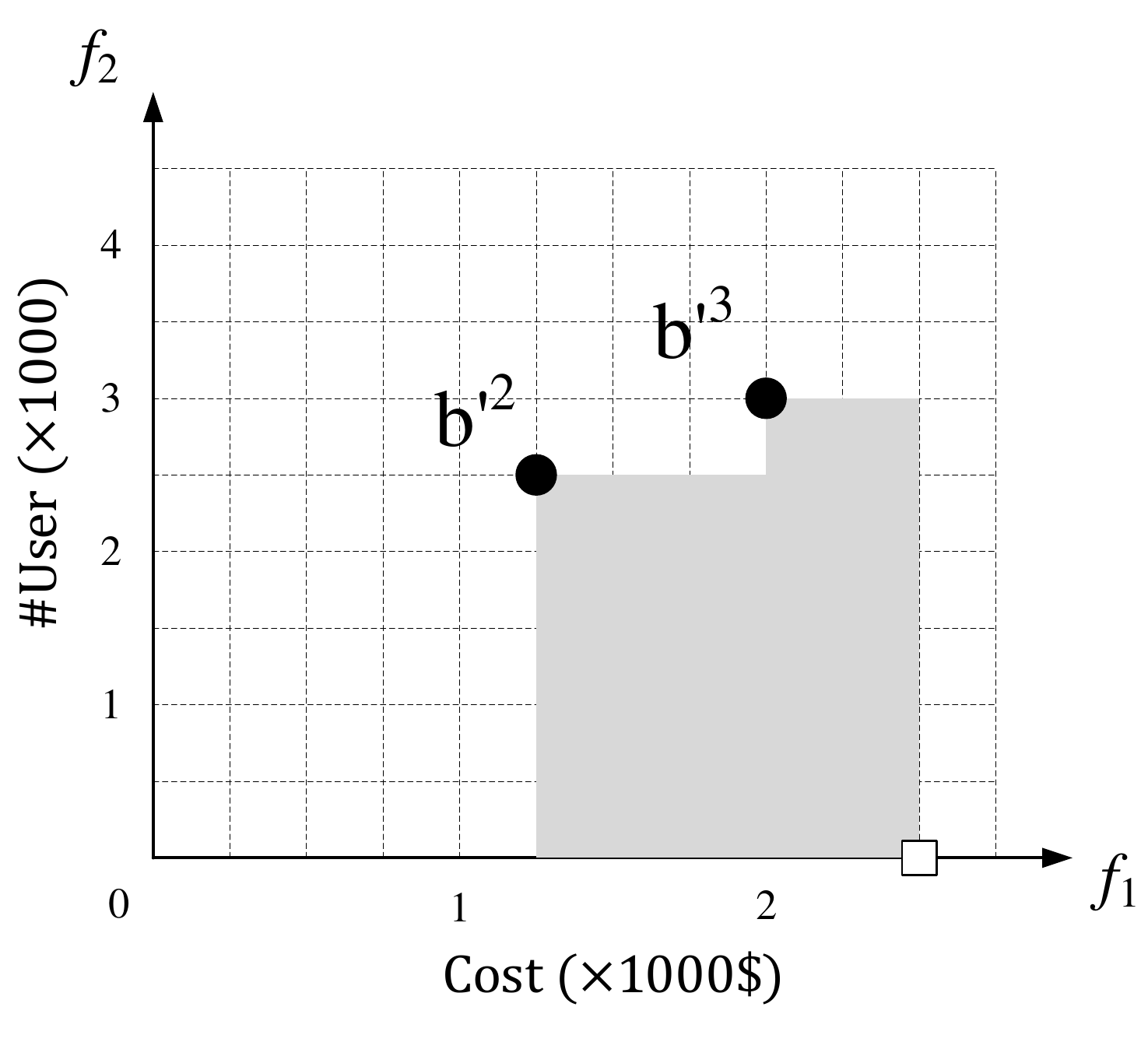}\\
				(a) $HV(\mathbf{A})=4,375,000$ &(b) $HV(\mathbf{B})=4,625,000$ & (c) $HV(\mathbf{A'})=4,375,000$ & (d) $HV(\mathbf{B'})=3,375,000$\\
			\end{tabular}
		\end{center}
		\caption{HV comparison of with/without integrating the DM's preferences ``\texttt{the cost shall be low while the product should be able to support at least 1500 simultaneous users and ideally reach 3000 users}'' into solutions on the basis of transformation function Equation~(\ref{eq:Transformation}). Without considering the preferences ((a) and (b)), the solution set $\mathbf{B}$ is evaluated by $HV$ better than $\mathbf{A}$ as it can provide more diverse solutions. Yet, when considering the preferences, the solution $\mathbf{b}^1$ in $\mathbf{B}$ will be no interest to the DM (thus discarded) as its supported user number is less than $1500$, and the number of users that $\mathbf{b}^3$ supports can be regarded down to $3000$ as $3000$ is the best expected value. After this transformation ((c) and (d)), the set $\mathbf{B'}$ is evaluated significantly worse than $\mathbf{A'}$.      
		}
		\label{Fig:PrefHV}
	\end{figure*}
	
	Another (perhaps more pragmatic) approach is to directly transform the original solutions 
	into new solutions which accommodate the preference information, 
	and then apply $HV$ (or other quality indicators) to the new solutions, 
	provided that such a transformation is in line with the selected indicator. 
	For instance,
	consider the above example that the cost shall be low while the product shall ideally support up to $3000$ users. 
	Let us say that there are two solutions $\mathbf{a} = (1500, 3000)$ and $\mathbf{b}=(2000, 4000)$ obtained for this problem. 
	Solution $\mathbf{a}$ has a lower cost while solution $\mathbf{b}$ can support more users. 
	However, 
	according to the preference information, 
	for the number of users anything beyond $3000$ can be deemed as equivalent.
	As such, 
	the user number of the solution $\mathbf{b}$ can be transformed to $3000$, 
	that is, now $\mathbf{b}=(2000, 3000)$, worse than (i.e., dominated by) $\mathbf{a}$.
	The $HV$ indicator can capture such dominance relation information --- a dominated solution is always evaluated worse by $HV$ than one dominating it. 
	Next, 
	we look at a case study based on this example to see how such transformation affects the evaluation results.
	
	Consider a situation of designing a product with the requirements that ``\texttt{the cost shall be low while the product should be able to support at least 1500 simultaneous users and ideally reach 3000 users}''.
	As can be seen, 
	the first objective cost is a normal one (i.e., the lower the better), 
	while for the second objective the number of simultaneous users,
	there are two types of preferences: clear one and vague one. 
	The statement ``\texttt{support at least 1500 users}'' is a clear one, 
	which means the product is useless if it cannot support 1500 users.
	The statement ``\texttt{ideally reach 3000 users}'' is a vague one, 
	which implies that despite the threshold, 
	it is acceptable to support less users and it will have the same level of satisfaction even if more users are supportable. 
	As such,
	we can have the following transformation function.
	
	\begin{equation}
	\mathbf{a}'_2 = \left \{ \begin{array}{@{}ll}
	3000, & \mathbf{a}_2>3000 \\
	\mathbf{a}_2, & 1500 \leq \mathbf{a}_2 \leq 3000 \\
	\textrm{to disgard}~\mathbf{a}, & \mathbf{a}_2<1500
	\end{array}\right.
	\label{eq:Transformation}
	\end{equation}
	where $\mathbf{a}'_i$ denotes the transformed value of solution $\mathbf{a}$ on the $i$th objective.
	
	Now let us assume two solution sets $\mathbf{A} = \{\mathbf{a}^1, \mathbf{a}^2, \mathbf{a}^3\}, \mathbf{B} = \{\mathbf{b}^1, \mathbf{b}^2, \mathbf{b}^3\}$ obtained by two search algorithms, 
	where $\mathbf{a}^1 = (750,1500), \mathbf{a}^2 = (1000,2000), \mathbf{a}^3=(1500,3000), \mathbf{b}^1=(500,1000), \mathbf{b}^2=(1250,2500), \mathbf{b}^3=(2000,4000)$, shown in Figure~\ref{Fig:PrefHV}(a) and (b).
	We want to evaluate and compare them under the circumstances with/without the preference information given above to see how transferring preferences into solutions affects the evaluation results.
	As seen in Figure~\ref{Fig:PrefHV}(a) and (b), 
	without considering the preferences
	$\mathbf{A}$ is evaluated worse than $\mathbf{B}$ ($HV(\mathbf{A}) = 4,375,000 < \mathbf{B} = 4,625,000$.
	This makes sense as the solutions of $\mathbf{B}$ spread more widely than those of $\mathbf{A}$.
	Yet, 
	when considering the preferences of the DM (transferred by Equation~(\ref{eq:Transformation})),
	while the set $\mathbf{A}$ stays unchanged ($\mathbf{A'} = \mathbf{A}$), 
	the solution $\mathbf{b}^1$ will be discarded and the solution $\mathbf{b}^3$ will become $(2000, 3000)$.
	As a result, 
	$\mathbf{A'}$ is evaluated significantly better than $\mathbf{B'}$ ($HV(\mathbf{A'}) = 4,375,000 > \mathbf{B'} = 3,375,000$),
	as shown in Figure~\ref{Fig:PrefHV}(c) and (d).
	This shows that the integration of the DM's preferences can completely change the evaluation results between solution sets.

	\subsection{When the DM's Interest Is in Some Specific Parts of the Pareto Front}
		\label{sec:p3}
	Sometimes, 
	the DM may be more interested in specific part/solutions of the Pareto front than others.
	Knee points are certainly among such solutions, 
	preferred in many situations, e.g., in~\cite{Sarro2017Adaptive}~\cite{Gueorguiev2009Software}~\cite{Ferrucci2013Not}~\cite{Wang2015Cost}~\cite{Fleck2017Model}~\cite{Li2010SLA}~\cite{Chen2017Self}~\cite{Chen2018FEMOSAA}~\cite{Mkaouer2016On}.
	Knee points are points on the Pareto front where a small improvement on one objective would lead to a large deterioration on at least one other objective. 
	They represent ``good'' trade-offs between conflicting objectives,
	thus naturally more of interest to the DM. 
	For example,
	on the cloud autoscaling problem~\cite{Chen2017Self}, 
	where different cloud tenants (users) may introduce conflicting objectives due to the interference and shared infrastructure.  
	From the perspective of the cloud vendor, 
	ensuring fairness among tenants of the same class is often the top priority 
	and thus the knee solutions are more of interest. 
	As we explained previously, 
	$HV$ is a good choice in such a situation, 
	alongside other indicators like the $\epsilon$-indicator~\cite{Zitzler2003}, $IGD^+$~\cite{Ishibuchi2015} and $PCI$~\cite{Li2015b},
	whereas unfortunately $IGD$~\cite{Coello2004} is not one of them, 
	despite being widely used, e.g., in~\cite{Fleck2017Model,Mkaouer2016On}.
	
	Another relatively common situation is that the DM may be more interested in the extreme solutions (e.g., in~\cite{Zhang2007The}~\cite{Wada2012E}~\cite{Tan2018Evolutionary}), 
	namely, solutions achieving the best on one objective or another.
	For example,
	for the service composition problem~\cite{Wada2012E}, 
	one may prefer the extreme solutions around the edges, 
	e.g., those with low latency but high cost, 
	or vice versa. 
	For this situation,
	$HV$ can also be a viable solution.
	As shown previously (Section~5.7),
	setting the reference point fairly distant from the combined nondominated solution set gives the extreme solutions bigger weighting on the evaluation results. 
	Besides, 
	one may directly compare solution sets through their best value on the corresponding objective(s).
	Such a $DOE$ measure, 
	in contrast to $HV$ which provides comprehensive evaluation results,
	returns the objective values which are straightforward for the DM to understand. 


	\subsection{When the DM's Preferences Are Completely Unavailable}
		\label{sec:p4}
	As can be seen in Table~\ref{table:papers-preferences},
	the majority of studies in Pareto-based SBSE effectively do not involve any preference. 
	For this situation,
	a solution set that well represents the whole Pareto front is preferred.  
	As aforementioned,
	the ``representation'' can be broken down to the quality aspects 
	convergence, diversity (i.e., spread and uniformity), and cardinality.
	Naturally,
	it is expected to consider quality indicators which (together) are able to cover all of them. 
	
	In general, 
	there are two ways to implement that in practice.  
	One is to consider several indicators, 
	each responsible for one specific aspect.
	For example,
	$GD^+$~\cite{Ishibuchi2015} is for a solution set's convergence,
	$\Delta$~\cite{Deb2002} for diversity (under the bi-objective circumstance), 
	and $UNFR$ for cardinality.
	The other one is to consider a comprehensive indicator to evaluate all the aspects. 
	Such indicators include $HV$, $IGD$, and $\epsilon$-indicator. 
	Today, there is a tendency to use comprehensive indicators. 
	Numerous recent studies used $HV$ and $IGD$.
	However, 
	as explained previously,
	$IGD$ may not be an ideal indicator in Pareto-based SBSE as a Pareto front representation with densely and uniformly distributed points is usually unavailable in practice.

	In addition,
	when using comprehensive indicators
	we suggest to consider multiple differently-behaving indicators if applicable.
	Each indicator has its own (explicitly or implicitly) preferences.
	A solution set evaluated better on an indicator is often evaluated better as well on another similar indicator; 
	this means nothing but the set favored under this type of preference.  
	When a solution set is evaluated better on all of the considered indicators whose preferences are quite different, 
	then that set certainly has a higher chance to be chosen by the DM.
	Unfortunately,
	many comprehensive indicators behave similarly as $HV$~\cite{Liefooghe2016,Ravber2017}, namely, preferring knee points of the Pareto front rather than a set of uniformly-distributed solutions on the Pareto front, 
	such as $R2$~\cite{Hansen1998} and $\epsilon$-indicator (except $IGD$ which, however, is not applicable typically). 
	Therefore,
	as a supplement to $HV$, 
	considering a quality indicator that can well evaluate the diversity of a solution set sounds reasonable.
	In this regard, 
	the indicator $\Delta$ ($Spread$)~\cite{Deb2002} may be chosen for the bi-objective case and $DCI$~\cite{Li2014d} for the more-objective case. 
	
	\begin{figure*}[tbp]
		\centering
		\includegraphics[width=2.1\columnwidth]{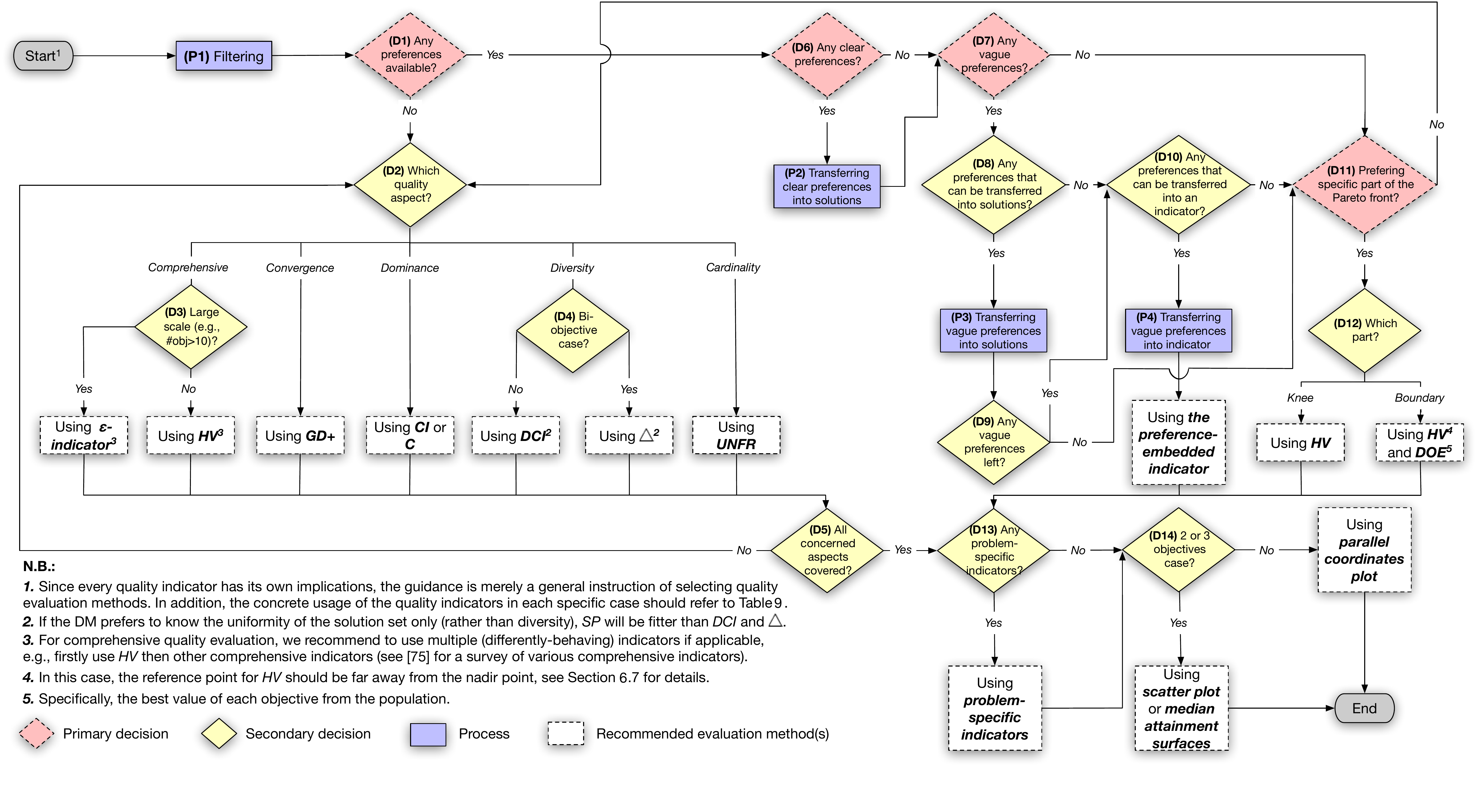}
		\caption{General procedure of quality evaluation in Pareto-based SBSE.}
		\label{Fig:Guidance}
	\end{figure*}

	\subsection{Aided Evaluation Methods}
	
	The above are four general cases of solution sets' quality evaluation on the basis of the DM's preferences.
	On top of those,
	there exist some quality indicators for specific SBSE scenarios (see Table~\ref{table:papers-preferences}), 
	which we call problem-specific indicators ($PSI$). 
	For example,
	in the library recommendation~\cite{Ouni2017Search},
	top-k accuracy, precision, and recall on history datasets are commonly used $PSI$ for evaluating recommendation systems.
	For another example, 
	in the software modularization problem the $PSI$ MoJoFM~\cite{Kumari2016Hyper}, 
	derived from the MoJo distance, 
	compares a produced solution to a given ``golden rule'' solution,
	which naturally represents the DM's preference over the objectives such as cohesion and coupling~\cite{wen2004effectiveness}.
	\rev{Another example can be seen in the test case generation problem where
	some works use multi-objectivization to improve the code coverage~\cite{Panichella2015Reformulating,main-sbse,panichella2018automated}.
	They reformulate the coverage criterion as a many-objective optimization problem,
	where the objectives to be optimized are different coverage targets (e.g., branches in \cite{Panichella2015Reformulating}), 
	but only the total coverage of all test cases in the produced test suite is of interest. 
	In this case, 
	such a total coverage criterion can be regarded as a problem-specific indicator, 
	as it does reflect the quality of a solution set in this specific problem 
	but not directly involved in the search-based optimization.}
	Overall, 
	such $PSI$ indicators not only represent more ``accessible'' quality evaluation of solution sets  
	(i.e., how they perform under the practical problem background), 
	but also usually imply some preferences from the DM.
	Therefore, 
	it is highly recommended to include them in the evaluation if existing.
	
	Nevertheless,
	it is worth noting that $PSI$ indicators usually need to work together with generic quality indicators (e.g., those in Table~\ref{Table:Summary}) to provide reliable evaluations 
	since they may be irrelevant to Pareto-based optimization (e.g., only focusing on particular objectives in evaluation). 
	For example, 
	the study~\cite{REMAP} mainly relies on APFD, 
	the average percentage of fault detected, 
	to evaluate the solution set of prioritized test cases. 
	Indeed, 
	APFD is a frequently used $PSI$ in the test case prioritization, 
	but it can only reflect the rate of fault detected, 
	not the reliance of test cases, 
	both of which are the objectives to be optimized for the problem.
	
	In addition, 
	plotting representative solution sets ($SSP$) is also desirable as an auxiliary evaluation, 
	as it empowers the SBSE researchers/practitioners to get a sense of what the solution set looks like. 
	This is very helpful not only for solution set comparison, 
	but also for the DM to understand the problem and then perhaps to refine their preferences further. 
	
	To use $SSP$,
	for an algorithm involving stochastic elements 
	we suggest plotting the solution set in a particular run which corresponds to the evaluation result 
	(obtained by a comprehensive quality indicator, e.g., $HV$) 
	that is the closest to the median value in all the runs. 
	Alternatively, 
	for optimization problems with two and three objectives, 
	median attainment surfaces~\cite{fonseca1996performance,knowles2005summary,Lopez2010} can be used to visualize the performance of the algorithm with respect to all the runs 
	(which have already been adopted in the literature~\cite{Chicano2011Using,Ferrer2012Evolutionary,Wu2015Deep,Zhang2007The}).  
	For problems with more objectives, 
	the parallel coordinates plot (instead of the scatter plot) is a helpful tool, 
	which can reflect the convergence and diversity of a solution set to some extent~\cite{li2017how}. 
	It has started to be used recently, 
	e.g., in~\cite{Mkaouer2016On,Mkaouer2014High,Xiang2018Configuring,Hierons2019Many,xiang2019}.

	\subsection{A General Procedure}

	Based on the above, 
	we now are in a good position to provide a general procedure of how to evaluate solution sets in Pareto-based SBSE in Figure~\ref{Fig:Guidance}. 
	At first, we suggest conducting some screening (\textbf{P1} in the figure) to filter out trivial solutions in the considered solution sets according to the nature of the optimization problem in SBSE.
	The trivial solutions can be seen as those which are straightforward to obtain and would never be of interest to the DM, but may affect the evaluation result. 
	e.g., the solution with zero cost and zero coverage in the example of Figure~\ref{Fig:FullCoverage}.

	After the filtering,
	it comes to evaluating solution sets according to the DM's preference information (\textbf{D1}).
	If there is no preference information available at all \rev{(e.g., the effort estimation and test case prioritization problem),} 
	we suggest to consider quality indicators that together are able to accurately reflect all the quality aspects (\textbf{D2}--\textbf{D5}). For example,
	one can consider separately evaluating distinct quality aspects of solution sets, 
	e.g., $GD^+$ for convergence (together with $CI$ if willing to know the dominance relation between sets), 
	$DCI$ for diversity (when involving $\geq 3$ objectives), and $UNFR$ for cardinality. 
	Alternatively, 
	one can consider evaluating the comprehensive quality of solution sets, e.g., $HV$ in most cases; 
	or even some mix (e.g., $HV$ plus $UNFR$) if it requires understanding on some specific quality aspects on top of solution sets' general quality.
	\rev{Note that the cardinality of solution sets tends to have more weight in the SBSE area 
	than some other areas such as evolutionary computation, 
	since many SBSE problems are combinatorial ones, 
	where their Pareto front size may be relatively small 
	and it is likely to have comparable solutions (e.g., dominated/duplicate solutions) from different sets.}
	In any case, 
	whatever indicators considered, 
	the way of using them needs to comply with their usage note and caveats (see Table~\ref{Table:Summary}).

	If there is preference information available, 
	we recommend first to see whether (part of) it belongs to clear preferences (\textbf{D6});
	if so, \rev{such as the software product line configuration problem,}
	one can transfer the clear preferences into the solutions (\textbf{P2}), as what we discussed in Section~\ref{sec:p1}.
	It is necessary to note that sometimes 
	after the transfer there is only one objective left to be considered 
	(e.g., in the example of Figure~\ref{Fig:FullCoverage}).  
	In this case,
	the best value on that remaining objective represents the quality of the solution set.
	
	After considering clear preferences, 
	one needs to see whether there exist some vague preferences (\textbf{D7}).
	if the answer is yes, \rev{e.g., the software configuration and adaptation problem,}
	then we transfer those that are transferable (\textbf{\textbf{D8}, \textbf{P3}, \textbf{D9}}), e.g., the example in Figure~\ref{Fig:PrefHV} and resolution from Section~\ref{sec:p2}.
	After that, 
	we recommend to check whether the rest preferences can be transferred into an indicator (\textbf{D10}); 
	if so, one can then transfer them (\textbf{P4}), e.g., transferring certain preferences into a weight distribution function in the weighted $HV$~\cite{Zitzler2007a}, 
	and use that indicator to evaluate the solution sets as shown in Section~\ref{sec:p2}.
	
	When the preference information cannot be accommodated into an indicator,
	the next step is to check if the DM prefers a specific part of the Pareto front (\textbf{D11}), \rev{such as the software modularization and service composition problem.} As we discussed in Section~\ref{sec:p3},
	if one prefers knee points on the Pareto front (i.e., well-balanced solutions between conflicting objectives), 
	then $HV$ is a good option.
	If one prefers boundary solutions,
	then $HV$ with an unusual configuration of its reference point can be used,
	alongside with reporting the best value of relevant objective(s) in the population.

	Note that the DM may present several types of preference information. 
	For example, 
	one may specify a clear threshold on one objective and at the same time be interested in knee points on the Pareto front --- \rev{a typical case when non-functional requirements of the software are involved.}  
	Another example has been seen in the situation of Figure~\ref{Fig:PrefHV}, 
	where the DM's preferences contain both clear and vague information.
	In addition,
	it is necessary to mention that there do exist some situations where the DM's preferences cannot be quantified/transferred properly, 
	e.g., the DM may state like ``\texttt{the cost should be reasonable}''.
	In such a situation, 
	we suggest proceeding to \textbf{D2} --- the general multi-objective optimization case (without specific preferences).

	After going through all possible cases of the DM's preferences, 
	it comes to check the last two quality evaluation methods, problem-specific evaluation, i.e., $PSI$ (\textbf{D13}), and solution set plotting, i.e., $SSP$ (\textbf{D14}). 
	These are two very helpful methods in reflecting the solution set's quality that may have not been captured by generic quality indicators.

	\section{Threats to Validity}
	\label{sec:tov}

  Threats to construct validity can be raised by the research methodology, 
  which may not serve the purpose of surveying the evaluation methods for Pareto-based optimization in existing SBSE studies. 
  We have mitigated such threats by following the systematic review protocol proposed by Kitchenham et al.~\cite{Kitchenham2009Systematic}, 
  which is a widely recognized search methodology for conducting a survey in the SE research. 
  \rev{Another threat is related to the citation count used in the exclusion criteria. 
  Indeed, 
  it is difficult to set a threshold for such, 
  as the citation count itself cannot well reflect the impact of work. 
  Since there is no metric that suffices to do so, 
  in this work
  we used the citation count and set a threshold by averaging the candidate studies. 
  It is however worth noting that we do not seek to provide a comprehensive review over the entire SBSE field, but to capture major trends on the evaluation of solution sets, which can at least provide some sources for analyzing and building the methodological guidance. Therefore it is necessary to reach a trade-off between the trend coverage and the efforts required for detailed data collections of the studies.}

  Threats to internal validity may be introduced by having inappropriate classification and interpretation of the SBSE papers, 
  their implied preferences, and used quality indicators/evaluation methods. 
  We have limited this by conducting three iterations of study reviews by the first two authors. 
  Error checks and investigations were also conducted to correct any issues found during the search procedure. 
  The key issues identified have also been resolved among the first two authors or by counseling external researchers.

  Threats to external validity may restrict the generalizability of the proposed guidance and considered cases. 
  We have mitigated such by conducting the survey more widely and deeply: 
  it covers 717 searched papers published between 2009 and 2019, on 36 venues from seven repositories; while at the same time, extracting 95 prominent primary studies following the exclusion and inclusion procedures. 
  This has included 21 most noticeable SBSE problems that spread across the whole SDLC. 
  The extracted assumptions of the DM's preferences, 
  together with rigorous analyses of the 12 representative quality indicators 
  (i.e., either used widely in SBSE or proposed herein for a more accurate evaluation), 
  have provided rich sources for us to establish a general methodological guidance for the community.

	\rev{Finally, 
	although our guidance has been designed in a way that it aims to cover a wide range of SBSE problems, 
	it is always possible that there are situations 
	which we have unfortunately missed; 
	for example, 
	the behavior of solutions in the decision space, 
	e.g., their diversity and robustness. 
	As different settings of parameters 
	(i.e., decision variables) 
	may lead to similar/same solutions' quality 
	(e.g., multiple points in the decision map to a single point in the objective space),
	the DM of course likes those which are easier to implement.
	Therefore, 
	a set of diverse solutions in the decision space are preferred,
	providing more options for the DM.  
	Another aspect that the DM may consider is robustness~\cite{Calinescu2017Designing}, 
	which is related to how fast the quality of solutions degrades when varying their parameters (decision variables).
	This issue is particularly important in an uncertain environment 
	where the solution may not be able to be deployed accurately and/or the objective functions estimated may be of a margin of error.
	Therefore, 
	robust solutions are preferred to sensitive ones 
	even if their quality is slightly lower in some circumstances.
	Overall,
	in those cases,  
	an evaluation of solution sets' quality both in the decision space and in the objective space is needed.}

	\rev{
	\section{Related Work}}	
	
	\rev{Various surveys on SBSE (e.g., \cite{Harman2012, Harman2014search, Mcminn2011search}) reveal intense interests 
	in developing computational search methods for complex optimization problems in SE. 
	Some of them focus on or are relevant to Pareto-based multi-objective optimization. 
	For example, Sayyad et al.~\cite{Sayyad2013b} performed a brief literature review of SBSE studies 
	that used Pareto-based evolutionary algorithms for multi-objective optimization problems;
	Boussa\"{\i}d et al.~\cite{Boussaid2017survey} conducted a comprehensive survey on 
	search-based model-driven engineering 
	and classified relevant search algorithms into single- and multi-objective ones;  
	Ram\'{\i}rez et al.~\cite{Ramirez2019survey} reviewed SBSE studies on a subarea 
	of multi-objective optimization, many-objective optimization, 
	where the number of objectives is larger than 3. 
	In general, 
	these papers concentrate on the development of search algorithms 
	for Pareto-based multi-objective optimization problems; very few touch on 
	the quality evaluation of the results obtained by search algorithms until recently.}
	
	\rev{Wang et al.~\cite{wang2016practical} proposed a practical guide for SBSE researchers/practitioners 
	to select quality indicators in Pareto-based optimization, 
	on the basis of the results of experimental studies evaluating eight quality indicators 
	in three industrial and real-world problems.  
	They firstly classified these indicators into four categories, 
	convergence, diversity, combination, and coverage, 
	and then they, 
	based on empirical observations, 
	have drawn several conclusions about the indicator selection.
	For example, 
	they have concluded that it does matter which indicator to select
	in the diversity category,
	but it does not matter which indicator to select within the same convergence or
	combination category.}

	\rev{Very recently, Ali et al.~\cite{Ali2020quality} substantially extended Wang et al.'s work~\cite{wang2016practical} and 
	provided a set of guidelines drawn from an extensive empirical evaluation 
	in nine SBSE problems from industrial, real-world and open-source projects.
	From these experiments, 
	they produced 22 observations based on statistical comparisons 
	between six multi-objective evolutionary algorithms.
	They have claimed that the differences in SBSE problems have
	high effect on the consistency of quality indicators' evaluation results, 
	whereas the effect of search algorithms is low. 
	A noticeable difference from \cite{wang2016practical} is that 
	the guidance provided did not build on a classification of indicators.}
	
	\rev{Li et al.~\cite{li2018critical} conducted a critical review of Wang et al.'s work~\cite{wang2016practical}. 
	They argued that some conclusions (e.g., it matters which indicator to select in the category diversity) 
	are actually caused by the inaccurate classification of the considered indicators.
	More importantly, 
	they argued that even if an accurate classification is made, 
	one still cannot draw any conclusions like it does not matter which indicator to select, 
	whether in the same category or across different categories.} 

	\rev{Indeed, 
	as can be seen in Section~\ref{sec:QI_revisit}, 
	each quality indicator has its own distinct quality implications. 
	A solution set being evaluated better by an indicator does not mean that it generally has higher quality, 
	but rather that it is preferred under the assumption that the indicator accurately reflects the DM's preferences.
	However, 
	different DMs may prefer different trade-off solutions between objectives, 
	even for the same problem. 
	For example,
	for the project scheduling problem, 
	in some scenarios, 
	the DM may prefer knee solutions~\cite{Sarro2017Adaptive}~\cite{Gueorguiev2009Software}~\cite{Ferrucci2013Not},
	in some other scenarios, 
	the DM may prefer widely distributed solutions~\cite{Chicano2011Using},
	in some other scenarios, 
	the DM may prefer specific solutions relying on the Analytic Hierarchy Process~\cite{Shen2016Dynamic}.
	Consequently, 
	observations on quality indicators  
	drawn from an empirical investigation on specific SBSE scenarios
	may not be well generalized. 
	This suggests a need of a general, methodological guidance on how to select and use indicators in SBSE.
	Such a guidance is not based upon empirical studies on specific problems 
	but upon the fundamental goal of multi-objective optimization --- 
	supplying the DM a set of solutions which are the most consistent with their preferences.}
	
	\rev{It is worth mentioning that 
	a recent survey paper~\cite{li2019quality} on quality evaluation in multi-objective optimization appeared, 
	albeit not specific for SBSE.  
	It systematically reviewed 100 quality indicators, 
	analyzed correlations between representative indicators, 
	discussed several important issues in designing indicators,
	and suggested a few future research directions.
	One key purpose of that work is about indicator design and development, i.e., 
	to inform the researchers and practitioners on what aspects to bear in mind when designing new indicators.
	In contrast, 
	our work here is about indicators (and other evaluation methods) selection and use, i.e.,
	to guide the researchers and practitioners on how to select/use existing indicators, or even specialize them, for evaluating solution sets in SBSE. 
	Another major difference between the two works is that 
	the work \cite{li2019quality} considered the general situation that the DM's preferences are not available,
	whereas our work considers the situations based precisely upon various DM's preferences.}
	
	\rev{As such, 
	these two works complement well with each other. 
	If the SBSE researchers/practitioners want to understand correlations between different quality indicators, 
	to know some important issues (e.g., scaling, normalization and effect of dominated/duplicate solutions) 
	when performing quality evaluation for a practical problem, 
	or even by themselves to design/develop new indicators, 
	they can refer to the general survey paper in \cite{li2019quality}. 
	In contrast, 
	if the SBSE researchers/practitioners want to select and use existing indicators in various optimization scenarios in SBSE, 
	or to adapt existing indicators to fit explicit/implicit preferences from the DM in the given SBSE problem, 
	this work can be well served.} 

	\rev{Overall, 
	in comparison with the existing works, 
	this work presents several additional contributions. 
	First, 
	we conduct a systematic literature review on quality evaluation for Pareto-based optimization in SBSE. 
	Second, 
	we, from that review, 
	present a variety of inappropriate/inadequate selection and
	inaccurate/misleading use of evaluation methods, 
	and identify five important but overlooked issues.
	Third, 
	from the perspective of the goal of multi-objective optimization,
	we discuss the reasons that quality indicators are needed, 
	carry out an in-depth analysis of frequently-used quality indicators in the area, 
	and explain the scope of their applicability.
	Finally, 
	we provide a methodological
	guidance and procedure of selecting and using
	evaluation methods in various SBSE scenarios.}

	\section{Conclusions}
	
	The nature of considering multiple (conflicting) objectives in many SBSE problems leads to a link between SE and multi-objective optimization.
	However,
	compared to the flourish of the use/design of multi-objective optimizers in SBSE, 
	the evaluation of the optimizers' outcome remains relatively ``casual''. 
	Existing SBSE researches often work by analogy, namely, 
	following popular (or previously used) quality evaluation methods without considering whether they are truly suitable for their specific situation. 
	In this paper,
	we have carried out a systematic and critical review of the quality evaluation in Pareto-based SBSE, covering 95 prominent studies published between 2009 and 2019 from 36 venues in seven repositories.
	We have found that in many studies the selection/use of evaluation methods is not appropriate and can even be misleading, 
	based on which we summarize five critical issues, namely:
	\begin{itemize}
	    \item Inadequacy of Solution Set Plotting.
	    \item Inappropriate use of Descriptive Objective Evaluation.
	    \item Confusion of the quality aspects covered by generic quality indicators.
	    \item Oblivion of context information.
	    \item Noncompliance of the DM’s preferences.
	\end{itemize}

	Through revisiting the pros and cons of widely used quality indicators in SBSE,
	we have provided a methodological guidance and procedure of selecting, adjusting and using quality evaluation methods
	on the basis of the following availability/types of the DM's preferences:
	
	\begin{itemize}
	    \item There are clear preferences between the objectives.
	    \item There are vague/rough preferences between the objectives.
	    \item There are preferences on some specific parts of the Pareto front.
	    \item There are no preferences available.
	\end{itemize}
	
	We hope that our guidance would help to mitigate the evaluation issues in future SBSE work, 
	and more importantly, would enable the quality evaluation of solution sets easier, 
	clearer and more accurate for SBSE researchers and practitioners.

	\input{Tables/all-qi-1}
    \input{Tables/all-qi-2}
\appendix
Table A1 specifies the evaluation methods and how they are used for all the 95 studies analyzed in this work.

\section*{Acknowledgment}
This work was supported by the Guangdong Provincial Key Laboratory
(Grant No. 2020B121201001), the Program
for Guangdong Introducing Innovative and Enterpreneurial
Teams (Grant No. 2017ZT07X386), Shenzhen
Science and Technology Program (Grant No.
KQTD2016112514355531), and the Program for University
Key Laboratory of Guangdong Province (Grant No.
2017KSYS008). We are grateful to the editor and anonymous reviewers for their constructive comments on the early version of this paper.

	\bibliographystyle{IEEEtranS}
	\bibliography{IEEEabrv,references}
	

\begin{IEEEbiography}[{\includegraphics[width=1in,height=1.25in,clip,keepaspectratio]{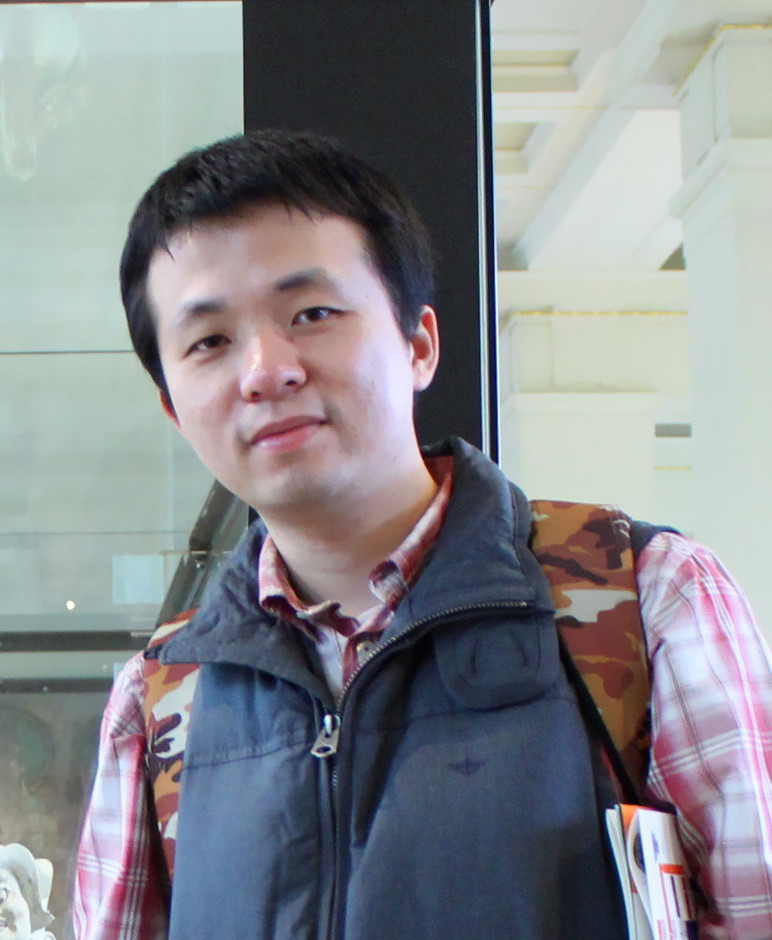}}]{Miqing Li} is currently a Lecturer at School of Computer Science at the University of Birmingham. His research is principally on multi-objective optimisation, where he focuses on developing population-based randomised algorithms (mainly evolutionary algorithms) for both general challenging problems (e.g. many-objective optimisation, constrained optimisation, robust optimisation, expensive optimisation) and specific challenging problems (e.g. those in software engineering, system engineering, product disassembly, post-disaster response, neural architecture search, reinforcement learning for games). Dr Li has published over 60 research papers in scientific journals and international conferences. Some of his papers, since published, have been amongst the most cited papers in corresponding journals such as \textsc{IEEE Transactions on Evolutionary Computation}, \textsc{Artificial Intelligence}, \textsc{ACM Transactions on Software Engineering and Methodology}, \textsc{IEEE Transactions on Parallel and Distribution Systems}, and \textsc{ACM Computing Surveys}. His work has received the Best Student Paper Award or Best Paper Award nomination in EC mainstream conferences, CEC, GECCO, and SEAL. Dr Li is the founding chair of the IEEE CIS' Task Force on Many-Objective Optimisation.
\end{IEEEbiography}

\begin{IEEEbiography}[{\includegraphics[width=1in,height=1.25in,clip,keepaspectratio]{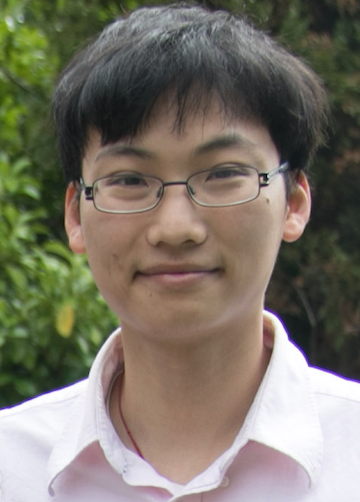}}]{Tao Chen} (M'15) received his Ph.D. from the School of Computer Science, University of Birmingham, United Kingdom, in 2016. He is currently a Lecturer (assistant professor) in Computer Science at the Department of Computer Science, Loughborough University, United Kingdom. He has broad research interests on software engineering, including but not limited to performance engineering, self-adaptive software systems, search-based software engineering, data-driven software engineering and computational intelligence. As the lead author, his work has been published in internationally renowned journals, such as \textsc{IEEE Transactions on Software Engineering}, \textsc{ACM Transactions on Software Engineering and Methodology}, \textsc{IEEE Transactions on Services Computing}, and \textsc{Proceedings of the IEEE}; and top-tier conferences, e.g., ICSE, ASE, and GECCO. Among other roles, Dr. Chen regularly serves as a PC member for various conferences in his fields and is an associate editor for the \textsc{Services Transactions on Internet of Things}. 
\end{IEEEbiography}

\begin{IEEEbiography}[{\includegraphics[width=1in,height=1.25in,clip,keepaspectratio]{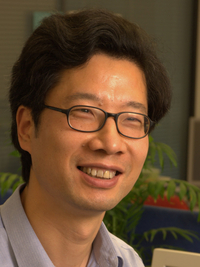}}]{Xin Yao} (F'03) received the BSc degree from the University of Science and Technology of China (USTC), Hefei, China, in 1982, the MSc degree from the North China Institute of Computing Technologies, Beijing, China, in 1985, and the PhD degree from USTC in 1990. He is a Chair Professor of Computer Science at the Southern University of Science and Technology (SUSTech), Shenzhen, China, and a part-time Professor of Computer Science at the University of Birmingham, UK. His current research interests include evolutionary computation, machine learning, and their real world applications, especially to software engineering. He started his work on search-based software engineering (SBSE) more than a decade ago, including ``Coevolving Programs and Unit Tests from Their Specification'' at ASE'07 and ``Software Module Clustering as a Multi-Objective Search Problem'' in March 2011's \textsc{IEEE Transactions on Software Engineering}. His latest work on SBSE includes "Software Effort Interval Prediction via Bayesian Inference and Synthetic Bootstrap Resampling" in January 2019's \textsc{ACM Transactions on Software Engineering and Methodology} and ``Synergizing Domain Expertise With Self-Awareness in Software Systems: A Patternized Architecture Guideline'' in July 2020's \textsc{Proceedings of the IEEE}. He was a recipient of the Royal Society Wolfson Research Merit Award in 2012, the IEEE Computational Intelligence Society (CIS) Evolutionary Computation Pioneer Award in 2013 and the IEEE Frank Rosenblatt Award in 2020. His work won the 2001 IEEE Donald G. Fink Prize Paper Award, the 2010, 2016, and 2017 \textsc{IEEE Transactions on Evolutionary Computation} Outstanding Paper Awards, the 2011 \textsc{IEEE Transactions on Neural Networks} Outstanding Paper Award, and many other best paper awards at conferences. He was the President of IEEE CIS from 2014 to 2015 and the Editor-in-Chief of \textsc{IEEE Transactions on Evolutionary Computation} from 2003 to 2008. He is an IEEE fellow since 2003 and was a Distinguished Lecturer of IEEE CIS.
\end{IEEEbiography}

\end{document}

%% file: Tables/data-item.tex
\begin{table}
\caption{Data collection items.}
\label{tb:items}
\centering

\begin{tabular}{lll}\toprule

\textbf{ID}&\textbf{Item}&\textbf{Questions}\\ 
\midrule

$I_1$&Author(s)&N/A\\
\rowcolor{steel!10}$I_2$&Year&N/A\\
$I_3$&Title&N/A\\
\rowcolor{steel!10}$I_4$&Venue (journal or conference)&N/A\\
$I_5$&Citation count&N/A\\
\rowcolor{steel!10}$I_6$&Indicator and method&RQ1\\
$I_7$&Stated quality aspects&RQ2\\
\rowcolor{steel!10}$I_8$&Reference point/front&RQ2\\
$I_9$&\# objectives&RQ3\\
\rowcolor{steel!10}$I_{10}$&SBSE problem&RQ3\\
$I_{11}$&DM's preferences and contextual information&RQ3\\

\bottomrule
\end{tabular}
\end{table}

%% file: Tables/venue.tex
	\begin{table}[t!]
		\caption{The reviewed studies counts and venues.}
		
		\label{table:papers-count} 
		\setlength\tabcolsep{4.5pt}
		\begin{tabularx}{\columnwidth}{X|P{0.4cm}|P{0.4cm}|P{0.2cm}}
			\toprule 
			\multicolumn{1}{l|}{\textbf{{Journal}}} & \rotatebox[origin=lt]{90}{\textbf{{Search}}} & \rotatebox[origin=lt]{90}{\textbf{{Candidate}}    } & \rotatebox[origin=lt]{90}{\textbf{{Primary}}}
			\\
			\midrule 
			ACM Transactions on Software Engineering and Methodology &\multirow{2}{*}{15} &\multirow{2}{*}{10} &\multirow{2}{*}{8} \\  
			\rowcolor{steel!10}Elsevier Information and Software Technology & 65 & 33 & 6 \\  
			Elsevier Applied Soft Computing & 17 & 4 & 0 \\  
			\rowcolor{steel!10}Springer Automated Software Engineering & 16 & 6 & 2 \\  
			IEEE Transactions on Software Engineering & 79 & 41 & 9 \\  
			\rowcolor{steel!10}Springer Empirical Software Engineering & 37 & 17 & 3 \\  
			Elsevier Future Generation Computing Systems & 3 & 3 & 1 \\  
			\rowcolor{steel!10}Springer Soft Computing & 11 & 4 & 0 \\  
			IEEE Transactions on Evolutionary Computation & 13 & 3 & 2 \\  
			\rowcolor{steel!10}IEEE Transactions on Services Computing & 6 & 3 & 3 \\  
			Elsevier Journal of Systems and Software & 70 & 35 & 8 \\  
			\rowcolor{steel!10}Elsevier Information Sciences & 18 & 7 & 3 \\  
			Springer Requirements Engineering & 5 & 3 & 1 \\  
			\rowcolor{steel!10}Springer Software Quality Journal & 12 & 7 & 2 \\  
			Wiley Software Testing, Verification and Reliability & 4 & 2 & 1 \\  
			\rowcolor{steel!10}Wiley Software: Practice and Experience & 8 & 4 & 1 \\  
			Springer Software and Systems Modeling & 2 & 1 & 1 \\  
			\rowcolor{steel!10}IEEE Transactions on Systems, Man, and Cybernetics & 2 & 2 & 1 \\  \midrule 
			\multicolumn{1}{l}{\textbf{{Conference, Symposium and Congress}}}&\multicolumn{3}{c}{} 
			\\ \midrule 
			
			IEEE/ACM Conference on Software Engineering & 30 & 12 & 8 \\  
			\rowcolor{steel!10}Springer Symposium on Search Based Software Engineering &\multirow{2}{*}{110} &\multirow{2}{*}{33} &\multirow{2}{*}{4} \\  
			IEEE Congress on Evolutionary Computation & 21 & 8 & 0 \\  
			\rowcolor{steel!10}IEEE/ACM Conference on Automated Software Engineering &\multirow{2}{*}{21} &\multirow{2}{*}{8} &\multirow{2}{*}{5} \\  
			ACM Conference and Symposium on the Foundations of Software Engineering &\multirow{2}{*}{13} &\multirow{2}{*}{3} & \multirow{2}{*}{0} \\  
			\rowcolor{steel!10}ACM Genetic and Evolutionary Computation Conference & \multirow{2}{*}{57} & \multirow{2}{*}{33} & \multirow{2}{*}{10} \\  
			IEEE Conference on Software Testing, Verification and Validation & \multirow{2}{*}{20} & \multirow{2}{*}{11} & \multirow{2}{*}{3} \\  
			\rowcolor{steel!10}ACM Symposium on Software Testing and Analysis & 8 & 5 & 3 \\  
			IEEE/ACM Conference on Empirical Software Engineering and Measurements & \multirow{2}{*}{1} & \multirow{2}{*}{0} & \multirow{2}{*}{0} \\  
			\rowcolor{steel!10}ACM Systems and Software Product Line Conference & \multirow{2}{*}{11} & \multirow{2}{*}{3} & \multirow{2}{*}{1} \\ 
			IEEE Conference on Web Services & 2 & 2 & 1 \\  
			\rowcolor{steel!10}IEEE Conference on Software Maintenance & 15 & 1 & 1 \\  
			IEEE Conference on Software Maintenance and Reengineering & \multirow{2}{*}{4} & \multirow{2}{*}{3} & \multirow{2}{*}{1} \\  
			\rowcolor{steel!10}IEEE Workshop on Combining Modelling and Search-Based Software Engineering & \multirow{2}{*}{5} & \multirow{2}{*}{3} & \multirow{2}{*}{1} \\  
			IEEE Conference on Requirements Engineering & 7 & 3 & 1 \\ 
			\rowcolor{steel!10}ACM Conference on Performance Engineering & 2 & 2 & 2 \\ 
			IEEE/ACM Conference on Program Comprehension & 3 & 2 & 1 \\ 
			\rowcolor{steel!10}IEEE Conference on Software Architecture & 4 & 2 & 1 \\ 	\midrule 
			\textbf{{Total}} & 717 & 319 & 95 \\ \bottomrule  
		\end{tabularx}
	\end{table}

%% file: Tables/acronyms.tex
\begin{table*}
\caption{Acronyms of the evaluation methods.}
\label{tb:acronym}
\setlength{\tabcolsep}{1mm}
\centering

\begin{tabular}{cc|cc|cc}\toprule

\textbf{Acronym}&\textbf{Full Name}&\textbf{Acronym}&\textbf{Full Name}&\textbf{Acronym}&\textbf{Full Name}\\ 
\midrule

$AS$~\cite{knowles2005summary}&Attainment Surface&$IGD^+$~\cite{Ishibuchi2015}&Inverted Generational Distance$^+$&$CI$~\cite{Meunier2000}&Contribution Indicator\\

\rowcolor{steel!10}$\mathcal{C}$ ($CS$)&$\mathcal{C}$ Metric&$GD$~\cite{Veldhuizen1998}&Generational Distance&$Spread$ ($\Delta$)~\cite{Deb2002}&Spread\\

$SP$~\cite{Schott1995}&Spacing&$NFS$&Nondominated Front Size&$\epsilon$~\cite{Zitzler2003}&$\epsilon$-Indicator\\

\rowcolor{steel!10}$IGD$~\cite{Coello2004}&Inverted Generational Distance&$HV$~\cite{Zitzler1998}&Hypervolume&$ED$&Euclidean Distance\\

$ER$&Error Rate&$SSP$&Solution Set Plotting&$DOE$&Descriptive Objective Evaluation\\

\bottomrule
\end{tabular}
\end{table*}


%% file: Tables/doe.tex
	\begin{table}[t!]
		\caption{Descriptive Objective Evaluation (DOE) methods used.}
		
		\label{table:DOE} 
		\begin{tabularx}{\columnwidth}{Xp{2.8cm}}
			\toprule 
			\textbf{{$DOE$}} & \textbf{{Used in}} \\ \midrule 
			Mean Fitness Value ($MFV$)&\textbf{2:} \cite{Li2017Zen} \cite{Wang2015Cost} \\  
			\rowcolor{steel!10}Analytic Hierarchy Process ($AHP$)&\textbf{2:} \cite{Shen2016Dynamic} \cite{shen2018q} \\  
			\multirow{7}{*}{\parbox{5.2cm}{Mean, best, worst, median and/or statistical result of each objective for solutions in the population$^\star$}} &\textbf{27:}  \cite{Minku2013Software} \cite{Bowman2010Solving} \cite{Wang2015Cost} \cite{Kumari2016Hyper} \cite{Bavota2012Putting} \cite{PubSub10849_Chen} \cite{Wada2012E} 
			\cite{Fleck2017Model} \cite{Kalboussi2013Preference}
			\cite{Durillo2014Multi} 
			[82]
			\cite{Praditwong2011Software}
			\cite{Wu2015Deep} 
			\cite{Li2014Robust}
			\cite{Sarro2017Adaptive}
			\cite{Simons2010Interactive} \cite{Simons2012Elegant} \cite{Sayyad2013Optimum} \cite{Ferrer2012Evolutionary} \cite{Pascual2015Applying} \cite{Boukharata2019} \cite{pradhan2019}
			\cite{Li2017Zen,Segura2016Multi}
			\cite{White2011Evolutionary,Wagner2012Multi,Abdeen2014Multi}\\ 
			\rowcolor{steel!10}Best of one objective over the population while another is below certain thresholds$^\star$ &\multirow{2}{*}{\textbf{1:} \cite{lee2019}} \\ \bottomrule
		\end{tabularx}
		\begin{tablenotes}
			\item $^\star$ The mean of all repeated runs are reported. Note that a study could involve more than one $DOE$ form.
		\end{tablenotes}
	\end{table}

%% file: Tables/qi-summary.tex
\begin{table}
\caption{Summary of how generic quality indicators are used to evaluate solution sets in the primary studies.}
\label{tb:qi-summary}
\centering

\begin{tabular}{lll}\toprule

\textbf{Indicator}&\textbf{\makecell[l]{Stated Quality \\Aspects to Measure}}&\textbf{\makecell[l]{Reference \\Point or Front}}\\ 
\midrule

$HV$&\makecell[l]{Unknown (25), \\$Q_1\cup Q_2\cup Q_3$ (15), \\$Q_1\cup Q_2$ (3), $Q_1$ (3), \\$Q_2\cup Q_3$ (2)}&\makecell[l]{Unknown (23), \\Worst values (10), \\Nadir point (9), \\Boundary (6)}\\


\rowcolor{steel!10}$IGD$&\makecell[l]{$Q_1\cup Q_2\cup Q_3$ (9), $Q_1$ (4), \\Unknown (2), $Q_1\cup Q_2$ (1)}&\makecell[l]{Best Pareto \\front found (14),~~ \\Unknown (2)}\\

$IGD+$&\makecell[l]{$Q_1\cup Q_2\cup Q_3$ (1)}&Unknown (1)\\

\rowcolor{steel!10}$GD$&Unknown (10), $Q_1$ (6)&\makecell[l]{Best Pareto \\front found (10),~~ \\Unknown (6)}\\

$Spread$&\makecell[l]{$Q_2\cup Q_3$ (9), $Q_2$ (5)}&N/A\\

\rowcolor{steel!10}$\epsilon$-Indicator&\makecell[l]{$Q_1$ (5), Unknown (2),~~~~~~~~~\\$Q_1\cup Q_2\cup Q_3$ (1)}&N/A\\

$NFS$&\makecell[l]{Unknown (4), $Q_2\cup Q_3$ (3)}&N/A\\

\rowcolor{steel!10}$CI$&Unknown (5), $Q_1$ (1)&N/A\\

$CS$&\makecell[l]{Unknown (4), $Q_1$ (1)}&N/A\\

\rowcolor{steel!10}$AS$&Unknown (4)&N/A\\

$SP$&\makecell[l]{$Q_3$ (2), $Q_2\cup Q_3$ (1)}&N/A\\

\rowcolor{steel!10}$ED$&Unknown (3)&N/A\\

$ER$&\makecell[l]{$Q_1$ (1)}&Unknown (1)\\


\bottomrule
\end{tabular}
\begin{tablenotes}
    \footnotesize
    \item $Q_1$=Convergence; $Q_2$=Spread; $Q_3$=Uniformity; $Q_4$=Cardinality. The number of primary studies is shown within the brackets.
   \end{tablenotes}
\end{table}

%% file: Tables/obj-summary.tex
\begin{table}
\caption{Summary of the \# objectives under which the generic quality indicators, $SPI$, $SSP$ and $DOE$ are used.}
\label{tb:obj-summary}
\centering

\begin{tabular}{ll}\toprule

\textbf{Method}&\textbf{\# Objectives}\\ 
\midrule

$HV$&\makecell[l]{2 (19), 3 (24), 4 (9), 5 (11), 6 (1), 7 (1), 8 (1), 9 (2)}\\

\rowcolor{steel!10}$IGD$&2 (4), 3 (5), 4 (2), 5 (5), 8 (1), 15 (1)\\

$IGD+$&\makecell[l]{2 (1), 3 (1), 4 (1)}\\

\rowcolor{steel!10}$GD$&2 (6), 3 (9), 4 (2), 5 (2)\\

$Spread$&\makecell[l]{2 (5), 3 (3), 4 (2), 5 (7)}\\

\rowcolor{steel!10}$\epsilon$-Indicator&2 (1), 3 (4), 5 (4)\\

$NFS$&\makecell[l]{2 (5), 3 (2), 5 (2)}\\

\rowcolor{steel!10}$CI$&2 (3), 3 (4), 4 (1), 5 (1)\\

$CS$&\makecell[l]{2 (3), 4 (1), 5 (1)}\\

\rowcolor{steel!10}$AS$&2 (2), 3 (2)\\

$SP$&\makecell[l]{4 (1), 5 (1), 6 (1), 7 (1), 8 (1), 9 (1)}\\

\rowcolor{steel!10}$ED$&2 (2), 3 (1)\\

$ER$&\makecell[l]{5 (1)}\\


\rowcolor{steel!10}$SSP$&2 (30), 3 (19), 4 (6), 5 (2), 7 (1), 8 (1)\\

$DOE$&\makecell[l]{2 (4), 3 (13), 4 (5), 5 (5), 7 (1), 9 (1)}\\

\rowcolor{steel!10}$PSI$&\makecell[l]{2 (17), 3 (14), 4 (4), 5 (5), 6 (1), 7 (1), 8 (1), 9 (2),~~ \\15 (1), $>$100 (2)}\\

\bottomrule
\end{tabular}
\begin{tablenotes}
    \footnotesize
    \item The bracket shows the number of problems under which the a pair of objective number and evaluation method is considered. Note that a study may consider problems with different \# objectives.
   \end{tablenotes}
\end{table}

%% file: Tables/problem.tex
	
	\begin{table*}[t!]
		\caption{Pareto-based SBSE problems in different SDLC phases.}
		
		\label{table:papers-sdlc} 
		\begin{tabularx}{\textwidth}{p{2cm}|p{3cm}Xp{2.5cm}}
			\toprule 
			\textbf{{SDLC Phase}} & \textbf{{SBSE Problem}} & \multicolumn{1}{l}{\textbf{{Description}}} & \textbf{{Primary Studies}}
			\\
			\midrule 
			
			\multirow{4}{*}{Planning} &\multirow{2}{*}{Effort Estimation} & Optimize, e.g., accuracy and confidence interval, by changing the number of measured samples. &\multirow{2}{*}{\textbf{2:} \cite{Minku2013Software}\cite{Sarro2016Multi}} \\  
			&\multirow{2}{*}{\cellcolor{steel!10}Project Scheduling} &\cellcolor{steel!10} Optimize, e.g., duration and cost, by assigning employee into the tasks of a software project. &\cellcolor{steel!10}\textbf{7:} \cite{shen2018q}\cite{Shen2016Dynamic}\cite{Sarro2017Adaptive}\cite{Gueorguiev2009Software}\cite{Ferrucci2013Not}\cite{Chicano2011Using}\cite{1600-7617} \\ \hline  
			
			\multirow{4}{*}{\parbox{2.2cm}{Requirement Analysis}} & Requirement Assignment &Optimize, e.g., completeness and familiarity, by assigning requirements to different stakeholders' for their reviews. &\multirow{2}{*}{\textbf{1:} \cite{Li2017Zen}}  \\  
			&\multirow{2}{*}{\cellcolor{steel!10}Next Release Problem} &\cellcolor{steel!10}Optimize, e.g., robustness and cost, by selecting stakeholders' requirements in the next release of software. &\cellcolor{steel!10}\textbf{8:} \cite{Finkelstein2009A}\cite{Durillo2009A}\cite{Harman2009Search}\cite{Li2014Robust}\cite{Zhang2013Empirical}\cite{Zhang2007The}\cite{1600-7617}\cite{zhang2018} \\ \hline 
			\multirow{6}{*}{Design} & Software Modeling and Architecting & Optimize, e.g., cohesion and coupling, by modeling the object-oriented concept of the software and its architecture using standard notations. &\textbf{7:} \cite{Ramirez2016Comparative}\cite{Bowman2010Solving}\cite{Simons2010Interactive}\cite{Simons2012Elegant}\cite{Heaven2011Simulating}\cite{busari2017radar}\cite{JournalIST}  \\  
			&\multirow{4}{*}{\cellcolor{steel!10}Software Product Line} &\multirow{4}{*}{\cellcolor{steel!10}\parbox{9cm}{Optimize, e.g., correctness and richness, by finding the concrete products from the feature model.}} &\cellcolor{steel!10}\textbf{15:} \cite{Sayyad2013Optimum}\cite{Sayyad2013Scalable}\cite{Sayyad2013On}\cite{Hierons2016SIP}\cite{Henard2015Combining}\cite{Olaechea2014Comparison} \cite{Xiang2018Configuring}\cite{Wang2015Cost}\cite{1600-7617}\cite{Guo2019}\cite{xiang2019}\cite{lee2019}\cite{saber2017IsSeeding}\cite{lian2017}\cite{Hierons2019Many} \\ \hline
			\multirow{7}{*}{Implementation} &\multirow{3}{*}{Library Recommendation} & Optimize, e.g., library linked-usage and semantic similarity, by prioritizing the libraries that meet the required functionality to be used in the codebase. &\multirow{3}{*}{\textbf{1:} \cite{Ouni2017Search}} \\  
			&\multirow{2}{*}{\cellcolor{steel!10}Program Improvement} &\cellcolor{steel!10} Optimize, e.g., execution time and number of instructions, by producing semantically preserved software code. &\multirow{2}{*}{\cellcolor{steel!10}\textbf{1:} \cite{White2011Evolutionary}} \\  
			& Software Modularization & Optimize, e.g., modularization quality, cohesion and coupling, by placing different classes of code into different clusters. &\textbf{5:} \cite{Kumari2016Hyper}\cite{Bavota2012Putting}\cite{Praditwong2011Software}\cite{Fleck2017Model}\cite{Boukharata2019} \\ \hline
			
			\multirow{13}{*}{Testing} &\multirow{3}{*}{\cellcolor{steel!10}Code Smell Detection} &\cellcolor{steel!10} Optimize, e.g., coverage of bad examples and detection of good examples, by identifying the code and modules that could potentially cause issues. &\multirow{3}{*}{\cellcolor{steel!10}\textbf{1:} \cite{Mansoor2016Multi}} \\  
			&\multirow{2}{*}{Defect Prediction} & Optimize, e.g., effectiveness and cost, by adjusting the source code components to be predicted by the model. &\multirow{2}{*}{\textbf{4:} \cite{Canfora2015Defect}\cite{Canfora2013Multi}\cite{chen2017}\cite{ni2019}}  \\  
			&\multirow{2}{*}{\cellcolor{steel!10}Test Case Prioritization} &\cellcolor{steel!10} Optimize, e.g., coverage of the code and cost of test, by ordering the test cases to be tested.  &\cellcolor{steel!10}\textbf{7:} \cite{Assun2014A}\cite{Panichella2015Improving}\cite{Mariani2016}\cite{Segura2016Multi}\cite{Epitropakis2015Empirical}\cite{REMAP}\cite{issta18main} \\  
			&\multirow{3}{*}{\parbox{3cm}{White Box Test Case Generation}} & Optimize, e.g., coverage of the code and cost of test, by identifying the test cases, inputs and test suits based on internal information of the software. &\textbf{9:} \cite{Abdessalem2016Testing}\cite{Yoo2010Using}\cite{Zheng2016Multi}\cite{Mao2016Sapienz}\cite{Kalboussi2013Preference}\cite{Ferrer2012Evolutionary}\cite{Panichella2015Reformulating}\cite{jakubovskifilho2019}\cite{main-sbse} \\  
			&\multirow{3}{*}{\cellcolor{steel!10}\parbox{3cm}{Black Box Test Case Generation}} &\cellcolor{steel!10} Optimize, e.g., length of the inputs, distance to the ideal inputs, and cost of test, by identifying the test cases, inputs and test suits without internal information about the software &\multirow{3}{*}{\cellcolor{steel!10}\textbf{3:} \cite{Shahbazi2016Black}\cite{pradhan2019}\cite{abdessalem2018}}  \\ \hline
			
			\multirow{16}{*}{\parbox{2.1cm}{Deployment and Maintenance}} &\multirow{2}{*}{Resource Management} & Optimize, e.g., response time and cost, by changing the supported software and hardware resources such as in the Cloud environment. &\multirow{2}{*}{\textbf{3:} \cite{Frey2013Search}\cite{Li2010SLA}\cite{Chen2017Self}} \\ 
			
			&\multirow{3}{*}{\parbox{3cm}{\cellcolor{steel!10}Software Configuration and Adaptation}} &\cellcolor{steel!10} Optimize, e.g., response time and energy consumption, by changing software specific configurations, structure and connectors at design time or runtime. &\multirow{3}{*}{\parbox{2.5cm}{\cellcolor{steel!10}\textbf{7:} \cite{Pascual2015Applying}\cite{Gerasimou2016Search}\cite{Martens2010Automatically} \cite{Letier2014Uncertainty}\cite{Abdeen2014Multi}\cite{Chen2018FEMOSAA}\cite{Calinescu2017Designing}}}  \\ 
			
			&\multirow{2}{*}{Program Manipulation} & Optimize, e.g., response time and memory consumption, by changing the parametrized variables within the program code. &\multirow{2}{*}{\textbf{1:} \cite{Wu2015Deep}} \\ 
			
			&\multirow{2}{*}{\cellcolor{steel!10}Service Composition} &\cellcolor{steel!10} Optimize, e.g., latency and cost, by mapping the concrete services into abstract services within a workflow. &\cellcolor{steel!10}\textbf{4:} \cite{Wagner2012Multi}\cite{Wada2012E}\cite{Tan2018Evolutionary}\cite{PubSub10849_Chen} \\ 
			
			& Log Template Identification & Optimize, e.g., the frequency and specificity of the log message matched to a log template. &\multirow{2}{*}{\textbf{1:} \cite{ICPC-2018}} \\ 
			
			&\multirow{2}{*}{\cellcolor{steel!10}Workflow Scheduling} &\cellcolor{steel!10} Optimize, e.g., makespan and energy consumption, by assigning activities into a given application workflow. &\multirow{2}{*}{\cellcolor{steel!10}\textbf{1:} \cite{Durillo2014Multi}} \\ 
			
			
			&\multirow{3}{*}{Software Refactoring} &\multirow{3}{*}{\parbox{9cm}{Optimize, e.g., number of defects found and semantics, by changing the design model or the program code.}} &\textbf{9:} \cite{Ouni2013Search}\cite{Mkaouer2014High}\cite{Ouni2013maintainability}\cite{Ouni2012Search}\cite{Ouni2013The}\cite{Ouni2016Multi}\cite{Mansoor2015Multi}\cite{Mkaouer2014Recommendation}\cite{Mkaouer2016On}  \\ \bottomrule
			
		\end{tabularx}
	\end{table*}

%% file: Tables/preference.tex
	\begin{table*}[t!]
		\caption{Assumptions of DM's preferences and contextual information in the Pareto-based SBSE problems and the corresponding evaluation methods used.}  
		
		\label{table:papers-preferences} 
		\begin{tabularx}{\textwidth}{p{3.5cm}|Xp{7.8cm}}
			\toprule 
			\textbf{{SBSE Problem}}  & \multicolumn{1}{l}{\textbf{{Assumptions}}}& \multicolumn{1}{l}{\textbf{{Evaluation Methods}}}
			\\
			\midrule 
			
			Effort Estimation  & Not specified~\cite{Minku2013Software}~\cite{Sarro2016Multi} &  $DOE$,  $SSP$,  $CI$, $GD$, $HV$ \\ \hline
			\multirow{5}{*}{Project Scheduling} &	\cellcolor{steel!10} Not specified~\cite{1600-7617}~\cite{shen2018q}&\cellcolor{steel!10}  $SSP$, $DOE$, $Spread$, $GD$, $IGD$, $NFS$, $HV$ \\ 
			& \textbf{(P)} Prefer solutions that favor certain objectives using Analytic Hierarchy Process~\cite{Shen2016Dynamic} &\multirow{2}{*}{$DOE$, $CS$,  $SSP$, $GD$, $SP$, $Spread$, $HV$}\\ 
			&	\cellcolor{steel!10} \textbf{(P)} Prefer knee solutions~\cite{Sarro2017Adaptive}~\cite{Gueorguiev2009Software}~\cite{Ferrucci2013Not} &\cellcolor{steel!10}  $SSP$,  $CI$, $GD$, $DOE$, $HV$ \\ 
			& \textbf{(P)} Prefer widely distributed solutions~\cite{Chicano2011Using} & $AS$, $HV$ \\ \hline
			Requirement Assignment &	\cellcolor{steel!10} Not specified~\cite{Li2017Zen} &\cellcolor{steel!10}  $DOE$,  $SSP$, $HV$ \\ \hline
			\multirow{3}{*}{Next Release Problem}  &\multirow{2}{*}{Not specified~\cite{Finkelstein2009A}~\cite{Durillo2009A}~\cite{Harman2009Search}~\cite{Li2014Robust}~\cite{Zhang2013Empirical}~\cite{1600-7617}~\cite{zhang2018}} &  $SSP$, $DOE$, $AS$, $NFS$, $GD$,  $CI$, $Spread$, $HV$, \% of included requirements$^\star$ \\ 
			&	\cellcolor{steel!10} \textbf{(P)} Prefer extreme solutions~\cite{Zhang2007The} &\cellcolor{steel!10}  $SSP$\\ \hline
			\multirow{8}{*}{\parbox{3cm}{Software Modeling and Architecting}} &\multirow{2}{*}{Not specified~\cite{Ramirez2016Comparative}~\cite{Bowman2010Solving}~\cite{Simons2010Interactive}} &  $DOE$,  $SSP$, $SP$, $HV$, \% of within-range solutions$^\star$, \% of equivalent solutions$^\star$ \\ 
			&	\cellcolor{steel!10} \textbf{(P)} Prefer knee solutions~\cite{JournalIST}&\cellcolor{steel!10}average correction$^\star$, manual correction$^\star$, recall$^\star$, precision$^\star$ \\  
			& \textbf{(P)} Prefer solutions that favor certain objectives as ranked by users~\cite{Simons2012Elegant}&\multirow{2}{*}{$DOE$} \\  
			&	\cellcolor{steel!10} \textbf{(P)} Prefer solutions that meet preferences in, e.g., the requirement documentations or the goal model~\cite{Heaven2011Simulating}~\cite{busari2017radar}&\multirow{3}{*}{\cellcolor{steel!10}$SSP$}\\ \hline
			\multirow{5}{*}{Software Product Line} & Not specified~\cite{xiang2019} & $HV$, $IGD^+$, $SSP$\\ 
			&	\cellcolor{steel!10} \textbf{(C)} Prefer solutions that favor correctness objective over the others~\cite{Sayyad2013Optimum}~\cite{Sayyad2013Scalable}~\cite{Sayyad2013On}~\cite{Hierons2016SIP} \cite{Henard2015Combining}~\cite{Olaechea2014Comparison}~\cite{Xiang2018Configuring}\cite{Hierons2019Many}~\cite{1600-7617}~\cite{Guo2019}~\cite{lee2019}~\cite{saber2017IsSeeding}~\cite{lian2017} &\cellcolor{steel!10}  $CS$, $Spread$, $NFS$,  $SSP$, $\epsilon$-indicator, $IGD$, $HV$, $DOE$, number of required evaluations to find a valid solution$^\star$, full coverage ratio$^\star$, \% of valid solutions$^\star$ \\ 
			& \textbf{(P)} Prefer balanced solutions~\cite{Wang2015Cost}&  $DOE$,  $SSP$\\ \hline
			Library Recommendation  &	\cellcolor{steel!10} Not specified~\cite{Ouni2017Search} &\cellcolor{steel!10} $GD$, $Spread$, $HV$,  $SSP$, accuracy$^\star$, precision$^\star$, recall$^\star$\\ \hline
			Program Improvement  & \textbf{(C)} Prefer the program validity~\cite{White2011Evolutionary} &  $DOE$, $SSP$\\ \hline
			\multirow{7}{*}{Software Modularization} &	\cellcolor{steel!10} \textbf{(C)} Prefer solutions that favor modularization quality objective over the others~\cite{Kumari2016Hyper}~\cite{Bavota2012Putting}~\cite{Praditwong2011Software} &\multirow{3}{*}{\cellcolor{steel!10}$DOE$,  $SSP$, MoJoFM$^\star$} \\ 
			&\multirow{4}{*}{\textbf{(P)} Prefer knee solutions~\cite{Fleck2017Model}~\cite{Boukharata2019}}&  $DOE$, $IGD$, $HV$, precision$^\star$, recall$^\star$, manual precision$^\star$, difficulty to perform task by human$^\star$, possibility of manually fix the bug in solution by human$^\star$, possibility of manually adapt the solution by human$^\star$\\ \hline
			Code Smell Detection  &	\cellcolor{steel!10} Not specified~\cite{Mansoor2016Multi} &\cellcolor{steel!10} $IGD$, $HV$, precision$^\star$, recall$^\star$\\ \hline
			\multirow{2}{*}{Defect Prediction} &\multirow{2}{*}{Not specified~\cite{Canfora2015Defect}~\cite{Canfora2013Multi}~\cite{chen2017}~\cite{ni2019}} &  $SSP$, $HV$, $ACC^\star$, $P_{opt}^\star$, precision$^\star$, recall$^\star$, AUC$^\star$, cost of code inspection$^\star$\\ \hline
			\multirow{2}{*}{Test Case Prioritization} &	\multirow{2}{*}{\cellcolor{steel!10}Not specified~\cite{Assun2014A}~\cite{Panichella2015Improving}~\cite{Mariani2016}~\cite{Segura2016Multi}~\cite{Epitropakis2015Empirical}~\cite{REMAP}~\cite{issta18main}} &\cellcolor{steel!10}  $SSP$,  $DOE$, $CS$, $ED$, $Spread$, $GD$, $\epsilon$-indicator, $IGD$, $HV$, $NFS$, $APFD^\star$, \% of detected faults$^\star$\\ \hline
			\multirow{6}{*}{\parbox{3cm}{ White Box Test Case Generation}} & Not specified~\cite{Abdessalem2016Testing}~\cite{Yoo2010Using}~\cite{Mao2016Sapienz}~\cite{Ferrer2012Evolutionary}~\cite{main-sbse}~\cite{Panichella2015Reformulating} &  $DOE$,  $SSP$,  $CI$, $GD$, $NFS$, $HV$, $AS$, total coverage$^\star$ \\ 	
			& 	\cellcolor{steel!10} \textbf{(P)} Prefer solutions that favor certain objectives as ranked by users, i.e., reference point~\cite{jakubovskifilho2019}~\cite{Kalboussi2013Preference} &\multirow{3}{*}{\parbox{7.8cm}{\cellcolor{steel!10}$ED$, $SSP$, $R$-$HV$, Average number of solutions in the region of interest$^\star$}}\\ 
			& \textbf{(C)} Prefer solutions that favor coverage objective over the others~\cite{Zheng2016Multi} &  \multirow{2}{*}{$SSP$, $HV$}\\ \hline
			\multirow{3}{*}{\parbox{3cm}{ Black Box Test Case Generation}} &	\cellcolor{steel!10} Not specified~\cite{pradhan2019}~\cite{abdessalem2018} &\cellcolor{steel!10} $DOE$, $HV$, $GD$, $Spread$ \\ 
			& \textbf{(C)} Prefer solutions that favor coverage objective over the others~\cite{Shahbazi2016Black} &\multirow{2}{*}{$SSP$, p-measure$^\star$} \\ \hline
			\multirow{2}{*}{Resource Management} &	\cellcolor{steel!10} Not specified~\cite{Frey2013Search} &\cellcolor{steel!10} $IGD$, $HV$ \\ 
			& \textbf{(P)} Prefer knee solutions~\cite{Li2010SLA}~\cite{Chen2017Self}&   $SSP$, $CS$, $GD$, elasticity$^\star$\\ \hline
			\multirow{7}{*}{\parbox{3cm}{Software Configuration and Adaptation}} &	\cellcolor{steel!10} Not specified~\cite{Pascual2015Applying}~\cite{Gerasimou2016Search} &\cellcolor{steel!10}  $DOE$, $GD$, $\epsilon$-indicator, $IGD$, $HV$,  $SSP$ \\ 
			& \textbf{(P)} Prefer solutions that meet preferences form the natural descriptions from the stakeholders~\cite{Martens2010Automatically}~\cite{Letier2014Uncertainty}~\cite{Abdeen2014Multi}&  \multirow{3}{*}{$DOE$, $SSP$, expected value of total perfect information$^\star$}\\ 
			&	\cellcolor{steel!10} \textbf{(P)} Prefer knee solutions~\cite{Chen2018FEMOSAA}&\cellcolor{steel!10}   $SSP$, $ED$, $HV$, \% of valid solutions$^\star$\\ 
			& \textbf{(P)} Prefer robust solutions around a given region~\cite{Calinescu2017Designing}&  $SSP$, modified $\epsilon$-indicator and $IGD$ according to problem nature \\ \hline
			Program Manipulation  &	\cellcolor{steel!10} Not specified~\cite{Wu2015Deep} &\cellcolor{steel!10}  $DOE$, $AS$,  $CI$, $HV$,  $SSP$\\ \hline
			\multirow{3}{*}{Service Composition} & Not specified~\cite{Wagner2012Multi}~\cite{PubSub10849_Chen} &  $DOE$, $HV$ $\epsilon$-indicator \\ 
			&\multirow{2}{*}{\cellcolor{steel!10}\textbf{(P)} Prefer extreme solutions~\cite{Wada2012E}~\cite{Tan2018Evolutionary}}&\cellcolor{steel!10}  $DOE$, $SSP$, $IGD$, $HV$, coefficient of variation of objective values$^\star$\\ \hline
			Log Template Identification  & \textbf{(P)} Prefer knee solutions~\cite{ICPC-2018} & $SSP$, precision$^\star$, recall$^\star$, f-measure$^\star$\\ \hline
			Workflow Scheduling  &	\cellcolor{steel!10} Not specified~\cite{Durillo2014Multi} &\cellcolor{steel!10}  $DOE$,  $SSP$, $HV$\\ \hline
			\multirow{7}{*}{Software Refactoring} &\multirow{4}{*}{\parbox{5.5cm}{Not specified~\cite{Ouni2013Search}~\cite{Mkaouer2014High}~\cite{Ouni2013maintainability}~\cite{Ouni2012Search}~\cite{Ouni2013The} \cite{Ouni2016Multi}~\cite{Mansoor2015Multi}~\cite{Mkaouer2014Recommendation}}} &  $SSP$, $IGD$, precision$^\star$, recall$^\star$, defect correction ratio$^\star$, reused refactoring$^\star$, usefulness by human$^\star$, \% of fixed code smells$^\star$, code change score$^\star$, manual precision$^\star$, quality gains$^\star$, medium value of refactoring$^\star$ \\   
			&\multirow{3}{*}{\cellcolor{steel!10}\textbf{(P)} Prefer knee solutions~\cite{Mkaouer2016On}}&\cellcolor{steel!10}  $SSP$, $IGD$, precision$^\star$, recall$^\star$, manual precision$^\star$, quality gain$^\star$, defect correction ratio$^\star$, number of suggested refactoring$^\star$, usefulness by human$^\star$\\ \bottomrule
		\end{tabularx}
		 \begin{tablenotes}
      \item All problem specific indicators are listed in full and marked as $\star$.
      \item \textbf{(P)} denotes DM's preferences; \textbf{(C)} denotes contextual information. 
    \end{tablenotes}
	\end{table*}
	

%% file: Tables/qi-revisit.tex
	\begin{table*}[tbp]
		\caption{A summary of representative quality indicators used in SBSE, their usage note/caveats and applicable conditions.}
		\setlength\tabcolsep{2pt}
		\centering
		\begin{scriptsize}{
				\begin{tabular}{l|c|c|c|c|c|l|l}\toprule
					\textbf{Indicator}&\textbf{Convergence}&\textbf{Spread}&\textbf{Uniformity}& \textbf{Cardinality}&\textbf{Pareto Compliant}&\textbf{Usage Note/Caveats} & \textbf{Applicable Conditions} \\\midrule
					
					\multirow{7}{*}{$CI$} & \multirow{7}{*}{$\bm{-}$} & & & \multirow{7}{*}{$\bm{-}$} & \multirow{7}{*}{$\bm{+}$} & \multicolumn{1}{p{4.5cm}|}{(i) not able to distinguish between sets if their solutions are nondominated to each other, which may happen frequently in many-objective optimization;
						
						(ii) binary indicator which evaluates relative quality of two sets and cannot be converted into a unary indicator.} 
					& \multicolumn{1}{p{4.5cm}@{}}{(i) when the DM wants to know the relative quality difference (in terms of dominance relation) between two sets, and
						(ii) when the Pareto front size is relatively small, e.g., on some low-dimensional combinatorial problems.} \\
					
					\rowcolor{steel!10}\multirow{9}{*}{$\mathcal{C}$ ($CS$)} & \multirow{9}{*}{$\bm{-}$} & & & \multirow{9}{*}{$\bm{-}$} & \multirow{9}{*}{$\bm{+}$} & \multicolumn{1}{p{4.5cm}|}{(i) not able to distinguish between sets if their solutions are nondominated to each other, which may happen frequently in many-objective optimization;
						
						(ii) binary indicator which evaluates relative quality of two sets and cannot be converted into a unary indicator;
						
						(iii) removing duplicate solutions before the calculation.} 
					& \multicolumn{1}{p{4.5cm}@{}}{(i) when the DM wants to know the relative quality difference (in terms of dominance relation) between two sets, and
						(ii) when the Pareto front size is relatively small, e.g., on some low-dimensional combinatorial problems.} \\

					\multirow{8}{*}{$GD$}& \multirow{8}{*}{$\bm{+}$} & & & & & \multicolumn{1}{p{4.5cm}|}{(i) additional problem knowledge: a reference set that represents the Pareto front (not necessarily a set of uniformly-distributed points);
						
						(ii) each objective needs normalization;
						
						(iii) may give misleading results due to not holding the Pareto compliance property.}
					& \multicolumn{1}{p{4.5cm}@{}}{(i) when the DM wants to know how close the obtained sets from the Pareto front,
						(ii) when the compared sets are nondominated to each other (i.e., no \textit{better} relation between the sets), and
						(iii) when the Pareto front range can be estimated properly (e.g., no DRS points in the reference set~\cite{li2019quality}).} \\

					\rowcolor{steel!10}\multirow{5}{*}{$GD^+$}& \multirow{5}{*}{$\bm{+}$} & & & & \multirow{5}{*}{$\bm{+}$} & \multicolumn{1}{p{4.5cm}|}{(i) additional problem knowledge: a reference set that represents the Pareto front (not necessarily a set of uniformly-distributed points);
						
						(ii) each objective needs normalization.} 
					& \multicolumn{1}{p{4.5cm}@{}}{(i) when the DM wants to know how close the obtained sets from the Pareto front.} \\

					\multirow{6}{*}{$Spread$}& & \multirow{6}{*}{$\bm{+}$} & \multirow{6}{*}{$\bm{+}$} & & & \multicolumn{1}{p{4.5cm}|}{(i) additional problem knowledge: extreme points of the Pareto front;
						
						(ii) each objective needs normalization;
						
						(iii) reliable only on bi-objective problems.} 
					& \multicolumn{1}{p{4.5cm}@{}}{(i) when the DM wants to know the diversity (including both spread and uniformity) of the obtained sets on bi-objective problems, and
						(ii) when the compared sets are nondominated to each other.} \\

					\rowcolor{steel!10}\multirow{6}{*}{$DCI$} & & \multirow{6}{*}{$\bm{+}$} & \multirow{6}{*}{$\bm{-}$} & \multirow{6}{*}{$\bm{-}$} & \multirow{6}{*}{$\bm{-}$} & \multicolumn{1}{p{4.5cm}|}{(i) additional problem knowledge: proper setting of the grid division;
						
						(ii) $M$-nary indicator that evaluates relative quality of $M$ sets, but can be converted into a unary one by comparing the obtained set with the Pareto front.} 
					& \multicolumn{1}{p{4.5cm}@{}}{(i) when the DM wants to know the diversity of the obtained sets.} \\

					\multirow{4}{*}{$SP$}& & & \multirow{4}{*}{$\bm{+}$} & & & \multicolumn{1}{p{4.5cm}|}{(i) each objective needs normalization;
						
						(ii) cannot reflect the spread of solution sets.} 
					& \multicolumn{1}{p{4.5cm}@{}}{(i) when the DM wants to know the uniformity of the obtained sets, and
						(ii) when the compared sets are nondominated to each other.} \\

					\rowcolor{steel!10}\multirow{3}{*}{$NFS$} & & & & \multirow{3}{*}{$\bm{+}$} & & \multicolumn{1}{p{4.5cm}|}{(i) not able to compare sets as it only counts the number of nondominated solutions in a set.} 
					& \multicolumn{1}{p{4.5cm}@{}}{(i) not reliable when the DM wants to compare sets.} \\

					\multirow{4}{*}{$UNFR$} & & & & \multirow{4}{*}{$\bm{+}$} & \multirow{4}{*}{$\bm{+}$} & \multicolumn{1}{p{4.5cm}|}{None} 
					& \multicolumn{1}{p{4.5cm}@{}}{(i) when the DM wants to compare the cardinality of sets, particularly how much they contribute the combined nondominated front.} \\

					\rowcolor{steel!10}\multirow{7}{*}{$IGD$}& \multirow{7}{*}{$\bm{+}$} & \multirow{7}{*}{$\bm{+}$} & \multirow{7}{*}{$\bm{-}$} & \multirow{7}{*}{$\bm{-}$} & & \multicolumn{1}{p{4.5cm}|}{(i) additional problem knowledge: a reference set that well represents the Pareto front (i.e., densely and uniformly distributed points);
						
						(ii) each objective needs normalization;
						
						(iii) may give misleading results due to the lack of Pareto compliance property.} 
					& \multicolumn{1}{p{4.5cm}@{}}{(i) when the DM wants to know how well the obtained sets can represent the Pareto front,
						(ii) when the compared sets are nondominated to each other, and 
						(iii) when there is a Pareto front with densely and uniformly distributed points.} \\

					\multirow{8}{*}{$HV$}& \multirow{8}{*}{$\bm{+}$} & \multirow{8}{*}{$\bm{+}$} & \multirow{8}{*}{$\bm{-}$} & \multirow{8}{*}{$\bm{+}$} & \multirow{8}{*}{$\bm{+}$} & \multicolumn{1}{p{4.5cm}|}{(i) additional problem knowledge: a reference point that worse than the nadir point of the Pareto front;
						
						(ii) exponentially increasing computational cost in objective dimensionality.
						
						(iii) the DM can specify the reference point according to their preference to extreme solutions or to inner ones.} 
					& \multicolumn{1}{p{4.5cm}@{}}{(i) when the DM wants to know comprehensive quality of the obtained sets, especially suitable if the DM prefers knee points of the problem, and
						(ii) when the objective dimensionality is not very high.} \\

					\rowcolor{steel!10}\multirow{6}{*}{$\epsilon$-indicator}& \multirow{6}{*}{$\bm{+}$} & \multirow{6}{*}{$\bm{+}$} & \multirow{6}{*}{$\bm{-}$} & \multirow{6}{*}{$\bm{-}$} & \multirow{6}{*}{$\bm{+}$} & \multicolumn{1}{p{4.5cm}|}{(i) each objective needs normalization;
						
						(ii) binary indicator, but can be converted into a unary indicator by comparing the obtained set with the Pareto front;
						
						(iii) differently-performed sets may have the same/similar evaluation results.} & \multicolumn{1}{p{4.5cm}@{}}{(i) when the DM wants to know maximum difference between two solution sets (or the obtained solution set from the Pareto front).} \\\bottomrule
					
				\end{tabular}
					\begin{tablenotes}
   \item 
			``$\bm{+}$'' generally means that the indicator can well reflect the specified quality (or meet the specified property). 
			``$\bm{-}$'' for convergence means that the indicator can reflect the convergence of a set to some extent; 
			e.g., indicators only considering the dominance relation as convergence measure.
			``$\bm{-}$'' for spread means that the indicator can only reflect the extensity of a set.
			``$\bm{-}$'' for uniformity means that the indicator can reflect the uniformity of a set to some extent; 
			i.e., a disturbance to an equally-spaced set may not certainly lead to a worse evaluation result.
			``$\bm{-}$'' for cardinality means that adding a nondominated solution into a set is not surely 
			but likely to lead to a better evaluation result and also it never leads to a worse evaluation result.
			``$\bm{-}$'' for Pareto compliance means that the indicator holds the property subject to certain conditions.
  \end{tablenotes}
		}\end{scriptsize}	
		\label{Table:Summary}
	\end{table*}

%% file: Tables/all-qi-1.tex
\begin{table*}[p!]
\caption*{TABLE A1: What and how the evaluation methods are used in each primary study.}
\label{tb:all-qi}
\setlength{\tabcolsep}{4.7pt}
\centering
\begin{tabular}{llllll}\toprule

\textbf{Study}&\textbf{Indicator/Method}&\textbf{Stated Quality Aspects to Measure}&\textbf{\# Objectives}&\textbf{Reference Point}&\textbf{Reference Front}\\ 
\midrule
\cite{Minku2013Software}&SSP, DOE&N/A&3&N/A&N/A\\

\rowcolor{steel!10}\cite{Sarro2016Multi}&HV, CI, GD&Unknown&2-5&Nadir point&Best Pareto front found\\

\cite{shen2018q}&DOE, HV, IGD, SP, Spread, SSP&\makecell[l]{HV=$Q_1\cup Q_2$, IGD=$Q_1\cup Q_2\cup Q_3$,\\SP=$Q_3$, Spread=$Q_2\cup Q_3$}&5&Worst values&Best Pareto front found\\

\rowcolor{steel!10}\cite{1600-7617}&GD, GS, NFS, HV&\makecell[l]{GD=$Q_1$, GS=$Q_2\cup Q_3$,\\NFS=$Q_2\cup Q_3$, HV=$Q_1\cup Q_2\cup Q_3$~~~~}&2&Unknown&Unknown\\

\cite{Finkelstein2009A}&SSP&N/A&4&N/A&N/A\\

\rowcolor{steel!10}\cite{zhang2018}&CS, CI, GD, HV, Spread&\makecell[l]{CS=$Q_1$, CI=$Q_1$, GD=$Q_1$\\ HV=Unknown, Spread=$Q_2\cup Q_3$~~~~~~~}&2&Worst values&Best Pareto front found\\

\cite{Durillo2009A}&PSI, HV, Spread, SSP, NFS&\makecell[l]{HV=$Q_1\cup Q_2\cup Q_3$\\  Spread=$Q_2$, NFS=Unknown}&2&Worst values&N/A\\

\rowcolor{steel!10}\cite{Harman2009Search}&GD, SSP&Unknown&2&N/A&Unknown\\

\cite{Li2014Robust}&GD, SSP, DOE&Unknown&3&N/A&Unknown\\

\rowcolor{steel!10}\cite{Zhang2013Empirical}&SSP, GD, AS, Spread, NFS&\makecell[l]{GD=$Q_1$, AS=Unknown\\Spread=$Q_2\cup Q_3$, NFS=Unknown~~~~~~}&2&N/A&Unknown\\

\cite{Li2017Zen}&HV, DOE, SSP&HV=$Q_1\cup Q_3$&3&Worst values&N/A\\

\rowcolor{steel!10}\cite{Ramirez2016Comparative}&HV, SP&HV=$Q_1$, SP=$Q_2\cup Q_3$&2-9&Boundary&N/A\\

\cite{Bowman2010Solving}&PSI, DOE, SSP&N/A&4&N/A&N/A\\

\rowcolor{steel!10}\cite{Simons2010Interactive}&DOE, SSP&N/A&3&N/A&N/A\\

\cite{xiang2019}&HV, IGD$+$, SSP&\makecell[l]{HV=$Q_1\cup Q_2\cup Q_3$\\IGD$+$=$Q_1\cup Q_2\cup Q_3$}&2-4&Unknown&Unknown\\

\rowcolor{steel!10}\cite{Ouni2017Search}&PSI, HV, Spread, GD, SSP&\makecell[l]{HV=$Q_1\cup Q_2\cup Q_3$\\Spread=$Q_2\cup Q_3$, GD=$Q_1$~~~~~~~~~~~~~~~~~~}\hfill&3&Nadir point&Best Pareto front found\\

\cite{White2011Evolutionary}&DOE, SSP&N/A&2&N/A&N/A\\

\rowcolor{steel!10}\cite{Frey2013Search}&HV, IGD&HV=$Q_1\cup Q_2$, IGD=$Q_1\cup Q_2$&3&Nadir point&Best Pareto front found\\

\cite{Pascual2015Applying}&GD, HV, DOE&Unknown&3&Worst values&Best Pareto front found\\

\rowcolor{steel!10}\cite{Gerasimou2016Search}&$\epsilon$-indicator, HV, IGD, SSP&\makecell[l]{$\epsilon$-indicator=$Q_1$ \\ HV=$Q_1\cup Q_2 \cup Q_3$,\\ IGD=$Q_1\cup Q_2 \cup Q_3$~~~~~~~~~~~~~~~~~~~~~~~~~~~~}&3&Nadir point&Best Pareto front found\\

\cite{Wu2015Deep}&AS, HV, CI, DOE, SSP&Unknown&3&Nadir point&N/A\\

\rowcolor{steel!10}\cite{Wagner2012Multi}&$\epsilon$-indicator, DOE&Unknown&3&N/A&Unknown\\

\cite{PubSub10849_Chen}&HV, DOE&HV=$Q_1\cup Q_2 \cup Q_3$&3&Worst values&N/A\\

\rowcolor{steel!10}\cite{Durillo2014Multi}&HV, DOE, SSP&Unknown&3&Boundary&N/A\\

\cite{Ouni2013Search}&PSI&N/A&3&N/A&N/A\\

\rowcolor{steel!10}\cite{Mkaouer2014High}&PSI, SSP, IGD&IGD=$Q_1\cup Q_2 \cup Q_3$&15&N/A&Best Pareto front found\\

\cite{Ouni2013maintainability}&PSI, SSP&N/A&2&N/A&N/A\\

\rowcolor{steel!10}\cite{Ouni2012Search}&PSI, SSP&N/A&2&N/A&N/A\\

\cite{Ouni2013The}&PSI&N/A&3&N/A&N/A\\

\rowcolor{steel!10}\cite{Ouni2016Multi}&PSI, SSP&N/A&4&N/A&N/A\\

\cite{Mansoor2015Multi}&PSI&N/A&3&N/A&N/A\\

\rowcolor{steel!10}\cite{Mkaouer2014Recommendation}&PSI&N/A&3&N/A&N/A\\

\cite{Mansoor2016Multi}&PSI, HV, IGD&IGD=$Q_1$, HV=Unknown&2&Unknown&Best Pareto front found\\

\rowcolor{steel!10}\cite{Canfora2015Defect}&PSI, SSP&N/A&2&N/A&N/A\\

\cite{chen2017}&PSI, HV&HV=$Q_1\cup Q_2 \cup Q_3$&2&Unknown&N/A\\

\rowcolor{steel!10}\cite{Canfora2013Multi}&PSI, SSP&N/A&2&N/A&N/A\\

\cite{ni2019}&SSP, PSI, HV&HV=$Q_1\cup Q_2 \cup Q_3$&2&Unknown&N/A\\

\rowcolor{steel!10}\cite{Abdessalem2016Testing}&HV, GD&Unknown&3&Unknown&Best Pareto front found\\

\textcolor{black}{\cite{main-sbse}}&PSI&N/A&$>$100&N/A&N/A\\

\cellcolor{steel!10} [82]&\cellcolor{steel!10}DOE&\cellcolor{steel!10}N/A&\cellcolor{steel!10}3&\cellcolor{steel!10}N/A&\cellcolor{steel!10}N/A\\

\cite{Ferrer2012Evolutionary}&DOE, HV, AS&Unknown&2&Worst values&N/A\\

\rowcolor{steel!10}\cite{Yoo2010Using}&SSP, NFS, CI&Unknown&2-3&N/A&N/A\\

\cite{pradhan2019}&HV,DOE&HV=$Q_1\cup Q_2 \cup Q_3$&4&Unknown&N/A\\

\rowcolor{steel!10}\cite{abdessalem2018}&HV, GD, Spread, PSI&\makecell[l]{HV=Unknown,\\ GD=Unknown, Spread=$Q_2$~~~~~~~~~~~~~~~~}&3&Unknown&Best Pareto front found\\

\cite{Assun2014A}&SSP, CS, ED, GD, IGD&\makecell[l]{CS=Unknown, ED=Unknown\\ GD=$Q_1\cup Q_2 \cup Q_3$, \\ IGD=$Q_1\cup Q_2 \cup Q_3$}&2&N/A&Best Pareto front found\\

\rowcolor{steel!10}\cite{issta18main}&PSI, GD, HV, Spread&\makecell[l]{GD=Unknown, HV=Unknown,~~~~~~~~~~\\Spread=$Q_2$}&3&Unknown&Unknown\\

\cite{Panichella2015Improving}&NFS, HV, SSP&Unknown&2-3&Nadir point&N/A\\

\rowcolor{steel!10}\cite{Mariani2016}&HV, SSP&Unknown&2&Unknown&N/A\\

\cite{parejo2016multi}&PSI, HV, SSP&Unknown&2-3&Unknown&N/A\\

\rowcolor{steel!10}\cite{Epitropakis2015Empirical}&PSI, $\epsilon$-indicator, IGD, HV&\makecell[l]{$\epsilon$-indicator=$Q_1$, \\ IGD=$Q_1$, HV=$Q_2 \cup Q_3$~~~~~~~~~~~~~~~~~~~~~~~}&3&Unknown&Best Pareto front found\\

\cite{REMAP}&PSI&N/A&2&N/A&N/A\\

\rowcolor{steel!10}\cite{Heaven2011Simulating}&SSP&N/A&2&N/A&N/A\\

\cite{busari2017radar}&SSP&N/A&2&N/A&N/A\\

\rowcolor{steel!10}\cite{Martens2010Automatically}&SSP&N/A&2-3&N/A&N/A\\

\cite{Letier2014Uncertainty}&PSI, SSP&N/A&2&N/A&N/A\\

\rowcolor{steel!10}\cite{Abdeen2014Multi}&SSP, DOE&N/A&2&N/A&N/A\\

\bottomrule
\end{tabular}
\end{table*}

%% file: Tables/all-qi-2.tex
\begin{table*}[t!]
\caption*{TABLE A1: What and how the evaluation methods are used in each primary study (continue).}

\setlength{\tabcolsep}{4.7pt}
\centering

\begin{tabular}{llllll}\toprule

\textbf{Study}&\textbf{Indicator/Method}&\textbf{Stated Quality Aspects to Measure}&\textbf{\# Objectives}&\textbf{Reference Point}&\textbf{Reference Front}\\ 
\midrule


\cite{Shen2016Dynamic}&DOE, SSP, HV, GD, SP, Spread, CS&\makecell[l]{HV=$Q_1\cup Q_2$, GD=$Q_1$,\\SP=$Q_3$, Spread=$Q_2\cup Q_3$\\ CS=Unknown}&4&Worst values&Best Pareto front found\\

\rowcolor{steel!10}\cite{Simons2012Elegant}&DOE&N/A&4&N/A&N/A\\

\cite{Xiang2018Configuring}&SSP, PSI, HV, IGD&\makecell[l]{HV=$Q_1\cup Q_2\cup Q_3$, \\ IGD=$Q_1\cup Q_2\cup Q_3$}&4&Boundary&Best Pareto front found\\

\rowcolor{steel!10}\cite{lee2019}&DOE, HV, PSI, SSP&Unknown&2&Unknown&N/A\\

\cite{Hierons2019Many}&SSP, HV, PSI&\makecell[l]{HV=$Q_1\cup Q_2\cup Q_3$}&9&Boundary&N/A\\

\rowcolor{steel!10}\cite{lian2017}&PSI, $\epsilon$-indicator, IGD, Spread, HV&\makecell[l]{$\epsilon$-indicator=$Q_1$, IGD=$Q_1$\\Spread=$Q_2\cup Q_3$, HV=$Q_2\cup Q_3\;\;\;\;\;\;\;$}&5&Unknown&Unknown\\

\cite{Sayyad2013Optimum}&HV, DOE, Spread, PSI&\makecell[l]{HV=Unknown, Spread=$Q_2$}&5&Unknown&N/A\\

\rowcolor{steel!10}\cite{Sayyad2013Scalable}&HV, PSI&Unknown&5&Unknown&N/A\\

\cite{Sayyad2013On}&HV, Spread&\makecell[l]{HV=Unknown,\\Spread=$Q_2$}&2-5&Unknown&N/A\\

\rowcolor{steel!10}\cite{Hierons2016SIP}&HV, PSI&HV=$Q_1\cup Q_2\cup Q_3$&5-9&Boundary&N/A\\

\cite{Henard2015Combining}&HV, IGD, $\epsilon$-indicator, NFS, Spread&\makecell[l]{HV=$Q_1$, IGD=$Q_1$,\\ $\epsilon$-indicator=$Q_1$, \\ NFS=$Q_2\cup Q_3$, Spread=$Q_2\cup Q_3$}&5&Nadir point&Best Pareto front found\\

\rowcolor{steel!10}\cite{Olaechea2014Comparison}&CS, HV&Unknown&2&Worst values&N/A\\

\cite{saber2017IsSeeding}&HV, IGD, $\epsilon$-indicator, NFS, Spread&\makecell[l]{HV=$Q_1$, IGD=$Q_1$,\\ $\epsilon$-indicator=$Q_1$, \\ NFS=$Q_2\cup Q_3$, Spread=$Q_2\cup Q_3$}&5&Nadir point&Best Pareto front found\\

\rowcolor{steel!10}\cite{Guo2019}&HV, IGD, $\epsilon$-indicator, ER, Spread&\makecell[l]{HV=$Q_1\cup Q_2\cup Q_3$, \\IGD=$Q_1\cup Q_2\cup Q_3$,\\ $\epsilon$-indicator=$Q_1\cup Q_2\cup Q_3$,~~~~~~~~~~~~~~~ \\ ER=$Q_1$, Spread=$Q_2\cup Q_3$}&5&Unknown&Unknown\\

\cite{Kumari2016Hyper}&DOE&N/A&5&N/A&N/A\\

\rowcolor{steel!10}\cite{Bavota2012Putting}&PSI, DOE&N/A&3&N/A&N/A\\

\cite{Praditwong2011Software}&DOE&N/A&5&N/A&N/A\\

\rowcolor{steel!10}\cite{jakubovskifilho2019}&HV, PSI, ED&Unknown&3&Nadir point&N/A\\

\cite{Zheng2016Multi}&HV, SSP&HV=$Q_1\cup Q_2\cup Q_3$&2-4&Boundary&N/A\\

\rowcolor{steel!10}\textcolor{black}{\cite{Panichella2015Reformulating}}&PSI&N/A&$>$100&N/A&N/A\\

\cite{Kalboussi2013Preference}&DOE, SSP&N/A&7&N/A&N/A\\

\rowcolor{steel!10}\cite{Shahbazi2016Black}&SSP, PSI&N/A&2&N/A&N/A\\

\cite{Chicano2011Using}&AS, HV&Unknown&3&Unknown&N/A\\

\rowcolor{steel!10}\cite{Sarro2017Adaptive}&DOE, CI, HV, GD&Unknown&3&Unknown&Best Pareto front found\\

\cite{Gueorguiev2009Software}&SSP&N/A&3&N/A&N/A\\

\rowcolor{steel!10}\cite{Ferrucci2013Not}&CI, HV, GD&Unknown&3&Unknown&Best Pareto front found\\

\cite{Zhang2007The}&SSP&N/A&2&N/A&N/A\\

\rowcolor{steel!10}\cite{JournalIST}&PSI&N/A&3&N/A&N/A\\

\cite{Wang2015Cost}&DOE, SSP&N/A&5&N/A&N/A\\

\rowcolor{steel!10}\cite{Boukharata2019}&DOE, PSI&N/A&9&N/A&N/A\\

\cite{Fleck2017Model}&DOE, HV, IGD&\makecell[l]{HV=$Q_1\cup Q_2\cup Q_3$, \\ IGD=$Q_1\cup Q_2\cup Q_3$}&4&Unknown&Best Pareto front found\\

\rowcolor{steel!10}\cite{Li2010SLA}&SSP&N/A&3&N/A&N/A\\

\cite{Chen2017Self}&PSI, CS, GD&Unknown&5&N/A&Unknown\\

\rowcolor{steel!10}\cite{Chen2018FEMOSAA}&HV, ED, SSP, PSI&Unknown&2&Unknown&N/A\\

\cite{Calinescu2017Designing}&SSP, $\epsilon$-indicator, IGD&Unknown&2-3&N/A&Best Pareto front found\\

\rowcolor{steel!10}\cite{Wada2012E}&HV, DOE, PSI&Unknown&3&Unknown&N/A\\

\cite{Tan2018Evolutionary}&HV, IGD, SSP&Unknown&2&Worst values&Best Pareto front found\\

\rowcolor{steel!10}\cite{ICPC-2018}&SSP, PSI&N/A&2&N/A&N/A\\

\cite{Mkaouer2016On}&SSP, IGD, PSI&IGD=$Q_1\cup Q_2\cup Q_3$&8&Unknown&Best Pareto front found\\

\bottomrule
\end{tabular}
 \begin{tablenotes}
    \footnotesize
    \item $Q_1$=Convergence; $Q_2$=Spread; $Q_3$=Uniformity; $Q_4$=Cardinality.
    \end{tablenotes}
\end{table*}